\documentclass[9pt]{article}
\usepackage{amssymb}
\usepackage{tikz}
\usetikzlibrary{arrows, calc, decorations.pathmorphing, backgrounds, positioning, fit, petri, automata}
\definecolor{grey1}{rgb}{0.5,0.5,0.5}
\usepackage[top=2.54cm, bottom=2.54cm, left=2.2cm, right=2.2cm]{geometry}
\usepackage{algorithmicx}
\usepackage{algcompatible}
\usepackage{algorithm, algpseudocode}

\usepackage{changebar}

\usepackage{hyperref}
\hypersetup{ 
  colorlinks,
  citecolor=blue,
  linkcolor=red,
  urlcolor=magenta}

\usepackage{multirow}

\usepackage{mathtools}
\usepackage{graphicx}
\usepackage{subfigure}
\usepackage{amsmath,amsthm,bm,bbm}
\usepackage{rotating}
\usepackage{mathrsfs}
\usepackage{chngcntr}
\usepackage{apptools}
\usepackage[titletoc,title]{appendix}
\usepackage{comment}

\usepackage{xcolor}         
\definecolor{grau}{rgb}{0.8,0.8,0.8}

\newcommand{\chen}[1]{\color{orange}}
\usepackage{setspace,lscape,longtable}
\usepackage{amsmath,amsthm,bm}
\usepackage{fullpage}  
\usepackage[shortlabels]{enumitem}
\usepackage{titlesec}
\usepackage[english]{babel}
\numberwithin{equation}{section}
\numberwithin{equation}{section}

\newtheorem{theorem}{Theorem}[section]

\newtheorem{lemma}[theorem]{Lemma}
\newtheorem{proposition}{Proposition}[section]
\theoremstyle{remark}
\newtheorem{remark}{Remark}
\newtheorem{definition}[theorem]{Definition}
\newtheorem{assumption}{Assumption}

\newtheorem{result}{Result}[section]

\usepackage{cleveref}

\crefname{assumption}{Assumption}{Assumptions}

\usepackage{iftex}
\ifPDFTeX
  \usepackage[T1]{fontenc}
  \usepackage[utf8]{inputenc}
  \usepackage{textcomp} 
\else 
  \usepackage{unicode-math}
  \defaultfontfeatures{Scale=MatchLowercase}
  \defaultfontfeatures[\rmfamily]{Ligatures=TeX,Scale=1}
\fi
\usepackage{lmodern}
\ifPDFTeX\else  
\fi
\IfFileExists{upquote.sty}{\usepackage{upquote}}{}
\IfFileExists{microtype.sty}{
  \usepackage[]{microtype}
  \UseMicrotypeSet[protrusion]{basicmath} 
}{}
\makeatletter
\@ifundefined{KOMAClassName}{
  \IfFileExists{parskip.sty}{%
    \usepackage{parskip}
  }{
    \setlength{\parindent}{0pt}
    \setlength{\parskip}{6pt plus 2pt minus 1pt}}
}{
  \KOMAoptions{parskip=half}}
\makeatother
\usepackage{xcolor}
\makeatletter
\ifx\paragraph\undefined\else
  \let\oldparagraph\paragraph
  \renewcommand{\paragraph}{
    \@ifstar
      \xxxParagraphStar
      \xxxParagraphNoStar
  }
  \newcommand{\xxxParagraphStar}[1]{\oldparagraph*{#1}\mbox{}}
  \newcommand{\xxxParagraphNoStar}[1]{\oldparagraph{#1}\mbox{}}
\fi
\ifx\subparagraph\undefined\else
  \let\oldsubparagraph\subparagraph
  \renewcommand{\subparagraph}{
    \@ifstar
      \xxxSubParagraphStar
      \xxxSubParagraphNoStar
  }
  \newcommand{\xxxSubParagraphStar}[1]{\oldsubparagraph*{#1}\mbox{}}
  \newcommand{\xxxSubParagraphNoStar}[1]{\oldsubparagraph{#1}\mbox{}}
\fi
\makeatother

\usepackage{longtable,booktabs,array}
\usepackage{calc} 
\usepackage{etoolbox}
\makeatletter
\patchcmd\longtable{\par}{\if@noskipsec\mbox{}\fi\par}{}{}
\makeatother
\IfFileExists{footnotehyper.sty}{\usepackage{footnotehyper}}{\usepackage{footnote}}
\makesavenoteenv{longtable}
\usepackage{graphicx}
\makeatletter
\def\maxwidth{\ifdim\Gin@nat@width>\linewidth\linewidth\else\Gin@nat@width\fi}
\def\maxheight{\ifdim\Gin@nat@height>\textheight\textheight\else\Gin@nat@height\fi}
\makeatother
\setkeys{Gin}{width=\maxwidth,height=\maxheight,keepaspectratio}
\makeatletter
\def\fps@figure{htbp}
\makeatother

\ifLuaTeX
  \usepackage{selnolig}  
\fi
\usepackage[]{natbib}
\usepackage{bookmark}

\newcommand{\RR}{\mathbb{R}}

\newcommand{\bds}{\boldsymbol}

\newcommand{\E}{\mathbb{E}}

\newcommand{\vertiii}[1]{{\vert\kern-0.25ex\vert\kern-0.25ex\vert #1 
    \vert\kern-0.25ex\vert\kern-0.25ex\vert}}

\newcommand{\F}{\mathrm{F}}

\newcommand{\prob}{{\mathbb{P}}}

\newcommand{\iid}{{\mathrm{i.i.d.}}}
\newcommand{\iidsim}{{\overset{\mathrm{i.i.d.}}{\sim}}}
\newcommand{\transpose}{^{\mathrm{T}}}

\newcommand{\calI}{{\mathcal{I}}}

\newcommand{\calL}{{\mathcal{L}}}

\newcommand{\bT}{{\mathbf{T}}}

\newcommand{\bP}{{\mathbf{P}}}

\newcommand{\bY}{{\mathbf{Y}}}
\newcommand{\bv}{{\mathbf{v}}}
\newcommand{\bx}{{\mathbf{x}}}
\newcommand{\bX}{{\mathbf{X}}}
\newcommand{\bA}{{\mathbf{A}}}
\newcommand{\bB}{{\mathbf{B}}}
\newcommand{\bC}{{\mathbf{C}}}
\newcommand{\bD}{{\mathbf{D}}}
\newcommand{\bE}{{\mathbf{E}}}

\newcommand{\bG}{{\mathbf{G}}}

\newcommand{\bI}{{\mathbf{I}}}

\newcommand{\bL}{{\mathbf{L}}}

\newcommand{\bR}{{\mathbf{R}}}
\newcommand{\bQ}{{\mathbf{Q}}}
\newcommand{\bS}{{\mathbf{S}}}
\newcommand{\bW}{{\mathbf{W}}}
\newcommand{\bZ}{{\mathbf{Z}}}

\newcommand{\br}{{\mathbf{r}}}

\newcommand{\bu}{{\mathbf{u}}}
\newcommand{\bs}{{\mathbf{s}}}
\newcommand{\by}{{\mathbf{y}}}
\newcommand{\bz}{{\mathbf{z}}}
\newcommand{\xb}{{\mathbf{x}}}

\newcommand{\be}{{\mathbf{e}}}
\newcommand{\bb}{{\mathbf{b}}}

\newcommand{\eye}{{\mathbf{I}}}

\newcommand{\zero}{{\bm{0}}}
\newcommand{\eps}{\epsilon}

\newcommand{\Var}{\mathrm{Var}}
\newcommand{\tOp}{\widetilde O_p}
\newcommand{\I}{\mathbf I_{d_+,d_-}}

\DeclareUnicodeCharacter{2212}{-}

\begin{document}

{
  \title{\bf Estimating contextual networks via Fisher-scoring surrogate likelihood}
  \author{Myung Jin Lee
    and 
    Fangzheng Xie 
    \thanks{The authors gratefully acknowledge the support from NSF grant DMS 2610696. This research was also supported in part by Lilly Endowment, Inc., through its support for the Indiana University Pervasive Technology Institute.}
    \thanks{Correspondence should be addressed to \href{mailto:fxie@iu.edu}{fxie@iu.edu}.}
    \\
    Department of Statistics, Indiana University}
  \maketitle
}

\maketitle

\allowdisplaybreaks
\begin{abstract}
When networks are accompanied by vertex-level covariates, effectively integrating and aligning the information underlying both data sources becomes a natural yet critical task. Current spectral approaches, despite their popularity, may yield suboptimal results and necessitate the specification of additional tuning parameters that are contingent upon unobservable signal-to-noise ratios. We propose a tuning-free, likelihood-based learning framework for estimating latent positions in contextual generalized random dot product graphs {\color{black}in the presence of high-dimensional vertex covariates with a low-rank structure}. Our method constructs a smooth surrogate likelihood and implements a Fisher-scoring algorithm {\color{black}to iteratively estimate} latent positions without introducing additional tuning parameters {\color{black}that depend on the network sparsity}. We establish the consistency and asymptotic normality of the Fisher-scoring estimator. Application to real-world data further demonstrates the method’s practical utility, offering a principled and scalable alternative to spectral approaches for analyzing contextual networks in high-dimensional settings.
\end{abstract}

\noindent%

\tableofcontents

\section{Introduction}\label{sec:introduction}
Statistical inference for network data has become a central topic in modern data science, as relational structures arise naturally across a wide range of scientific domains. In social sciences, network models are used to study social interactions, community formation, and diffusion processes \citep{hoff2002latent, girvan2002community}. In neuroscience, brain connectivity networks derived from fMRI or diffusion imaging data are analyzed to understand functional and structural organization of the brain \citep{bullmore2009complex}. In genomics and systems biology, gene co-expression and protein–protein interaction networks provide insights into regulatory mechanisms \citep{barabasi2004network}. In computer science and information retrieval, large-scale networks such as citation graphs, recommendation systems, and online social platforms are central objects of study \citep{newman2018networks}. Similar network structures arise in epidemiology (contact networks), international trade (country-to-country trade flows), economics (financial contagion networks), ecology (food webs), and communication systems.

To analyze such relational data, spectral methods have emerged as a dominant paradigm {\color{black}because of} their computational scalability and strong theoretical guarantees. In particular, adjacency and Laplacian spectral embeddings provide consistent estimators of latent positions under random dot product graph and stochastic block model frameworks \citep{athreya2016limit, rohe2011spectral, sarkar2015role, sussman2012consistent,  tang2017semiparametric, tang2018limit}. More recently, extensions that integrate Bernoulli-likelihood information have been developed to improve statistical efficiency. These include one-step estimators (OSE) \citep{xie2023efficient} and surrogate likelihood estimators \citep{wu2025statistical}, which aim to achieve asymptotic efficiency by refining spectral starts through likelihood-based updates. These methods are widely adopted because they offer an algorithmically simple and scalable route to latent structure recovery while capturing the probabilistic nature of edge formation more accurately than purely spectral approaches.

However, in many modern applications, network observations are accompanied by vertex-level {\color{black}contextual} covariates {\color{black}that are sometimes high-dimensional}, yielding multi-source relational data. Formally, in addition to the adjacency matrix, each vertex is associated with a feature vector that encodes demographic, economic, biological, or behavioral information. For instance, in a social network, individuals may be connected through friendship ties while simultaneously possessing attributes such as age, education level, or political affiliation. These covariates may carry complementary or partially overlapping signals relative to the latent relational structure. The central statistical challenge is therefore not merely latent position estimation from the graph alone, but joint inference under heterogeneous data sources whose relative signal strengths are unknown. Developing methods that coherently integrate  and {\color{black}align} structural and contextual information while maintaining statistical interpretability remains a fundamental problem in contextual network analysis.

Several approaches have been proposed to incorporate node covariates into network modeling.
{\color{black}
Within the scope of contextual stochastic block models,  learning discrete community labels has been studied extensively \citep{Binkiewicz_Vogelstein_Rohe_2017_Biometrika_CovariateSpectralClustering,Deshpande_Sen_Montanari_2018_NeurIPS_ContextualSBM,Hu_Wang_2024_Biometrika_Network-AdjustedCovariates,Lu_Sen_2023_ContextualSBM,Ma_Nandy_2023_TIT_ContextualMLSBM,Shen_Amini_Josephs_Lin_2025_BayesianClusteringNetworkCovariates,Yan_Sarkar_2021_JASA_CovariateRegularizedClustering}, and the main theoretical results along this line focus on misclustering rates or detectability thresholds, while inference for vertex-level parameters is not addressed. Also, the balance between graph and covariate information in \cite{Binkiewicz_Vogelstein_Rohe_2017_Biometrika_CovariateSpectralClustering,Yan_Sarkar_2021_JASA_CovariateRegularizedClustering} is set by a tuning constant that depends on unobservable signal-to-noise ratios rooted in the network sparsity level. Following a separate route, \cite{James_Yuan_Gaynanova_Arroyo_2024_arXiv_JointIndividualNetwork} propose a spectral method that extracts shared and individual subspaces when the network and vertex covariate matrix are available. These approaches, nonetheless, fail to justify the uncertainty quantification for vertex-level parameters, which is exactly the gap our framework aims to fill.
}


{\color{black} In this work, we propose a tuning-free} Fisher-scoring surrogate likelihood method {\color{black}that offers}
a principled yet computationally agile framework tailored for the Generalized Random Dot Product Graph (GRDPG) framework \citep{rubin2022statistical} {\color{black}with high-dimensional vertex covariates}. 
Our approach replaces {\color{black}the awkward} joint likelihoods with a strategically constructed surrogate likelihood, which enables the use of an efficient Fisher-scoring algorithm. By utilizing a spectral initialization, our method requires only a few iterations to converge, thereby achieving a computational profile comparable to spectral methods while inheriting the probabilistic precision of generative models.

Most importantly, our framework is tuning-free: it naturally weights the information from the network and covariates through the Fisher information, without requiring a balancing parameter that depends on prior knowledge of network sparsity. Furthermore, we provide rigorous theoretical guarantees showing that our estimator achieves both asymptotic normality and asymptotic efficiency. This not only ensures the optimal recovery of latent positions but also facilitates valid statistical inference, even in relatively sparse regimes.
This combination of algorithmic simplicity, the lack of manual tuning and the rigorous theoretical foundation provides a superior alternative for the analysis of modern contextual networks.

The remainder of this paper is organized as follows.
Section~\ref{sec:preliminaries} introduces the contextual generalized
random dot product graph model and the surrogate likelihood.
Section~\ref{sec:computation_algorithm} presents the computational
algorithm.
Section~\ref{sec:theoretical_properties} establishes the theoretical
properties of the proposed estimator, including asymptotic normality.
Section~\ref{sec:simulation_studies} evaluates its performance through
simulation studies, and Section~\ref{sec:Last.fm} presents an application to the Last.fm social network.
Section~\ref{sec:discussion} concludes with a discussion.

\textbf{\textit{Notations}}:
For any positive integer $n$, let $[n]\coloneqq \{1,2,\ldots,n\}$.
The set of all positive integers is denoted by $\mathbb{N}_+$.
For any $a,b\in\mathbb{R}$, we denote $a\wedge b\coloneqq \min(a,b)$ and
$a\vee b\coloneqq \max(a,b)$.
For any two non-negative sequences $(a_n)_{n\in\mathbb{N}_+}$,
$(b_n)_{n\in\mathbb{N}_+}$, we write $a_n\lesssim b_n$ ($a_n\gtrsim b_n$, resp.),
if there exists some absolute constant $C>0$, such that
$a_n\le C b_n$ ($a_n\ge C b_n$, resp.) for all $n\in\mathbb{N}_+$.
If the constant $C$ also depends on another parameter $c$ that is independent of
$n\in\mathbb{N}_+$, then we write $a_n\lesssim_c b_n$
($a_n\gtrsim_c b_n$, resp.).
We use $K_c, N_c,\ldots$ to denote constants that may depend on another parameter
$c$ but are independent of the varying index $n\in\mathbb{N}_+$.
Absolute constants are usually hidden using notations $\lesssim$ and $\gtrsim$,
and, when necessary, we use $C_0$ and $c_0$ to denote generic absolute constants
that may vary from line to line.
We use the notation $a_n\asymp b_n$ to indicate that
$a_n\lesssim b_n$ and $a_n\gtrsim b_n$.
If $a_n/b_n$ stays bounded away from $+\infty$, we write $a_n=O(b_n)$ and
$b_n=\Omega(a_n)$, and if $a_n/b_n\to 0$, we denote $a_n=o(b_n)$ and
$b_n=\omega(a_n)$.
For any symmetric positive semidefinite matrices $\Sigma$ and $\Gamma$, we denote
$\Sigma\succeq \Gamma$ ($\Sigma\preceq \Gamma$, resp.), if
$\Sigma-\Gamma$ ($\Gamma-\Sigma$, resp.) is positive semidefinite.
When $\Sigma-\Gamma$ ($\Gamma-\Sigma$, resp.) is strictly positive definite,
we use the notation $\Sigma\succ \Gamma$ ($\Sigma\prec \Gamma$, resp.).
For any $d\in\mathbb{N}_+$, we use $\mathbf I_d$ to denote the $d\times d$ identity
matrix and $\mathbf 0_d$ to denote the zero vector in $\mathbb{R}^d$.
For $n,d\in\mathbb{N}_+$, $n\ge d$, let
$\mathbb{O}(n,d)\coloneqq \bigl\{\mathbf U\in\mathbb{R}^{n\times d}:\mathbf U^\top \mathbf U=\mathbf I_d\bigr\}$
denote the set of all orthonormal $d$-frames in $\mathbb{R}^n$.
When $n=d$, we simply write $\mathbb{O}(d)=\mathbb{O}(d,d)$.
For an $n\times d$ matrix $\mathbf M$, we denote by $\sigma_k(\mathbf M)$ the $k$th largest singular value of $\mathbf M$, $k\in [\min(n,d)]$.
For any $j\in [n]$ and $k\in [d]$, we use $[\mathbf M]_{j*}$ to denote its $j$th row, $[\mathbf M]_{*k}$ to denote its $k$th column, and $[\mathbf M]_{jk}$ to denote its $(j,k)$th entry.
When $\mathbf M\in\mathbb{R}^{n\times n}$ is a square symmetric matrix, we use
$\lambda_k(\mathbf M)$ to denote the $k$th largest eigenvalue of $\mathbf M$,
namely, $\lambda_1(\mathbf M)\ge \lambda_2(\mathbf M)\ge \cdots \ge \lambda_n(\mathbf M)$,
and $\mathrm{tr}(\mathbf M)$ the trace of $\mathbf M$ (the sum of its diagonal elements).
If $\mathbf M\in\mathbb{R}^{d\times d}$ is positive definite, then we let
$\kappa(\mathbf M)\coloneqq \lambda_1(\mathbf M)/\lambda_d(\mathbf M)$ denote the
condition number of $\mathbf M$.
The spectral norm of a rectangular matrix $\mathbf M$, denoted by $\|\mathbf M\|_2$,
is defined as the largest singular value of $\mathbf M$.
The Frobenius norm of a rectangular matrix $\mathbf M$, denoted by $\|\mathbf M\|_{\mathrm{F}}$,
is defined as $\|\mathbf M\|_{\mathrm{F}}=\sqrt{\mathrm{tr}(\mathbf M^\top \mathbf M)}$.
We use $\|\mathbf M\|_{2\to\infty}$ to denote the two-to-infinity norm of a matrix
$\mathbf M=[M_{jk}]_{n\times d}$, defined as $\|\mathbf M\|_{2\to\infty}=\max_{j\in[n]}\sqrt{\sum_{k=1}^d M_{jk}^2},$
and $\|\mathbf M\|_\infty$ to denote the matrix infinity norm $\|\mathbf M\|_\infty=\max_{j\in[n]}\sum_{k=1}^d |M_{jk}|.$
Given $d$ real numbers $a_1,\ldots,a_d\in\mathbb{R}$, we let
$\mathrm{diag}(a_1,\ldots,a_d)$ denote the $d\times d$ diagonal matrix whose
$(k,k)$th element is $a_k$ for $k\in[d]$.
For a Euclidean vector $\mathbf x=[x_1,\ldots,x_d]^\top\in\mathbb{R}^d$,
$\|\mathbf x\|_2$ denotes the Euclidean norm of $\mathbf x$ given by $\|\mathbf x\|_2=\sqrt{\sum_{k=1}^d x_k^2},$
and $\|\mathbf x\|_\infty$ denotes the infinity norm of $\mathbf x$ defined as $\|\mathbf x\|_\infty=\max_{k\in[d]} |x_k|.$
When the dimension of the underlying Euclidean space is clear, we use $\mathbf e_i$
to denote the unit basis vector whose $i$th coordinate is one and the rest of the
coordinates are zeros. Given two sequences of non-negative random variables
$(X_n)_{n=1}^\infty$, $(Y_n)_{n=1}^\infty$
and a deterministic diverging non-negative sequence
$(M_n)_{n=1}^\infty$,
we denote by
$X_n = \tilde{O}_p(Y_n; M_n),$
if for any constant $c>0$, there exists a $c$-dependent constant
$K_c$ such that
$\mathbb{P}(X_n \le K_c Y_n) \ge 1 - O(M_n^{-c}).$
When $M_n = n$, we simply write
$X_n = \tilde{O}_p(Y_n).$

\section{Preliminaries}
\label{sec:preliminaries}

\subsection{Background and Setup}
\label{sub:background_and_setup}

Network data are commonly modeled through latent space models
\citep{hoff2002latent}, in which each node is associated with an
unobserved position in a low-dimensional space, and the probability of
an edge between two nodes is determined by their latent positions. Let
$\mathcal{X} \subset \mathbb{R}^d$ denote the space of latent positions,
and let $\mathbf{x}_1, \ldots, \mathbf{x}_n \in \mathcal{X}$ be the
latent positions of the $n$ nodes. {\color{black}The focus of this work is on the estimation and inference for these latent positions.} Given a symmetric link function
$\kappa : \mathcal{X} \times \mathcal{X} \to [0,1]$, the observed
network is represented by a symmetric and hollow adjacency matrix
$\mathbf{A} = [A_{ij}]_{n \times n} \in \{0,1\}^{n \times n}$ whose
entries are generated as
\begin{align*}
    A_{ij} \overset{\text{ind}}{\sim}
    \mathrm{Bernoulli}\big(\kappa(\mathbf{x}_i, \mathbf{x}_j)\big),
    \quad i < j, \quad i, j \in [n] := \{1, \ldots, n\}.
\end{align*}
This general formulation encompasses a number of widely used models
depending on the choice of $\kappa$: the logistic link function
$\kappa(\mathbf{x}_i, \mathbf{x}_j) = 1/(1 +
e^{-\mathbf{x}_i^{\top}\mathbf{x}_j})$ \citep{hoff2002latent}, the
random dot product graph (RDPG) with $\kappa(\mathbf{x}_i,
\mathbf{x}_j) = \mathbf{x}_i^{\top}\mathbf{x}_j$
\citep{young2007random}, and the graphon model with a nonparametric
$\kappa$ and {\color{black}independent and identically distributed} (i.i.d.)\ $\mathrm{Unif}(0,1)$ latent positions. 

Building on this framework, we now review the signal-plus-noise
formulation within the generalized random dot product graph (GRDPG) \citep{rubin2022statistical} and adjacency spectral embedding. We consider a symmetric adjacency matrix $\mathbf{A} = [A_{ij}]_{n \times n}$ with zero diagonal, where the entries are independently distributed as $A_{ij} \sim \text{Bern}(\mathbf{x}_i^\top \mathbf{I}_{d_+,d_-} \mathbf{x}_j)$ for all $i, j \in [n]$, $i < j$, where $\mathbf{I}_{d_+,d_-} = \text{diag}(\mathbf I_{d_+}, -\mathbf I_{d_-})${\color{black}, and $A_{ij} = A_{ji}$ for all $i > j$}. {\color{black}We write $\bA\sim\mathrm{GRDPG}_{(d_+,d_-)}(\bX)$ as the shorthand notation, where $\bX = [\bx_1,\ldots,\bx_n]\transpose$ is the latent position matrix. }
Here, 
$d_+$ and $d_-$ denote the numbers of positive and negative eigenvalues, respectively, with $d_+ + d_- = d$. 
The observed network can be viewed as a noisy version of an underlying true edge probability matrix $\mathbf{P} = [P_{ij}]_{n \times n},$ 
{\color{black}and this perspective can be}
formalized through the signal-plus-noise matrix model 
$\mathbf{A} = \mathbf{P} + \mathbf{E}$,
where the off-diagonal entries of $\bE$ are centered and independent random variables up to symmetry, whereas its diagonal entries are deterministic, with $\bE_{ii} = −\bP_{ii}$ because $\bA_{ii} = 0$. Throughout, we retain the diagonal terms in the surrogate objective, the score, and the Fisher-scoring matrix; their contribution is accounted for in the supplementary proofs. A fundamental {\color{black}feature} of this model is that the signal matrix $\mathbf{P}$ is low-rank with rank $d \ll n$, {\color{black}and we assume that $d$ is constant and known throughout the analysis below. When $d$ is unknown, one can estimate it by minimizing a model selection criterion such as the Bayesian information criterion (BIC)}. 

{\color{black}Given an observed network adjacency matrix $\bA \in \mathbb{R}^{n\times n}$ and the rank $d$ of $\bP$, the adjacency spectral embedding of $\bA$ into $\mathbb{R}^d$ \citep{sussman2012consistent} is defined by taking the truncated spectral decomposition of $\bA$. Formally, }let $\mathbf{A} = \sum_{k=1}^n \widehat{\lambda}_k \widehat{\mathbf{u}}_k \widehat{\mathbf{u}}_k^\top$ be the spectral decomposition of $\mathbf{A}$, where $
\color{black}
\widehat{\lambda}_1 \ge \widehat{\lambda}_2\ge \dots \ge \widehat{\lambda}_n$ are the eigenvalues in non-increasing order. 
{\color{black}Then, }the adjacency spectral embedding (ASE) into $\mathbb{R}^d$ is defined as
\[
\widehat{\mathbf{X}}_{\mathbf{A}} =
[|\widehat{\lambda}_{1}|^{1/2}\widehat{\mathbf{u}}_{1}, \ldots, |\widehat{\lambda}_{d_+}|^{1/2}\widehat{\mathbf{u}}_{d_+}, |\widehat{\lambda}_{n-d_-+1}|^{1/2}\widehat{\mathbf{u}}_{n-d_-+1}, \ldots ,|\widehat{\lambda}_{n}|^{1/2}\widehat{\mathbf{u}}_{n}].\] 
In the case where $d_+, d_-$ are unknown, one can take $\widehat{\lambda}_{i_1}, \ldots, \widehat{\lambda}_{i_n}$ as the eigenvalues of $\mathbf A$ sorted in non-increasing absolute values, i.e., $|\widehat{\lambda}_{i_1}| \ge \cdots \ge |\widehat{\lambda}_{i_n}|,$ and estimate $\widehat{d}_+ = \sum_{k=1}^d \mathbbm{1}(\widehat{\lambda}_{i_k} > 0), \ \widehat{d}_- = d-\widehat{d}_+.$  To account for network sparsity, $\mathbf{P}$ is associated with a scaling factor $\rho_n \in (0,1]$ such that the {\color{black}average} expected degree of the generated random graph {\color{black}is bounded from above and below by constant multiples of $n\rho_n$, namely, $\max_i\sum_{j = 1}^nP_{ij} = O(n\rho_n)$ and $\min_i\sum_{j = 1}^nP_{ij} = \Theta(n\rho_n)$.}

In addition to the adjacency matrix $\mathbf A$, {\color{black}practitioners also} observe a covariate matrix $\mathbf B = [B_{li}]_{p \times n} = \mathbf Z \mathbf X^\top + \mathbf W \in \mathbb R^{p \times n}$, where each column corresponds to a vertex. Here, $\mathbf Z \in \mathbb R^{p \times d}$ denotes the coefficient matrix, $\mathbf W = [W_{li}]_{p \times n}$ is an error matrix whose entries $W_{li}$ are $\iid \ \text{N}(0,1)$, 
{\color{black}and the latent position matrix $\bX$ contains the shared information across both the contextual covariates and the network data source. }
The {\color{black}overarching aim}
is to {\color{black}learn} the latent positions $\{\xb_i \}_{i=1}^n$ by integrating contextual information with the network structure. 

\subsection{The surrogate likelihood function}\label{sub:surrogate likelihood}

In this section, we {\color{black}briefly review} the surrogate likelihood function for the generalized random dot product graph model originally due to \cite{Wu_Xie_2025_arXiv_Sanvi}. {\color{black}First,} consider the exact log-likelihood function {\color{black} of $\bA\sim\mathrm{GRDPG}_{(d_+,d_-)}(\bX)$}:
\begin{equation}
\label{eqn:GRDPG_exact_loglikelihood}
    \ell_{\mathbf{A}}(\mathbf{X}) = \sum_{1 \le i < j \le n} \{ A_{ij} \log(\mathbf{x}_i^\top \mathbf{I}_{d_+,d_-} \mathbf{x}_j) + (1 - A_{ij}) \log(1 - \mathbf{x}_i^\top \mathbf{I}_{d_+,d_-} \mathbf{x}_j) \}.
\end{equation}
As observed in \cite{Wu_Xie_2025_arXiv_Sanvi,wu2025statistical}, the exact likelihood \eqref{eqn:GRDPG_exact_loglikelihood}
is {\color{black}structurally cumbersome for the analysis and optimization.}
{\color{black}Also, GRDPG falls into the category of a curved exponential family, }and the {\color{black}theory of maximum likelihood} estimation {\color{black}remains generally challenging due to the curved exponential family nature} (see Section 2.3 of \citealp{bickel2015mathematical}). 

To address these issues, {\color{black}we leverage the idea in \cite{Wu_Xie_2025_arXiv_Sanvi} and consider the following form of} the surrogate log-likelihood.
Let $\bY = \bX\bI_{d_+,d_-}$ represent the {\color{black}signature-adjusted} latent position {\color{black}matrix}. Let $\widehat{\mathbf{Y}} = [\widehat{\by}_1,\ldots,\widehat{\by}_n]\transpose$ and $\widehat{\mathbf{Z}} = [\widehat{\bz}_1,\ldots,\widehat{\bz}_p]\transpose$ be {\color{black}some preliminary estimators}. {\color{black}These estimators can be} obtained from the adjacency spectral embedding and {\color{black}a plug-in estimator of }the coefficient matrix $\bZ$, respectively. The surrogate log-likelihood is defined as
\begin{align*}
    \widetilde{\ell}_n(\mathbf{X}; \widehat{\mathbf{Y}}, \widehat{\mathbf{Z}}) = \sum_{i=1}^n \sum_{j=1}^n \{(A_{ij} - \mathbf{x}_i^\top \widehat{\mathbf{y}}_j)\psi(\mathbf{x}_i^\top \widehat{\mathbf{y}}_j) + \Psi(\mathbf{x}_i^\top \widehat{\mathbf{y}}_j)\} - \frac{1}{2}\sum_{i=1}^n \sum_{l=1}^p (B_{li} - \widehat{\mathbf{z}}_l^\top \mathbf{x}_i)^2,
\end{align*}
where $\psi(\cdot)$ and $\Psi(\cdot)$ are functions defined as follows.
For a small threshold $\tau \in (0, 1/2]$, we define {\color{black}$\psi_1(x), \psi_2(x):x\in\mathbb{R}\mapsto\mathbb{R}$} as the twice continuously differentiable quadratic extensions {\color{black}of $x\mapsto \log(x)$ and $x\mapsto \log(1 - x)$} outside the interval $[\tau, 1-\tau]$
    \begin{align*}
        \psi_1(x) &= \begin{cases} -\frac{x^2}{2\tau^2} + \frac{2x}{\tau} + \log(\tau) - \frac{3}{2}, & \text{if } x < \tau, \\ \log(x), & \text{if } x \ge \tau, \end{cases}\\
        \psi_2(x)& = \begin{cases} -\frac{x^2}{2\tau^2} + \frac{(1-2\tau)x}{\tau^2} + \log(\tau) + \frac{(3\tau-1)(1-\tau)}{2\tau^2}, & \text{if } x > 1-\tau, \\ \log(1-x), & \text{if } x \le 1-\tau. 
        \end{cases}
    \end{align*}
Then, we define $\psi(x) = \psi_1(x) - \psi_2(x)$ and $\Psi(x) = \int_0^x \psi(t) dt$. Both are continuously differentiable over the real line. This construction removes the requirement that the bilinear forms lie between zero and one, thereby facilitating computation and theoretical analysis.

\section{Computation algorithm}
\label{sec:computation_algorithm}


\subsection{Fisher-scoring algorithm}
\label{sub:fisher_scoring_algorithm}

{\color{black}We now formally describe the proposed Fisher scoring algorithm. 
}
Recall that the surrogate log-likelihood can be written as a sum of individual vertex-{\color{black}specific components}:
\begin{align*}
    \widetilde{\ell}_n(\mathbf{X}; \widehat{\mathbf{Y}}, \widehat{\mathbf{Z}})
    = \sum_{i=1}^n \widetilde{\ell}_{in}(\mathbf{x}_i; \widehat{\mathbf{Y}}, \widehat{\mathbf{Z}}),
\end{align*}
where for each $i \in [n]$, the local objective function is defined as
\begin{align*}
    \widetilde{\ell}_{in}(\mathbf{x}_i; \widehat{\mathbf{Y}}, \widehat{\mathbf{Z}})
    \coloneqq \sum_{j=1}^n \Big\{ (A_{ij} - \mathbf{x}_i^\top \widehat{\mathbf{y}}_j) \psi(\mathbf{x}_i^\top \widehat{\mathbf{y}}_j) + \Psi(\mathbf{x}_i^\top \widehat{\mathbf{y}}_j) \Big\} - \frac{1}{2} \sum_{l=1}^p (B_{li} - \widehat{\mathbf{z}}_l^\top \mathbf{x}_i)^2.
\end{align*}
The separability of the surrogate log-likelihood function allows the practitioners to implement computation for each vertex separately. Namely, for given preliminary estimators $(\widehat{\bY}, \widehat{\bZ})$,
each $\widetilde{\ell}_{in}(\mathbf{x}_i)$ depends exclusively on the $i$th latent position $\mathbf{x}_i$ while all other quantities remain fixed. 
{\color{black}
Then, for each fixed vertex $i\in[n]$, given some initial guess $\widehat{\bX}^{(0)} = [\widehat{\bx}_1^{(0)},\ldots,\widehat{\bx}_n^{(0)}]\transpose$ and preliminary estimators $\widehat{\bY}^{(0)}$, $\widehat{\bZ}^{(0)}$, consider the Fisher scoring update for $\bx_i$ at the $t$th iteration:}
\begin{equation}
\label{Fisherscoring_update}
    \widehat{\mathbf{x}}_i^{(t+1)} = \widehat{\mathbf{x}}_i^{(t)} + \left\{ \mathcal{I}_{in}(\widehat{\bx}_i^{(t)}; \widehat{\mathbf{Y}}^{(t)}, \widehat{\mathbf{Z}}^{(t)}) \right\}^{-1} \frac{\partial \widetilde\ell_{in}}{\partial \mathbf{x}_i} (\widehat{\mathbf{x}}_i^{(t)}; \widehat{\mathbf{Y}}^{(t)}, \widehat{\mathbf{Z}}^{(t)}),
\end{equation}
where we denote the Fisher information matrix with regard to $\bx_i$ as
\begin{equation}
\label{eqn:fisher_information_matrix}
    \mathcal{I}_{in}(\mathbf{x}_i; \mathbf{Y}, \mathbf{Z}) = -\mathbb{E} \left\{ \frac{\partial^2 \widetilde\ell_{in}}{\partial \mathbf{x}_i \partial \mathbf{x}_i^\top} (\mathbf{x}_i; \mathbf{Y}, \mathbf{Z}) \right\}.
\end{equation}
{\color{black}More specifically, a direct calculation shows that
\begin{equation}\label{Fisherinfo}
{\color{black}
\mathcal{I}_{in}(\bx_i; \bY, \bZ)
}
=
\sum_{j=1}^n \psi'(\bx_i^\top \by_j)\,\by_j\by_j^\top
+
\bZ^\top \bZ
\end{equation}
}
{\color{black}
and the score function takes the form
\begin{align}
\label{eqn:score_function}
    \frac{\partial\widetilde{\ell}_{in}}{\partial\bx_i}(\bx_i; \bY, \bZ) = \sum_{j=1}^n(A_{ij} - \bx_i\transpose\by_j)\psi'(\bx_i\transpose\by_j)\by_j + \bZ\transpose(\mathbf{b}_i - \bZ\bx_i),
\end{align}
where $\mathbf{b}_i$ denotes the $i$th column of $\bB$. 
Note that both the Fisher information matrix and the score function contain contributions corresponding to the network and contextual components of the model. 

Next, the covariate coefficient matrix estimator is updated by maximizing the surrogate likelihood function with respect to $\widehat{\bZ}$ with $\bX$ and $\widehat{\bY}$ being fixed at the most recently updated values, yielding
\begin{equation}
\label{eqn:update_for_Z}
\widehat{\mathbf{Z}}^{(t + 1)} = \mathbf{B}\widehat{\mathbf{X}}^{(t + 1)} \left( \widehat{\mathbf{X}}^{(t + 1)\mathrm{T}} \widehat{\mathbf{X}}^{(t + 1)} \right)^{-1},
\end{equation}
where we denote $\widehat{\bX}^{(t + 1)} = [\widehat{\bx}_1^{(t + 1)},\ldots,\widehat{\bx}_n^{(t + 1)}]\transpose$. The closed-form solution \eqref{eqn:update_for_Z} originates from the fact that the likelihood of the covariates component is equivalent to an ordinary least squares (OLS) objective function with a known $\bX$. 
Note that with a small embedding dimension $d$, the inversion of the $d \times d$ Gram matrix is computationally efficient, ensuring that the global update does not become a bottleneck.
}

{\color{black}The separable nature of the estimation task based on the surrogate likelihood} naturally allows for a parallel implementation across vertices. In {\color{black}practice,} the Fisher-scoring updates for different vertices are distributed across multiple computational cores, which significantly reduces the total runtime and enables the method to scale effectively to large-scale networks. The overall estimation procedure alternates between updating the latent position matrix $\mathbf{X}$ via these parallelized Fisher-scoring sweeps and updating the contextual loading matrix $\mathbf{Z}$ until numerical convergence is achieved.


\subsection{Backtracking line search}
\label{sub:backtracking_linesearch}

While the Fisher-scoring update in \eqref{Fisherscoring_update} {\color{black}enjoys theoretical properties as can be seen in Section \ref{sec:theoretical_properties}},
the full Newton step {\color{black}with a constant stepsize}
may not always guarantee an increase in the objective function {\color{black}and may suffer from} numerical {\color{black}instability} in finite samples \citep{jennrich1976newton,lange1999numerical,nocedal2006numerical}. To {\color{black}enhance practical numerical performance,}
we incorporate a backtracking line search based on the Armijo condition {\color{black}to learn an adaptive stepsize} \citep{armijo1966minimization}.

Let $\mathbf{p}_i^{(t)} = \mathcal{I}_{in}(\widehat{\bx}_i^{(t)}, \widehat{\bY}^{(t)}, \widehat{\bZ}^{(t)})^{-1} \partial\widetilde{\ell}_{in}/\partial\bx_i(\widehat{\mathbf{x}}_i^{(t)}, \widehat{\bY}^{(t)}, \widehat{\bZ}^{(t)})$ be the Fisher search direction. Starting with an initial step size $\eta = 1$, the algorithm repeatedly shrinks $\eta$ by a factor $\beta \in (0, 1)$ until the following Armijo condition is satisfied:
\begin{align}
\label{eqn:armijo_condition}
{\color{black}
\widetilde{\ell}_{in}(\widehat{\mathbf{x}}_i^{(t)} + \eta \mathbf{p}_i^{(t)}, \widehat{\bY}^{(t)}, \widehat{\bZ}^{(t)}) \ge \widetilde{\ell}_{in}(\widehat{\mathbf{x}}_i^{(t)}, \widehat{\bY}^{(t)}, \widehat{\bZ}^{(t)}) + c \eta 
\bigg\langle\frac{\partial\widetilde{\ell}_{in}}{\partial\bx_i}(\widehat{\mathbf{x}}_i^{(t)}, \widehat{\bY}^{(t)}, \widehat{\bZ}^{(t)}),
\mathbf{p}_i^{(t)}\bigg\rangle,
}
\end{align}
where $c \in (0, 1)$ is a small constant that ensures the update provides a sufficient increase relative to the gradient, and $\langle\cdot,\cdot\rangle$ denotes the Euclidean inner product between Euclidean vectors. In practice, we take $\beta = 1/2$, but the users are free to select other values based on their respective expertise. 

To ensure robust performance, we implement the following numerical safeguards:
\begin{itemize}
    \item \textbf{Ascent Direction Check}: If the Fisher direction is not an ascent direction (i.e., {\color{black}the inner product of the Fisher direction and the score is non-positive}
    ), the algorithm falls back to the gradient and replaces the search direction by the gradient $\color{black}{\partial\widetilde{\ell}_{in}}/{\partial\bx_i}(\widehat{\bx}_i^{(t)}, \widehat{\bY}^{(t)}, \widehat{\bZ}^{(t)})
    $ to maintain the ascent property.
    \item \textbf{\color{black}Probability thresholding}: To prevent singular {\color{black}Fisher information} matrices and numerical overflow near the boundaries of the parameter space, we employ a stabilization technique by thresholding the estimated probabilities. Formally, 
    let $\widetilde{p}_{ij}^{(t)} = \max\{ \min( \widehat{\mathbf{x}}_i^{(t)\mathrm{T}} \widehat{\mathbf{y}}_j^{(t)}, 1 - \tau ), \tau \}$ be the thresholded probability for $\bx_i\transpose\by_j$ at the $t$th iteration. Here, $\tau$ is also the threshold used to define the surrogate functions. For numerical stabilization, we evaluate the Fisher information weights at the clipped inner products $\widetilde p_{ij}^{(t)}$, while the score is evaluated at the original inner products. Specifically, we compute
    \begin{align}
    \color{black}
    \label{eqn:stabilized_score}
    \bS_i^{(t)} &\color{black}= \sum_{j=1}^n(A_{ij} - \widehat{\bx}_i^{(t)\mathrm{T}}\widehat{\by}_j^{(t)})\psi'\!\left(\widehat{\mathbf x}_i^{(t)\top}
\widehat{\mathbf y}_j^{(t)}\right)\widehat{\by}_j^{(t)} + \widehat{\bZ}^{(t)\mathrm{T}}(\bb_i - \widehat{\bZ}^{(t)}\widehat{\bx}_i^{(t)}),\\
    \label{eqn:stabilized_Fisher_information_matrix}
    \color{black}\bG_{in}^{(t)} &\color{black}= \sum_{j=1}^n\psi'(\widetilde{p}_{ij}^{(t)})\widehat{\by}_j^{(t)}\widehat{\by}_j^{(t)\mathrm{T}} + \widehat{\bZ}^{(t)\mathrm{T}}\widehat{\bZ}^{(t)}.
    \end{align}
\end{itemize}

{\color{black}At each iteration $t$,} once the vertex-wise updates for the latent positions $\widehat{\mathbf{X}}^{(t + 1)}$ are completed within a sweep {\color{black} over $i\in[n]$}, the model synchronizes the contextual information by refining the coefficient matrix $\widehat{\mathbf{Z}}^{(t + 1)}$. Unlike the latent positions, which are updated separately for
each vertex, the covariate coefficient matrix
$\widehat{\mathbf{Z}}^{(t+1)}$ is updated jointly in closed form
using \eqref{eqn:update_for_Z}, with all other quantities held
fixed at their most recently updated values.
The full procedure is presented in Algorithm~\ref{alg:fisher_scoring}.

\begin{algorithm}[t]
\caption{Iterative Fisher-scoring for Surrogate Likelihood Optimization}
\label{alg:fisher_scoring}
\begin{algorithmic}[1]
\Require Adjacency matrix $\mathbf{A}$, Covariates $\mathbf{B}$, Threshold $\tau$, Tolerance $\epsilon$, Max iterations $T_{\max}$
\Ensure Refined latent positions $\widehat{\mathbf{X}}$, Contextual coefficients $\widehat{\mathbf{Z}}$

\Statex \textbf{Step 1: Initialization}
\State Obtain $\widehat{\mathbf{X}}^{(0)} \in \mathbb{R}^{n \times d}$ via Adjacency Spectral Embedding (ASE) of $\mathbf{A}$
\State $\widehat{\mathbf{Z}}^{(0)} \leftarrow \mathbf{B}\widehat{\mathbf{X}}^{(0)} \left( (\widehat{\mathbf{X}}^{(0)})^\top \widehat{\mathbf{X}}^{(0)} \right)^{-1}$ \Comment{Initial OLS projection}
\State Set $t \leftarrow 0$ and $\widehat{\bX}^{(-1)} = \mathbf{O}_{n\times d}$

\Statex \textbf{Step 2: Iterative Optimization (Block Coordinate Ascent)}
\While{$t < T_{\max}$ \textbf{and} $\|\widehat{\mathbf{X}}^{(t)} - \widehat{\mathbf{X}}^{(t-1)}\|_F > \epsilon$}
    
    \Statex \quad \textit{// Block 1: Update Latent Positions $\mathbf{X}$ in Parallel}
    \For{each vertex $i \in \{1, \dots, n\}$ \textbf{in parallel}}
        \State Initialize $\eta = 1$
        \State {\color{black}Compute $\mathbf{S}_i^{(t)}$ and $\mathbf{G}_{in}^{(t)}$ using formulas \eqref{eqn:stabilized_score}--\eqref{eqn:stabilized_Fisher_information_matrix}.}
        \State Solve $\mathbf{p}_i^{(t)} \leftarrow [\mathbf{G}_{in}^{(t)}]^{-1} \mathbf{S}_i^{(t)}$
        \If{$\mathbf{S}_i^{(t)\top} \mathbf{p}_i^{(t)} \le 0$} \Comment{Ascent direction check}
            \State $\mathbf{p}_i^{(t)} \leftarrow \mathbf{S}_i^{(t)}$ \Comment{Fallback to gradient}
        \EndIf
        \State {\color{black}\textbf{while} condition \eqref{eqn:armijo_condition} does not hold}
        
        \State\quad {\color{black}Set $\eta\longleftarrow \eta/2$.}

        \State {\color{black}\textbf{end while}}

        \State $\widehat{\mathbf{x}}_i^{(t+1)} \leftarrow \widehat{\mathbf{x}}_i^{(t)} + \eta \mathbf{p}_i^{(t)}$
    \EndFor
    
    \Statex \quad \textit{// Block 2: Global Update of Contextual Coefficients $\mathbf{Z}$}
    \State $\widehat{\mathbf{Z}}^{(t+1)} \leftarrow \mathbf{B}\widehat{\mathbf{X}}^{(t+1)} \left( (\widehat{\mathbf{X}}^{(t+1)})^\top \widehat{\mathbf{X}}^{(t+1)} \right)^{-1}$
    
    \State $t \leftarrow t + 1$
\EndWhile

\State \Return $\widehat{\mathbf{X}}^{(t)}, \widehat{\mathbf{Z}}^{(t)}$
\end{algorithmic}
\end{algorithm}

\section{Theoretical properties}
\label{sec:theoretical_properties}

We are now in a position to establish the theoretical properties of the proposed Fisher-scoring estimator. In preparation for doing so, 
we first introduce the following conditions on the signal-to-noise ratios (SNR) of both the network model and the contextual covariate {\color{black}model. Let $\bX_0 = [\bx_{01},\ldots,\bx_{0n}]\transpose$ and $\bZ_0 = [\bz_{01},\ldots,\bz_{0p}]\transpose$ denote the true values of $\bX$ and $\bZ$, respectively. Also, below, the score and Fisher information matrix for a given vertex $i\in[n]$ are computed by taking the sum over all $j\in[n]$ because it is asymptotically equivalent to the sum over $j\in[n]\backslash\{i\}$. }

\begin{assumption}[SNR of GRDPG]
\label[assumption]{assump:snr_network} 
    There exists an $n$-dependent quantity $\rho_n \in (0, 1]$ such that $\sigma_d(\mathbf{X}_0) \ge c\sqrt{n\rho_n}$, $\|\mathbf{X}_0\|_{2\to\infty} \le \rho_n^{1/2}$, 
    for some constant $c>0$, and $n\rho_n = \Omega(\log n)$. 
\end{assumption} 
   
\begin{assumption}[SNR of Contextual Covariates]
\label[assumption]{assump:snr_covariates} 
    The true coefficient matrix satisfies $\|\mathbf{Z}_0\|_{2 \to \infty} = O(1)$ and the number of covariates satisfies $p = O(n)$. 
\end{assumption}

\cref{assump:snr_network} ensures that the latent structure of the graph is sufficiently strong to be recovered from the noisy adjacency matrix, and \cref{assump:snr_covariates} guarantees that the signal from the contextual covariates does not dominate the inherent network signal, allowing for balanced integration during the Fisher-scoring update.

{\color{black}Below}, we begin by characterizing the distribution of the one-step {\color{black}Fisher-scoring estimator.}

\begin{theorem}\label{thm:OnestepFisher} Under \cref{assump:snr_network,assump:snr_covariates}, suppose $\tau = \delta \rho_n$ for some constant $\delta \in (0,1/2)$ and $\tau\leq P_{ij}\leq 1 - \tau$ for all $i,j\in[n]$. 
There exists a random orthogonal matrix
$\mathbf{Q}\in\mathbb{O}(d)$ satisfying
$\mathbf{Q}\mathbf{I}_{d_+,d_-}
=\mathbf{I}_{d_+,d_-}\mathbf{Q}$
such that, for each fixed $i\in[n]$,
\begin{align*}
\bQ\transpose \widehat \bx_i^{(1)} - \bx_{0i}
=
\bG_{in}^{-1} \frac{1}{n+p}
\frac{\partial \widetilde{\ell}_{in}}{\partial \bx_i}(\bx_{0i}; \bY_0, \bZ_0)
+
\br_{\bx_i}^{(1)},    
\end{align*}
where
\begin{equation}
\label{eqn:Fisher_information_matrix}
{\color{black}\bG_{in}
=\frac{1}{n + p}\bigg\{\sum_{j=1}^n 
\psi'(\bx_{0i}^\top \by_{0j}) \by_{0j}\by_{0j}^\top
+
\sum_{l=1}^p \bz_{0l} \bz_{0l}^\top\bigg\},
}
\end{equation}
and the remainder satisfies
\begin{align*}
    \| \br_{\bx_i}^{(1)} \|_2 =
    \widetilde{O}_p
    \left\{
    \frac{\log m_n + (\log n)^{1/2}}{n\rho_n^{1/2}}; m_n
    \right\}
\end{align*}
for any diverging nonnegative sequence $(m_n)_{n=1}^{\infty}$ with $m_n = O(n).$ In particular, when $n\rho_n = \omega(\log n)$, we have
\begin{align*}
\sqrt{n + p}\bG_{in}^{1/2}(\bQ\transpose\widehat{\bx}_i^{(1)} - \bx_{0i})&
\overset{\calL}{\to} \mathrm{N}(\zero, \eye_d)\quad\text{as }n\to\infty,\\
\| \widehat{\bX}^{(1)}\bQ - \bX_0 \|_{2 \rightarrow \infty}& = \widetilde{O}_p \bigg\{\frac{(\log n)^{1/2}}{\sqrt{n}}\bigg\}.
\end{align*}
\end{theorem}


{\color{black}Note that when $\tau \leq P_{0ij}\leq 1 - \tau$ for all $i,j\in[n]$, $\bG_{in}$ coincides with the Fisher information matrix with respect to $\bx_i$ \citep{xie2023efficient}. Consequently, }\Cref{thm:OnestepFisher} implies that our estimator is asymptotically efficient, with its variance reaching the inverse of the Fisher information matrix.
This allows for rigorous statistical inference, such as constructing confidence intervals and performing hypothesis tests on the latent positions.

\begin{assumption}[Signal strength for linearization]\label[assumption]{assump:3}
For every $a>0$, $\rho_n^{-1}=O\{(\log n)^a\}$.
\end{assumption}

Assumption \ref{assump:3} above is a stronger condition on the network sparsity compared to $n\rho_n = \omega(\log n)$. It permits us to develop the asymptotic theory of a general iterative Fisher scoring estimator, and guarantees that the signal-to-noise ratio is large enough so that noise does not accumulate across iterations. However, we also remark that the asymptotic theory of the one-step Fisher scoring estimator does not require this condition and the sparse regime $n\rho_n = \omega(\log n)$ is sufficient.



{\color{black}Next, we extend the distributional characterization of the one-step Fisher-scoring estimator established in \Cref{thm:OnestepFisher} to the general iterates. }
\begin{theorem}[Analysis of $\widehat{\mathbf{X}}^{(t+1)}$] \label{thm:efficiency}
Under \cref{assump:snr_network,assump:snr_covariates,assump:3}, if $\tau = \delta \rho_n$ for some constant $\delta \in (0, 1/2)$, 
then there exists a random orthogonal matrix
$\mathbf{Q}\in\mathbb{O}(d)$ satisfying
$\mathbf{Q}\mathbf{I}_{d_+,d_-}
=\mathbf{I}_{d_+,d_-}\mathbf{Q}$
such that, 
\begin{align*}
\widehat{\mathbf{Z}}^{(t)}\mathbf{Q} - \mathbf{Z}_0
&= \mathbf{W}\mathbf{X}_0(\mathbf{X}_0^\top \mathbf{X}_0)^{-1} + \mathbf{R}_{\mathbf{Z}}^{(t)},\\
\mathbf{Q}^\top \widehat{\mathbf{x}}_i^{(t+1)} - \mathbf{x}_{0i}
&= \mathbf{G}_{in}^{-1} \frac{1}{n+p} \frac{\partial \widetilde{\ell}_{in}}{\partial \mathbf{x}_i}(\mathbf{x}_{0i}; \mathbf{Y}_0, \mathbf{Z}_0) + \mathbf{r}_{\mathbf{x}_i}^{(t+1)},
\end{align*}
where
\begin{align*}
\|\mathbf{R}_{\mathbf{Z}}^{(t)}\|_{2 \to \infty} = \widetilde{O}_p \left\{ \frac{(\log n)^{2+\epsilon}}{n \rho_n} \right\},\quad
\|\mathbf{r}_{\mathbf{x}_i}^{(t+1)}\|_2 = \widetilde{O}_p \left\{ \frac{(\log n)^{2+\epsilon}}{n \rho_n^{1/2}} \right\}
\end{align*}
for any $\epsilon > 0$.
\end{theorem}

\Cref{thm:efficiency} {\color{black}and \Cref{thm:OnestepFisher} jointly} confirm that {\color{black}one-step and multi-step} Fisher-scoring {\color{black}estimators achieve} the Cram\'er--Rao lower bound asymptotically {\color{black}under mild conditions. Although the one-step estimator already enjoys nice theoretical properties such as asymptotic normality and efficiency, iterating the Fisher-scoring algorithm multiple times upon stabilization may improve practical performance, and \Cref{thm:efficiency} guarantees that the latter estimator shares the same theoretical features. }


\section{Simulation studies}\label{sec:simulation_studies}

Our simulation studies are organized around three complementary {\color{black}setups}.
First, under a correctly specified GRDPG model, we assess the finite-sample performance of the {\color{black}proposed estimates} in terms of {\color{black}the following aspects: }{\color{black}empirical coverage rates of the vertex-wise confidence sets 
}, the sum of squared errors (SSE) {\color{black}of the estimated latent positions}, and average computation time (\Cref{sub:correctly_specified}). Second, to isolate the contribution of the covariate information itself, we compare estimation performance when the covariates are informative ($\bz_1,\ldots,\bz_p \iidsim \mathrm{N}(\zero,\bI)$) against a baseline in which they carry no signal ($\bZ_0 = \mathbf{0}_{p\times d}$), thereby quantifying the gain that the MSLE derives from the covariate matrix
(\Cref{sub:covariate_comparison}). Third, to examine the robustness of
our framework to model misspecification, we generate networks from a {\color{black}latent space model with a}
logistic link function rather than the GRDPG, and
evaluate how accurately each method reconstructs the underlying edge probability matrix
(\Cref{sub:reconstruction_logi}). {\color{black}This} setting reflects practical situations in which the true link mechanism is unknown, and tests whether the proposed estimator retains its advantage when the working model is misspecified.

\subsection{Well-specified model}\label{sub:correctly_specified}
\subsubsection{A latent curve example}\label{ssub:example-1d}

In this example, we consider a random dot product graph with $n = 500$ vertices and latent dimension $d = 1$, together with a covariate matrix $\mathbf{B} \in \mathbb{R}^{p \times n}$ with $p = n/2 = 250$. For each vertex $i \in [n]$, the ground truth latent position is given by $x_{0i} = 0.75 \sin\!\left\{\pi i/(n-1)\right\} + 0.1$,
and we denote $\mathbf{X}_0 = [x_{01}, \dots, x_{0n}]^\top$. Unless otherwise specified, we generate the covariate matrix as
$\mathbf{B}=\mathbf{Z}_0\mathbf{X}_0^\top+\mathbf{W}$,
where the entries of $\mathbf{Z}_0$ and $\mathbf{W}$ are mutually
independent standard normal random variables, and
$(\mathbf{Z}_0,\mathbf{W})$ is independent of the network
conditional on $\mathbf{X}_0$. The {\color{black}proposed Fisher-scoring} algorithm is run for at most 3 iterations with convergence tolerance $\epsilon = 0.01$, {\color{black}and we assess convergence by computing} the maximum row-wise change between successive iterates.
The smoothing threshold is set to $\tau = 0.001$.

Given $\bA \sim \text{RDPG}(\bX_0)$, we investigate the performance using several metrics. First, we compute the empirical coverage probabilities of the vertex-wise confidence intervals derived from the maximum surrogate likelihood estimate (MSLE). These 95\% confidence intervals are constructed as follows: letting $z_{1-\alpha/2}$ be the $1−\alpha/2$ quantile of the standard normal distribution, \Cref{thm:efficiency} establishes that the $(1-\alpha)$ confidence interval for $x_{0i}$ is $(\widehat{x}_i - z_{1-\alpha/2}\{ (n+p)\widehat{\bG}_{in}(\widehat{x}_i)\}^{-1/2}, \ 
\widehat{x}_i + z_{1-\alpha/2}\{ (n+p)\widehat{\bG}_{in}(\widehat{x}_i)\}^{-1/2})$, where $\widehat{\bG}_{in}(\widehat{x}_i)
= \frac{1}{n+p} \{
\sum_{j=1}^n \psi'(\widehat{x}_i^\top \widehat{y}_j)\,\widehat{y}_j\widehat{y}_j^\top
+
\widehat{\bZ}^\top \widehat{\bZ} \}$, {\color{black}$\widehat{\bX} = [\widehat{\bx}_1,\ldots,\widehat{\bx}_n]\transpose$ is the MSLE, and $\widehat{\bY} = \widehat{\bX}$ in this example. }
Second, to evaluate the proximity of the estimates to the true latent position matrix $\bX_0$, we employ the sum of squared errors (SSE) as the evaluation metric. For a generic estimate $\bar{\bX}$, the SSE is defined as $\min_{w \in \{\pm 1\}} \| w \bar{\bX} - \bX_0 \|_{\text{F}}^2 $. This numerical experiment is conducted over 100 Monte Carlo replicates. {\color{black}For comparison, we consider the ASE from \cite{sussman2012consistent} and the covariate-free one-step estimator from \cite{xie2023efficient} as alternative methods. }

\begin{figure}[htbp]
  \centering
    \includegraphics[width=0.55\textwidth]{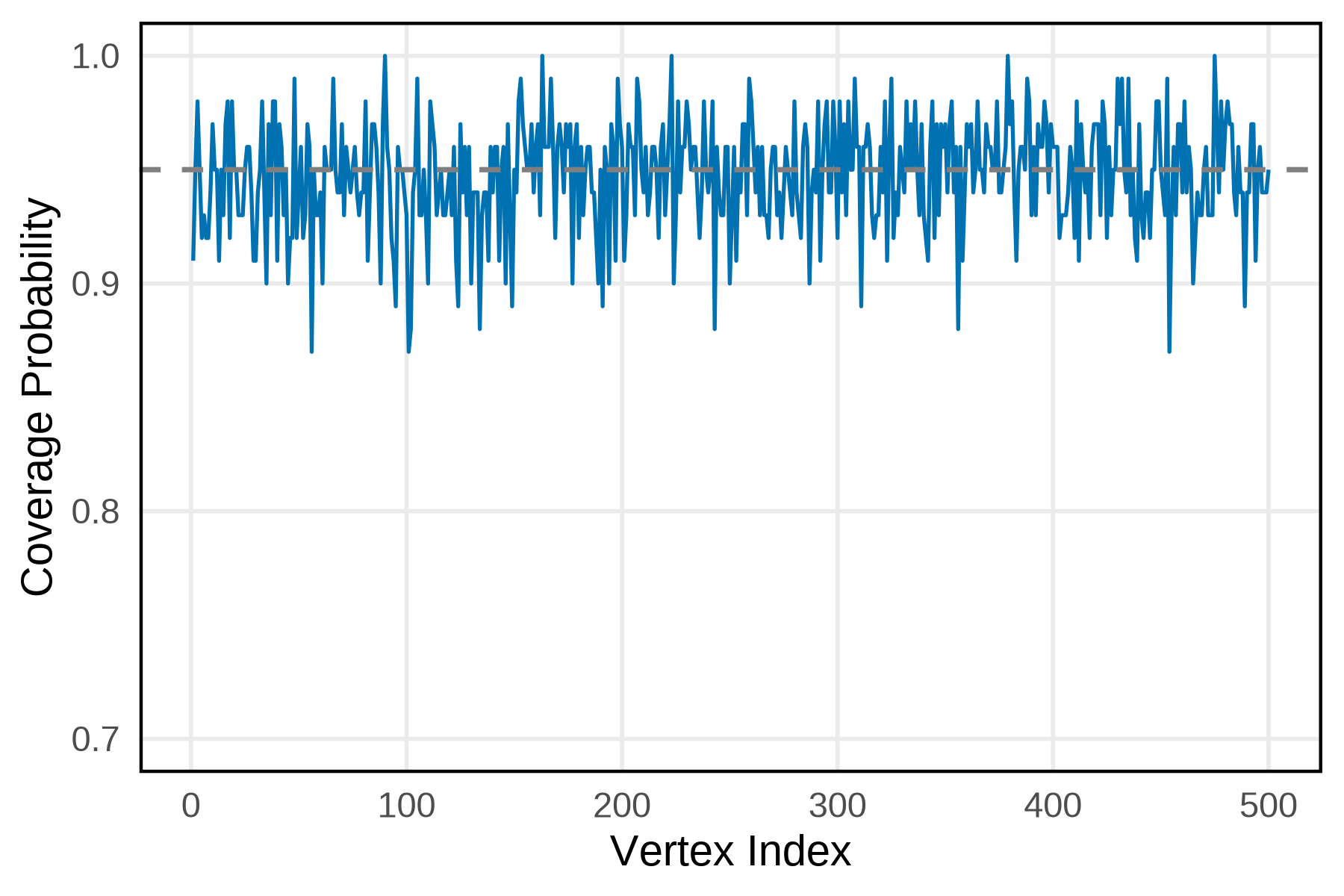}
  
  \caption{Comparison of empirical coverage probabilities for the 1D latent curve scenario ($n=500$). The dashed horizontal lines indicate the nominal 95\% coverage level.}
  \label{fig:msle_coverage_comparison}
\end{figure}

{\color{black}The coverage rates of the vertex-wise confidence intervals for the MSLE are shown}
in Figure \ref{fig:msle_coverage_comparison}, where the {\color{black}empirical} coverage probabilities {\color{black}concentrate} around the 0.95 target across all $500$ vertex indices. This confirms that the estimator is robust to the substitution of the true covariance matrix with its empirical counterpart. 

The {\color{black}SSEs and the computation time per experiment} are summarized in Table \ref{tab:1d_results}. The proposed MSLE significantly improves estimation precision, yielding an average SSE of $0.4150$, which is approximately $22\%$ lower than the ASE ($0.5348$) and the OSE ($0.5259$). 
While the iterative updates require {\color{black}substantially more computational effort,}
the significant gains in {\color{black}estimation} accuracy
justify the computational cost over near-instantaneous spectral methods.


\begin{table}[ht]
\centering
{\color{black}\begin{tabular}{lccccc}
\toprule
Method  & SSE (Mean) & SSE (SD) & Time (sec) & Time (SD) \\
\midrule
ASE  & 0.5348 & 0.03517 & 0.0363 & 0.0048 \\
OSE  & 0.5259 & 0.03499 & 0.0639 & 0.0106 \\
MSLE & 0.4150 & 0.03019 & 609.3 & 144.3 \\
\bottomrule
\end{tabular}
\caption{Simulation results for \Cref{ssub:example-1d}: SSEs and the computation time per experiment for the ASE, OSE, and the proposed MSLE based on the Fisher-scoring algorithm, together with standard deviations (SDs). }
\label{tab:1d_results}
}
\end{table}


\subsubsection{A three-dimensional latent curve example}
\label{ssub:example_3D_CGRDPG}

In this example, we consider a rank-three generalized random dot product graph (GRDPG) with signature $(2,1)$, whose latent positions are drawn from a latent curve in $\mathbb{R}^3$, parameterized as $\bx_{0i} = [0.42i/n + 0.46, 0.27\sin(2\pi i/n) + 0.46, 0.20\cos(4\pi i/n)]^\top$ for $i \in [n]$. The resulting edge probability matrix {\color{black}is an indefinite matrix} $\mathbf{P} = \mathbf{X}_0 \mathbf{I}_{2,1} \mathbf{X}_0^\top$, where $\mathbf{X}_0 = [\mathbf{x}_{01}, \ldots, \mathbf{x}_{0n}]^\top$ and $\mathbf{I}_{2,1} = \mathrm{diag}(1, 1, -1)$. 
{\color{black}In our Fisher-scoring algorithm,} the smoothing threshold {\color{black}in the surrogate likelihood} is set to $\tau = 0.001$, and the algorithm is run for a maximum of $T_{\max} = 30$ iterations with a convergence tolerance $\epsilon = 0.01$. {\color{black}We also use ASE and OSE as comparators for the proposed method. }

Table~\ref{tab:3d_results} summarizes the {\color{black}empirical coverage rates of the $95\%$ confidence sets implied by \Cref{thm:efficiency} (for MSLE) and \cite{xie2023efficient} (for ASE and OSE)}, together with SSE results for $n=1000$ and $n=2000$. {\color{black}The reported coverage rates are based on the averages of the coverage rates of vertex-wise confidence sets over $i\in[n]$.} Across both sample sizes, MSLE consistently achieves the highest coverage rates (93--94\%), closely approaching the nominal 95\% level, while also yielding the lowest SSE by a substantial margin. In contrast, ASE 
{\color{black}has} coverage noticeably below $95\%$, indicating that the plug-in variance estimator underestimates the true sampling variability. OSE exhibits {\color{black}the worst coverage rates,}
and its SSEs are highly unstable, with a mean SSE several orders of magnitude larger than those of MSLE and ASE, driven by {\color{black}the numerical instability of the one-step Fisher information matrix estimates}. 

\begin{table}[ht]
\color{black}
\centering
\setlength{\tabcolsep}{4.5pt}
\begin{tabular}{llccc}
\toprule
$n$ & Method & Coverage & SSE (Mean) & SSE (SD) \\
\midrule
1000 & ASE   & 0.8203 & 23.0506     & 1.6874       \\
     & OSE   & 0.6125 & 30696.3583  & 305845.3426  \\
     & MSLE  & 0.9333 & 3.7156      & 0.2355       \\
\midrule
2000 & ASE  & 0.8961 & 19.0920     & 0.5373       \\
     & OSE  & 0.8533 & 346.2035    & 3174.8322    \\
     & MSLE & 0.9368 & 3.7422      & 0.2094       \\
\bottomrule
\end{tabular}
\caption{Simulation results for \Cref{ssub:example_3D_CGRDPG}: Empirical coverage rates of the vertex-wise $95\%$ confidence sets (averaged over vertices and repeated experiments) and SSEs for the ASE, OSE, and the proposed MSLE based on the Fisher-scoring algorithm. }
\label{tab:3d_results}
\end{table}


\subsection{Information gain from covariates}
\label{sub:covariate_comparison}

To assess the value of incorporating covariate information, we compare two scenarios:
$\bZ_0 = 0$ (no covariate
information) versus $\color{black}\bz_1,\ldots,\bz_p \iidsim \mathrm{N}(\zero, \eye_d)$ (with covariate
information). 
{\color{black}We consider the following two setups for the true network generating distributions:
\begin{itemize}
    \item One-dimensional latent positions: the true latent positions $\bX_0$ are identical to those used in \Cref{ssub:example-1d}.

    \item Two-dimensional latent positions: the true latent positions $\mathbf{X}_0 =[\mathbf{x}_{01}, \ldots, \mathbf{x}_{0n}]^\top$ are generated according to the curve $\mathbf{x}_{0i} = [0.28 \sin(\pi i /(n-1)) + 0.42, \,\, 0.28 \cos(\pi i/(n-1)) + 0.42]^\top$ for each $i \in [n]$. We take $d_+ = 2$, $d_- = 0$, and let the sample size $n$ vary in $\{500,1000\}$. 
\end{itemize}
}
We evaluate the performance using 
{\color{black}SSEs} ($\|\widehat{\bX} - \bX_0\|_F^2$) across ASE, OSE, {\color{black}and the proposed MSLE}.

In the {\color{black}one-dimensional example} ($n=500$, $d=1$, $p=250$ {\color{black}as in \Cref{ssub:example-1d}}), 
the SSE results {\color{black}reveal a}
more nuanced story {\color{black}beyond the coverage rates in \Cref{sub:correctly_specified}}: while ASE and OSE show negligible change between
scenarios, MSLE {\color{black}generates SSEs that decrease} from $0.5292$ ({\color{black}noninformative} covariate) to $0.4151$ ({\color{black}informative} covariate), a modest but consistent improvement (Table~\ref{tab:sse_1d_n500}). 


\begin{table}[ht]
\color{black}
\centering
\begin{tabular}{llcc}
\toprule
Scenario & Method & SSE Mean & SSE SD \\
\midrule
\multirow{3}{*}{Noninformative Covariates}
  & MSLE & 0.5292 & 0.0352 \\
  & ASE    & 0.5379 & 0.0351  \\
  & OSE    & 0.5286 & 0.0350  \\
\midrule
\multirow{3}{*}{Informative Covariates}
  & MSLE & 0.4151 & 0.0301  \\
  & ASE    & 0.5348 & 0.0352  \\
  & OSE    & 0.5259 & 0.0350  \\
\bottomrule
\end{tabular}
\caption{Simulation results for \Cref{sub:covariate_comparison} in the one-dimensional example. }
\label{tab:sse_1d_n500}
\end{table}



The {\color{black}two-dimensional} setting above reveals a substantially {\color{black}larger gap. The numerical results are tabulated in Table~\ref{tab:sse_n500}.} At $n=500$
($d=2$, $p=250$),  {\color{black}under the noninformative covariate} scenario 
MSLE {\color{black}produces SSEs that are} substantially
higher (mean $14.56$) than {\color{black}those in the scenario with informative} covariates.
{\color{black}OSE is dramatically unstable
(mean $44.80$, SD $301.82$)}. With covariate information, {\color{black}MSLE generates SSEs that drop} to $2.28$ with a much lower {\color{black}standard deviation}, while OSE remains comparatively unstable 
and ASE remains essentially unchanged.

At the larger sample size $n=1000$ ($d=2$, $p=500$), the
same pattern persists, {\color{black}as can be seen by the numerical results reported in Table~\ref{tab:sse_n1000}}: 
{\color{black}MSLE has SSEs dropping} from $11.44$ to $2.12$ once covariates {\color{black}become informative,}
while the SSE values for ASE and OSE remain essentially unchanged. {\color{black}Throughout, we see that the numerical experiments and the outcomes are consistent with the overarching spirit that aligning the contextual covariates with the network signals strengthens the estimation quality, and the performance scales with larger sample size. }



  \begin{table}[t]
  \centering
  \color{black}
  \begin{tabular}{llcc}
  \toprule
  Scenario & Method & SSE Mean & SSE SD \\
  \midrule
  \multirow{3}{*}{Noninformative Covariates}
    & MSLE & 14.5562  &   2.4730 \\
    & ASE  & 13.9834  &   2.9905 \\
    & OSE  & 44.8022  & 301.8156 \\
  \midrule
  \multirow{3}{*}{Informative Covariates}
    & MSLE &  2.2832  &  0.2206 \\
    & ASE  & 13.2997  &  2.1721 \\
    & OSE  & 14.8076  & 13.0558 \\
  \bottomrule
  \end{tabular}
  \caption{Simulation results for \Cref{sub:covariate_comparison} in the two-dimensional example with $n=500$. }
  \label{tab:sse_n500}
  \end{table}



  \begin{table}[ht]
  \centering
  \color{black}
  \begin{tabular}{llcc}
  \toprule
  Scenario & Method & SSE Mean & SSE SD \\
  \midrule
  \multirow{3}{*}{Noninformative Covariates}
    & MSLE & 11.4379 & 0.6462  \\
    & ASE  & 10.9855 & 0.5990  \\
    & OSE  & 11.0250 & 0.6058  \\
  \midrule
  \multirow{3}{*}{Informative Covariates}
    & MSLE &  2.1211 & 0.1468 \\
    & ASE  & 10.9258 & 0.5855 \\
    & OSE  & 10.9694 & 0.5864  \\
  \bottomrule
  \end{tabular}
  \caption{Simulation results for \Cref{sub:covariate_comparison} in the two-dimensional example with $n = 1000$. }
  \label{tab:sse_n1000}
  \end{table}


\subsection{Misspecified latent space network model}
\label{sub:reconstruction_logi}
In this simulation setup, we {\color{black}inspect the} reconstruction accuracy of the network {\color{black} with respect to network model misspecification}. Instead of generating an adjacency matrix $\bA$ from the GRDPG edge probability matrix $\bP = \bX_0 \bI_{d_+,d_-}\bX_0^\top$, we generate an adjacency matrix from the logistic link edge probability matrix $\bP = \sigma(\bX_0 \bI_{d_+,d_-} \bX_0^\top)$, where $\sigma(x) = 1/\{1+\exp(-x)\}$ is applied elementwise, i.e.,  $P_{ij} = 1/[1+\exp\{-(\bX_0 \bI_{d_+,d_-}\bX_0^\top)_{ij}\}]$. We then estimate the latent positions through our CGRDPG framework and compute the estimated edge probability matrix $\widehat{\bP} = \widehat{\bX} \widehat{\bI}_{d_+,d_-}\widehat{\bX}^\top$. The network reconstruction performance is assessed with the relative Frobenius norm $\|\bP_0 - \hat{\bP}\|_F^2 / \|\bP_0\|_F^2$. {\color{black}Similarly, we consider ASE and OSE as alternative methods. Below are the specific simulation designs and experiments.}

\subsubsection{One-dimensional latent positions}
\label{ssub:one_dimensional_misspecified_latent_space_network}
{\color{black}Consider
$n=500$ nodes, $p=250$ covariates, and $d = 1$ latent position dimension}. The latent {\color{black}positions $\bX_0 = [x_{01},\ldots,x_{0n}]\transpose$ are} generated as $x_{0i} = 1.75 \{0.15
\sin(\pi i/n) + 0.6\}$, where the scale factor $1.75$ controls the overall
magnitude of the latent positions. Under this setting, the resulting
inner product matrix $\mathbf{X}_0\mathbf{X}_0\transpose\color{black}$ {\color{black}has elements ranging} from $1.1025$ to
$1.7227$, which translate into edge probabilities
ranging from $0.7507$ to $0.8485$ (mean $0.8139$) {\color{black}via the logistic link function}.
{\color{black}This results in} a moderately dense,
non-saturated network with reasonable probability diversity.
The covariates are generated as follows: First, we generate a covariate coefficient matrix $\mathbf{Z}_0 \in \mathbb{R}^{p \times d}$ with entries drawn independently from the standard normal distribution $N(0,1)$. Then, the covariate matrix $\mathbf{B} \in \mathbb{R}^{p \times n}$ is constructed as $\mathbf{B} =
\mathbf{Z}_0 \mathbf{X}_0^\top + \bW$, where
$\bW$ is a $p \times n$ noise matrix with entries independently drawn from $N(0,1)$. This structure ensures that the covariates contain information about the latent positions while
including realistic noise.
The {\color{black}Fisher-scoring algorithm for the proposed CGRDPG model} uses {\color{black}the following hyperparameters}:
$T_{\max} = 3$ (maximum number of iterations), $\epsilon = 0.01$ (convergence tolerance), and $\tau = 0.001$ (smoothing threshold). {\color{black}The} results are averaged over $100$ {\color{black}independent} replications {\color{black}and reported in Table \ref{tab:misspecified_1D} below}.

\begin{table}[ht]
      \centering
      \begin{tabular}{lcc}
      \toprule
      Method & Mean $\pm$ SD & As Percentage \\
      \midrule
      MSLE & $0.000875 \pm 0.000054$ & $0.087\% \pm 0.005\%$ \\
      ASE    & $0.000913 \pm 0.000055$ & $0.091\% \pm 0.006\%$ \\
      OSE    & $0.000910 \pm 0.000055$ & $0.091\% \pm 0.006\%$ \\
      \bottomrule
      \end{tabular}
      \caption{Simulation results for \Cref{ssub:one_dimensional_misspecified_latent_space_network}: Relative Frobenius norms $\|\bP_0 - \hat{\bP}\|_F^2 / \|\bP_0\|_F^2$. }
      \label{tab:misspecified_1D}
      \end{table}
Our proposed MSLE achieves $0.087\%$ relative error, and all three methods perform comparably well in this setting. 
Despite the misspecified link function, the underlying edge probability matrix is recovered {\color{black}with small relative errors}, 
suggesting that the proposed framework is robust to link misspecification in the {\color{black}one-dimensional latent position scenario.}

\subsubsection{Two-dimensional latent positions}
\label{ssub:two_dimensional_misspecified_latent_space_network}

We next consider a two-dimensional logistic link scenario, again with
$n=500$ nodes and $p=250$ covariates, but now with latent
position dimension $d=2$. The {\color{black}latent position matrix $\bX_0 = [\bx_{01},\ldots,\bx_{0n}]\transpose$ is}
generated as
$\bx_{0i} = [1.25\{0.15 \sin(\pi i/n) + 0.6\}, 1.00 \{0.15 \cos(\pi i/n) + 0.6\}]\transpose$ for $i\in[n]$. 
The resulting inner product {\color{black}matrix $\mathbf{X}_0\mathbf{X}_0\transpose$ has elements ranging}
from $0.7668$ to $1.2881$, which map to edge
probabilities ranging from $0.6828$ to $0.7838$.
{\color{black}This also results in }a non-saturated network with {\color{black}a well-behaved} probability diversity. 
The covariate matrix $\mathbf{B} \in \mathbb{R}^{p \times n}$ is generated as $\mathbf{B} = \mathbf{Z}_0 \mathbf{X}_0^\top + \mathbf{W}$, where $\mathbf{Z}_0 \in \mathbb{R}^{p \times d}$ is the covariate coefficient matrix and $\mathbf{W} \in \mathbb{R}^{p \times n}$ is a noise matrix, both with entries independently drawn from the standard normal distribution $N(0,1)$.
{\color{black}The algorithm hyperparameters are set to be the same as those in \Cref{ssub:one_dimensional_misspecified_latent_space_network},}
with results averaged over 100 {\color{black}independent replicates. The results are reported in Table \ref{tab:misspecified_2D}}.
      \begin{table}[ht]
      \centering
      \begin{tabular}{lcc}
      \toprule
      Method & Mean $\pm$ SD & As Percentage \\
      \midrule
      MSLE & $0.003164 \pm 0.000534$ & $0.316\% \pm 0.053\%$ \\
      ASE    & $0.003868 \pm 0.000097$ & $0.387\% \pm 0.010\%$ \\
      OSE    & $0.003764 \pm 0.000146$ & $0.376\% \pm 0.015\%$ \\
      \bottomrule
      \end{tabular}
      \caption{Simulation results for \Cref{ssub:two_dimensional_misspecified_latent_space_network}: Relative Frobenius norms $\|\bP_0 - \hat{\bP}\|_F^2 / \|\bP_0\|_F^2$. }
      \label{tab:misspecified_2D}
      \end{table}

MSLE achieves $0.32\%$ relative error
{\color{black}and} is 18\% better than ASE and
{\color{black}around} $16\%$ better than OSE. Taken together, these findings indicate that CGRDPG is robust to model misspecification, maintaining reliable network reconstruction performance across both the {\color{black}one-dimensional and two-dimensional latent position} scenarios.

\section{Real-World Application: Last.fm Social Graph Analysis}\label{sec:Last.fm}

To evaluate the practical utility of the {\color{black}proposed methodology}, we applied the Fisher-scoring {\color{black}algorithm} to a real-world social networking dataset of Asian users on the online music streaming site Last.fm collected from the public API by \citet{rozemberczki2020characteristic}. This dataset represents a user-user social graph from the Last.fm music service, consisting of $n=1{,}699$ users. Each user is associated with a 600-dimensional covariate vector ($p=600$) representing their listening history across various musical artists. The primary objective is to embed these users into a latent space to determine if their social connections and musical preferences can accurately recover their geographical origins, represented by five distinct country labels. To find an appropriate latent space dimension, we note that under a stochastic block model, the rank of the edge probability matrix is at most the number of blocks, so the latent dimension is bounded above by the number of communities \citep{rubin2022statistical}. Since the Last.fm network consists of five communities, we considered candidate dimensions $d \in \{1, \ldots, 5\}$.
For each candidate dimension, we ran the Fisher-scoring algorithm
for a maximum of 50 iterations, with a stopping tolerance of $0.1$
on the maximum row change. Specifically, the stopping criterion was
\[
\max_{i \in [n]}
\left\|
\widehat{\bx}_i^{(t+1)}-\widehat{\bx}_i^{(t)}
\right\|_2 \leq 0.1.
\]
We computed BIC using the surrogate likelihood evaluated at the
final iterate for each candidate dimension.

As shown in Table~\ref{tab:lastfm_dimsel}, the algorithm satisfied
the stopping criterion for $d=1,2,3$, whereas the runs for $d=4,5$
reached the iteration limit without satisfying this criterion.
Theorems~\ref{thm:OnestepFisher} and~\ref{thm:efficiency}
establish that, under the stated assumptions, statistical guarantees
can be obtained after one or finitely many Fisher-scoring updates,
without requiring numerical convergence.
We therefore report the final iterates for all candidate dimensions,
including those that reached the iteration limit.
Among the reported fits, BIC identifies $d=2$ as the optimal dimension, which is illustrated in Table~\ref{tab:lastfm_dimsel}.

\begin{table}[ht]
\centering
\begin{tabular}{crrrcc}
\toprule
$d$ & Neg-Log-Lik & \# Params & BIC & Converged & Iterations \\
\midrule
1 & 106,390 & 2,298 & 229,872 & \checkmark & 4 \\
2 & 93,778 & 4,594 & 221,726 & \checkmark & 5 \\
3 & 88,354 & 6,888 & 227,939 & \checkmark & 18 \\
4 & 84,710 & 9,180 & 237,698 & -- & 50 \\
5 & 81,105 & 11,470 & 247,522 & -- & 50 \\
\bottomrule
\end{tabular}
\caption{Dimension selection for the Last.fm network using the surrogate likelihood.}
\label{tab:lastfm_dimsel}
\end{table}

The Last.fm social graph presents a significant challenge for statistical methods due to its extreme sparsity, with a network density of only 0.49\%. In such a sparse environment, the adjacency matrix
contains {\color{black}scarce} signal, and {\color{black}it is desirable for a method to integrate covariate information and improve network estimation. For comparison, we also consider ASE and OSE as the alternative methods. }
Table~\ref{tab:cluster_lastfm} reports clustering performance at
embedding dimension $d=2$, as measured by the Silhouette score,
the Adjusted Rand Index (ARI), and the ratio of between-cluster
sum of squares to total sum of squares (BSS/TSS).
For a partition $\{C_1,\ldots,C_K\}$ of the embedded users, this ratio
is defined as
\[
\frac{\mathrm{BSS}}{\mathrm{TSS}}
=
\frac{\sum_{k=1}^{K}|C_k|
\|\bar{\bx}_k-\bar{\bx}\|_2^2}
{\sum_{i=1}^{n}\|\widehat{\bx}_i-\bar{\bx}\|_2^2},
\qquad
\bar{\bx}_k=\frac{1}{|C_k|}\sum_{i\in C_k}\widehat{\bx}_i,
\quad
\bar{\bx}=\frac{1}{n}\sum_{i=1}^{n}\widehat{\bx}_i,
\]
where $\widehat{\bx}_i$ denotes the embedding of user $i$ and
$C_k$ denotes the set of users assigned to cluster $k$.
Larger values indicate that a greater proportion of the total
variation in the embeddings is explained by differences between
cluster means.

Table~\ref{tab:cluster_lastfm} compares the clustering performance of
ASE, OSE, and MSLE at the BIC-selected dimension $d=2$.
MSLE achieves the highest Silhouette score ($-0.074$) and ARI ($0.100$),
indicating better cluster separation and greater agreement with the
country labels than ASE and OSE. Nevertheless, the negative Silhouette
scores and relatively low ARI values across all three methods suggest
that recovering geographical groups remains challenging, likely due
in part to the high sparsity of the data. Overall, despite its slightly
lower BSS/TSS ratio ($0.840$), MSLE provides the strongest evidence
of geographically meaningful clustering among the three methods.

\begin{table}[htbp]
\centering
\begin{tabular}{lrrr}
\toprule
Method & Silhouette & ARI & BSS/TSS \\
\midrule
ASE    & $-0.239$ & 0.022 & 0.868 \\
OSE    & $-0.207$ & 0.032 & 0.855 \\
MSLE   & $-0.074$ & 0.100 & 0.840 \\
\bottomrule
\end{tabular}
\caption{Clustering performance for the Last.fm network with $d=2$ (BIC-optimal
dimension).}
\label{tab:cluster_lastfm}
\end{table}

A critical technical finding in our algorithmic implementation is the
success of the modified clipping strategy rooted in the construction of the surrogate likelihood. In this extremely sparse network, the initial latent positions obtained via ASE yielded edge probabilities where approximately 84.71\% of the upper-triangular entries of the estimated edge probability matrix violate the standard numerical bounds $[\tau, 1-\tau]$ with $\tau = 0.001$ required for stable computation. While the original Fisher-scoring algorithm would fail under these conditions due to numerical instability, our modified approach, which clips the edge probabilities \textit{only} within the Fisher information denominator (but not in the score
function), enables the optimization to proceed effectively. This
modification significantly improved the model fit: the surrogate log-likelihood increased from $-105{,}701$ to $-93{,}778$ over $5$ iterations, representing an $11.3\%$ reduction in the numerical discrepancy between the proposed model and the observed adjacency structure, demonstrating the efficacy of the clipping-based Fisher-scoring approach.

In summary, the proposed surrogate optimization algorithm demonstrates remarkable empirical robustness, successfully achieving convergence within $5$ iterations with the final maximum row change of $0.0833$ falling well below the convergence threshold of $0.1$. The algorithm successfully navigated a highly sparse and non-convex likelihood surface to reach an improved local optimum. These findings suggest that even when initial spectral estimates provide a challenging starting point, the modified surrogate optimization approach remains a powerful tool for refining latent embeddings in highly sparse, real-world social networks.

\section{Discussion}
\label{sec:discussion}

Our primary contribution lies in the development of a framework that seamlessly integrates node-level covariate information $\mathbf{B}$ with the observed network adjacency matrix $\mathbf{A}$ through a surrogate likelihood function. Unlike traditional spectral embedding methods, such as Adjacency Spectral Embedding (ASE) or the One-Step Estimator (OSE), which rely primarily on the connectivity patterns of the graph, our approach utilizes the covariates to guide the curvature of the log-likelihood surface. By incorporating these features into the Fisher information matrix and the score function, the {\color{black}covariate term in the objective function} acts as a structural regularizer that is particularly vital in sparse regimes where the network signal is weak. In such cases, the covariates provide a necessary bridge to stabilize the estimation of latent positions $\mathbf{X}$ and reduce the {\color{black}variations} inherent in pure graph-based embeddings.

Theoretically, we have demonstrated that the {\color{black}proposed Fisher-scoring estimator} maintains desirable properties, including consistency and asymptotic normality, {\color{black}thereby} providing a robust foundation for statistical inference. These theoretical strengths {\color{black}are} validated through extensive numerical simulations, where the {\color{black}proposed estimator} consistently {\color{black}yields vertex-wise confidence sets that have} stable {\color{black}empirical} coverage rates. The empirical utility of this method {\color{black}is} most notably highlighted in the analysis of the Last.fm dataset: despite the extreme sparsity of the user-user social graph, the {\color{black}proposed estimate achieves} a {\color{black}significant} improvement in the Adjusted Rand Index (ARI) over ASE. This improvement suggests that the modified Fisher-scoring update is exceptionally capable of capturing the underlying homophily, where musical preferences drive social ties, to recover geographical clusters that spectral methods failed to resolve. By {\color{black}viewing the Fisher-scoring estimate} as a refinement step, we demonstrate that incorporating side information {\color{black}in contextual covariates} can substantially enhance the precision of latent space representations where traditional embeddings reach their information limits.

However, applying the surrogate likelihood comes with significant caveats, particularly regarding numerical stability and uncertainty quantification. Our simulations reveal that the empirical coverage rate is highly sensitive to the distribution of estimated edge probabilities $\hat{P}_{ij}$. When these probabilities are crowded within a very narrow range, the variances of the scores become exceptionally small, causing the condition number of the Fisher information matrix to increase {\color{black}rapidly}. An ill-conditioned information matrix represents a severe hurdle for reliable inference and {\color{black}implementation}, as the matrix inversion process required for Fisher-scoring amplifies minute floating-point errors. 

\section{Disclosure statement}\label{disclosure-statement}

The authors declare that there is no conflict of interest. 

\section{Data Availability Statement}\label{data-availability-statement}

The data that support the findings of this study are openly available in Kaggle at \url{https://www.kaggle.com/datasets/harshal19t/lastfm-dataset}.

\section*{Declaration of generative AI use}
The authors acknowledge the interactive use of generative AI (Anthropic models
Claude Opus 5.5 and Claude Fable 5.1--5.2) for checking the mathematical correctness, including the calculations and proofs of the propositions, theorems, and auxiliary lemmas. The authors have derived and verified the mathematical contents, and assume full responsibility for the contents of the work. 

\vspace*{5ex}

\begin{center}
\Large Supplementary Material for  ``Estimating Contextual Networks via Fisher-Scoring Surrogate Likelihood''
\end{center}

This supplementary material contains the proofs of the theoretical results in Section 4 of the manuscript.    


\appendix

\section{Preliminaries and Conventions}\label{sec:notation}

\paragraph{Oracle quantities.}
Throughout, ``oracle'' refers to the true coordinates
$(\bX_0,\bY_0,\bZ_0)$ and quantities evaluated using them.
An oracle score is evaluated at $(\bx_{0i},\bY_0,\bZ_0)$ while retaining
the observed data $\bA$ and $\bB$; it therefore remains random.
The term does not mean that the score has been replaced by its expectation.

\paragraph{Diagonal correction for the hollow graph.}
All network sums in this supplement over edges include $j=i$, as in the update.
Conditional on the oracle coordinates, only the off-diagonal
entries $E_{jk}$, $j<k$, are independent centered Bernoulli errors;
$E_{kj}=E_{jk}$ and $E_{jj}=-P_{jj}$ are deterministic. In particular,
$|E_{jj}|\le\rho_n$ and $\operatorname{Var}(E_{jj})=0$.
Every concentration argument for a full network sum is understood through
\[
 \sum_j E_{ij}\bv_j=\sum_{j\ne i}E_{ij}\bv_j-P_{ii}\bv_i,
 \qquad
 \sum_{j\le k}E_{jk}\bC_{jk}
 =\sum_{j<k}E_{jk}\bC_{jk}-\sum_jP_{jj}\bC_{jj}.
\]
Bernstein's inequality is applied to the off-diagonal sum; the displayed
diagonal corrections are bounded directly. Algebraic expansions retain
the full matrix $\bE$, including its diagonal.

\paragraph{Surrogate likelihood revisited.} With threshold $\tau=\delta\rho_n$
($\delta\in(0,1/2)$ fixed), $\psi_1,\psi_2$ the quadratic extensions of
$\log(x)$ and $\log(1-x)$ defined in the main text, $\psi=\psi_1-\psi_2$
and $\Psi(x)=\int_0^x\psi(t)\,\mathrm{d}t$, the vertex-wise surrogate log-likelihood is
\[
\widetilde\ell_{in}(\bx_i;\widehat\bY,\widehat\bZ)
=\sum_{j=1}^n\bigl\{(A_{ij}-\bx_i^\top\widehat\by_j)\psi(\bx_i^\top\widehat\by_j)
+\Psi(\bx_i^\top\widehat\by_j)\bigr\}
-\frac12\sum_{l=1}^p(B_{li}-\widehat\bz_l^\top\bx_i)^2 .
\]
We write $\ell_{in}=\widetilde\ell_{in}$ for this surrogate likelihood.
Its unnormalized score function is
\[
 \bS_i(\bx_i;\bY,\bZ)
 \coloneqq\frac{\partial\ell_{in}}{\partial\bx_i}(\bx_i;\bY,\bZ).
\]
When the nuisance-coordinate arguments are clear, we write $\bS_i(\bx_i)$.
Throughout, $\bs_{in}$ denotes its normalized oracle evaluation:
\[
 \bs_{in}\coloneqq\frac1{n+p}\bS_i(\bx_{0i};\bY_0,\bZ_0).
\]
Thus $\bS_i$ is a function of its coordinate arguments, whereas
$\bs_{in}$ is the random vector obtained by evaluating that function
at the true coordinates and dividing by $n+p$.
For general $(\bx_i,\bY,\bZ)$, the expected negative Hessian is
\[
\calI_{in}(\bx_i;\bY,\bZ)=-\E\left\{\frac{\partial^2\ell_{in}}{\partial\bx_i\partial\bx_i^\top}(\bx_i;\bY,\bZ)\right\}.
\]
We retain the full-sum working Fisher matrix used in the update:
\[
\bG_{in}=\frac1{n+p}\left\{\sum_{j=1}^n\psi'(P_{ij})\by_{0j}\by_{0j}^\top
+\sum_{l=1}^p\bz_{0l}\bz_{0l}^\top\right\}.
\]
For the hollow graph this differs from the expected negative Hessian
of the full-sum surrogate likelihood. Keeping each $\by_{0j}$ fixed
when differentiating the vertex-wise likelihood gives
\begin{align*}
 \E\{\bS_i(\bx_{0i})\}
 &=-P_{ii}\psi'(P_{ii})\by_{0i},\\
 \frac{\calI_{in}(\bx_{0i};\bY_0,\bZ_0)}{n+p}
 &=\bG_{in}+\frac{P_{ii}\psi''(P_{ii})}{n+p}
                 \by_{0i}\by_{0i}^{\top}.
\end{align*}
Thus $\bG_{in}$ denotes the working normalized Fisher matrix, rather
than exactly the expected negative Hessian under the hollow model.
The bounds $|P_{ii}|\le\rho_n$, $\|\by_{0i}\|_2\le\sqrt{\rho_n}$,
$|\psi'(P_{ii})|\lesssim_\delta\rho_n^{-1}$ and
$|\psi''(P_{ii})|\lesssim_\delta\rho_n^{-2}$ imply
\[
 \frac{\|\E\{\bS_i(\bx_{0i})\}\|_2}{n+p}
 \lesssim_\delta\frac{\sqrt{\rho_n}}n,
 \qquad
 \left\|\frac{\calI_{in}(\bx_{0i};\bY_0,\bZ_0)}{n+p}-\bG_{in}\right\|_2
 \lesssim_\delta\frac1n.
\]
The perturbation is $O(n^{-1})$. Once the inverse bounds established
below hold, the inverse identity also gives
\[
 \left\|\left\{\frac{\calI_{in}(\bx_{0i};\bY_0,\bZ_0)}{n+p}\right\}^{-1}
       -\bG_{in}^{-1}\right\|_2\lesssim_\delta n^{-1}.
\]
Moreover, the normalized diagonal score bias is $O(\sqrt{\rho_n}/n)$,
so its product with $\sqrt n$ tends to zero. It does not alter the
leading centered fluctuation at the $n^{-1/2}$ scale.
The update and all score decompositions below use the full-sum working
matrix $\widehat\bG_{in}^{(t)}$ and retain the diagonal score contribution.
No self-pair is deleted from the likelihood or update.

\paragraph{Coordinates and alignment.}
We work in oracle spectral coordinates compatible with the signature:
$\bX_0=[\bX_{0+},\bX_{0-}]$, where the two blocks correspond to the
positive and negative eigenspaces of $\bP$, respectively, and
$\bX_{0+}^{\top}\bX_{0-}=\mathbf 0$.
The columns of the adjacency spectral embedding are ordered with the
same positive/negative partition. We align the two blocks separately,
using the orthogonal alignment obtained from the corresponding spectral
perturbation argument for $\widehat\bX^{(0)}$. Thus, throughout the
proofs, the alignment has the form
\[
 \bQ=\operatorname{diag}(\bQ_+,\bQ_-),
 \qquad \bQ_+^{\top}\bQ_+=\bI_{d_+},
 \qquad \bQ_-^{\top}\bQ_-=\bI_{d_-}.
\]
Consequently,
\[
 \bQ\bI_{d_+,d_-}=\bI_{d_+,d_-}\bQ,
 \qquad \bQ^{\top}\bI_{d_+,d_-}\bQ=\bI_{d_+,d_-}.
\]
The same alignment is used for the subsequent iterates and their
covariate estimates. In particular, with
$\widehat\bY^{(t)}=\widehat\bX^{(t)}\bI_{d_+,d_-}$, we have
\[
 \bQ^{\top}\widehat\by_j^{(t)}-\by_{0j}
 =\bI_{d_+,d_-}\bigl(\bQ^{\top}\widehat\bx_j^{(t)}-\bx_{0j}\bigr).
\]
Whenever approximate linearization is used below, its alignment is
required to satisfy this block structure. These identities rely on the
specified coordinates and alignment; orthogonality alone does not
imply commutation with $\bI_{d_+,d_-}$.

\begin{definition}[Approximate linearization \citep{xie2023efficient}]\label{def:linearization}
$\widehat\bX=[\widehat\bx_1,\ldots,\widehat\bx_n]^\top$ is
\emph{approximately linearized} with orthogonal alignment $\bQ\in\mathbb O(d)$
and remainder $\bR_\bX=[\br_{\bx_1},\ldots,\br_{\bx_n}]^\top$ if there
exist deterministic $\{\bds\alpha_{nij}\}$, $\{\bds\beta_{nil}\}$ with
$\max_{i,j}\|\bds\alpha_{nij}\|_2=O\{(n\rho_n^{1/2})^{-1}\}$,
$\max_{i,l}\|\bds\beta_{nil}\|_2=O(1/n)$, such that
\[
\bQ^\top\widehat\bx_i-\bx_{0i}=\sum_{j=1}^n\bds\alpha_{nij}E_{ij}+\sum_{l=1}^p\bds\beta_{nil}W_{li}+\br_{\bx_i},
\qquad \max_{i\in[n]}\|\br_{\bx_i}\|_2=\widetilde O_p\{(\log n)^{2+\eps}/(n\rho_n^{1/2})\}\ \forall \eps>0.
\]
\end{definition}

\medskip
\noindent We use $\lesssim,\gtrsim,\asymp$ with the usual meaning
and let $C,c,C_0,c_0,C_\delta,K_c,\ldots$ denote generic constants that
may change from line to line but never depend on $n$.

\section{Proof Roadmap}\label{sec:roadmap}
The proofs proceed from initialization to the score expansion and its
remainder bounds, followed by the one-step and iterative stochastic
expansions. \Cref{lem:X0,lem:Z0} provide the initialization bounds.
\Cref{res:linearization-rate} converts approximate linearization into
an entrywise bound. The score decomposition \eqref{lem:taylor} and
\eqref{eq:term1}--\eqref{eq:term8} identify the terms controlled in
\Cref{sec:taylor}. These bounds are assembled in \Cref{sec:onestep}, which proves \Cref{thm:OnestepFisher};
\Cref{prop:first-step} then verifies the linearization conditions
for the first iterate.
Finally, \Cref{sec:iterative} proves \Cref{thm:efficiency} and
gives the iterative expansion.
Initialization and leave-one-out spectral estimates are used in the
remainder bounds as well as in the initialization analysis.

\section{Stochastic Expansions for the Initialization}\label{sec:init}

\subsection{Adjacency spectral embedding \texorpdfstring{$\widehat\bX^{(0)}$}{X0}}\label{sec:X0}

\begin{lemma}[Analysis of $\widehat\bX^{(0)}$]
\label[lemma]{lem:X0}
Under Assumption~\ref{assump:snr_network}, there exists
$\bQ\in\mathbb{O}(d)$ such that
\[
\widehat{\bX}^{(0)}\bQ-\bX_0
=
\bE\bX_0(\bX_0^\top\bX_0)^{-1}\bI_{d_+,d_-}
+\bR_{\bX}^{(0)},
\]
where
\[
\|\bR_{\bX}^{(0)}\|_{\mathrm{F}}
=
\widetilde O_p\!\left\{
\frac{(1\vee\rho_n\log n)^{1/2}}
     {(n\rho_n)^{1/2}}
\right\}.
\]
For any $i\in[n]$ and any sequence $(m_n)$ with $m_n=O(n)$,
\[
\|\be_i^\top\bR_{\bX}^{(0)}\|_2
=
\widetilde O_p\!\left\{
\frac{\log m_n}{n\rho_n^{1/2}};\,m_n
\right\},
\qquad
\|\bQ^\top\widehat{\bx}_i^{(0)}-\bx_{0i}\|_2
=
\widetilde O_p\!\left\{
\frac{(\log m_n)^{1/2}}{\sqrt n};\,m_n
\right\}.
\]
Furthermore,
\[
\|\widehat{\bX}^{(0)}\bQ-\bX_0\|_2
=
\widetilde O_p(1),
\qquad
\|\widehat{\bX}^{(0)}\bQ-\bX_0\|_{2\to\infty}
=
\widetilde O_p\!\left\{
\frac{(\log n)^{1/2}}{\sqrt n}
\right\}.
\]
\end{lemma}

The linear expansion and rowwise bounds in \Cref{lem:X0} follow from
\citep[Theorem~3.2]{xie2024entrywise}.

\subsection{Initial covariate estimator \texorpdfstring{$\widehat\bZ^{(0)}$}{Z0}}\label{sec:Z0}

\begin{lemma}[Analysis of $\widehat\bZ^{(0)}$]\label[lemma]{lem:Z0}
Under \Cref{assump:snr_network,assump:snr_covariates},
\[
\widehat\bZ^{(0)}\bQ-\bZ_0=\bW\bX_0(\bX_0^\top\bX_0)^{-1}+\bR^{(0)}_\bZ,
\]
where for each fixed $l\in[p]$ and any deterministic nonnegative
diverging sequence $(m_n)_{n=1}^\infty$ with $m_n=O(n)$,
\[
\|\be_l^\top\bR^{(0)}_\bZ\|_2=\widetilde O_p\!\left\{\frac{(\log n)^{1/2}}{n\rho_n}+\frac{(\log m_n)^{1/2}}{n\rho_n};m_n\right\}.
\]
In particular
\[
\|\widehat\bZ^{(0)}\bQ-\bZ_0\|_{2\to\infty}=\widetilde O_p\!\left\{\frac{(\log n)^{1/2}}{(n\rho_n)^{1/2}}\right\},
\qquad
\|\widehat\bZ^{(0)}\bQ-\bZ_0\|_2=\widetilde O_p\!\left\{\frac{(p+n)^{1/2}}{(n\rho_n)^{1/2}}\right\}.
\]
\end{lemma}

\begin{proof}
Throughout write $\widehat\bX_\bA\coloneqq\widehat\bX^{(0)}$ for brevity,
and recall from \Cref{sec:notation} that
$\bB=\bZ_0\bX_0^\top+\bW$ and
$\widehat\bZ^{(0)}=\bB\widehat\bX_\bA(\widehat\bX_\bA^\top\widehat\bX_\bA)^{-1}$.
We record three consequences of
\Cref{assump:snr_network,assump:snr_covariates} used repeatedly below:
\begin{equation}\label{eq:Z0-prelim}
\|\bX_0\|_2\le\|\bX_0\|_\F\lesssim(n\rho_n)^{1/2},
\;
\|(\bX_0^\top\bX_0)^{-1}\|_2=\sigma_d(\bX_0)^{-2}\lesssim\frac1{n\rho_n},
\;
\|\bZ_0\|_2\le\sqrt p\,\|\bZ_0\|_{2\to\infty}\lesssim\sqrt p,
\end{equation}
together with $\|\widehat\bX_\bA\|_2\lesssim(n\rho_n)^{1/2}$ and
$\|\widehat\bX_\bA\|_{2\to\infty}\lesssim\rho_n^{1/2}$, which follow from
\Cref{lem:X0} and \eqref{eq:Z0-prelim} since
$\|\widehat\bX_\bA\bQ-\bX_0\|_2=\widetilde O_p(1)$ and
$\|\widehat\bX_\bA\bQ-\bX_0\|_{2\to\infty}=\widetilde O_p\{(\log n)^{1/2}/\sqrt n\}$.

\emph{Step 1: exact decomposition.}
Since $\bQ$ is orthogonal,
$\widehat\bX_\bA(\widehat\bX_\bA^\top\widehat\bX_\bA)^{-1}\bQ
=\widehat\bX_\bA\bQ(\bQ^\top\widehat\bX_\bA^\top\widehat\bX_\bA\bQ)^{-1}$,
so that
\begin{align}
\widehat\bZ^{(0)}\bQ-\bZ_0
&=(\bZ_0\bX_0^\top)\widehat\bX_\bA\bQ(\bQ^\top\widehat\bX_\bA^\top\widehat\bX_\bA\bQ)^{-1}
+\bW\widehat\bX_\bA\bQ(\bQ^\top\widehat\bX_\bA^\top\widehat\bX_\bA\bQ)^{-1}-\bZ_0
\nonumber\\
&=(\bZ_0\bX_0^\top)\widehat\bX_\bA\bQ(\bX_0^\top\bX_0)^{-1}
+(\bZ_0\bX_0^\top)\widehat\bX_\bA\bQ\bigl\{(\bQ^\top\widehat\bX_\bA^\top\widehat\bX_\bA\bQ)^{-1}-(\bX_0^\top\bX_0)^{-1}\bigr\}
\nonumber\\
&\quad+\bW\widehat\bX_\bA\bQ(\bX_0^\top\bX_0)^{-1}
+\bW\widehat\bX_\bA\bQ\bigl\{(\bQ^\top\widehat\bX_\bA^\top\widehat\bX_\bA\bQ)^{-1}-(\bX_0^\top\bX_0)^{-1}\bigr\}-\bZ_0 .
\label{eq:Z0-split}
\end{align}
For the first summand, insert \Cref{lem:X0}:
\begin{align}
(\bZ_0\bX_0^\top)\widehat\bX_\bA\bQ(\bX_0^\top\bX_0)^{-1}
&=(\bZ_0\bX_0^\top)\bigl\{\bX_0+(\widehat\bX_\bA\bQ-\bX_0)\bigr\}(\bX_0^\top\bX_0)^{-1}\nonumber\\
&=\bZ_0
+(\bZ_0\bX_0^\top)\bigl\{\bE\bX_0(\bX_0^\top\bX_0)^{-1}\bI_{d_+,d_-}+\bR^{(0)}_\bX\bigr\}(\bX_0^\top\bX_0)^{-1},
\label{eq:Z0-cancel}
\end{align}
the term $\bZ_0\bX_0^\top\bX_0(\bX_0^\top\bX_0)^{-1}=\bZ_0$ cancelling
the $-\bZ_0$ of \eqref{eq:Z0-split}. For the third summand, likewise,
\begin{equation}\label{eq:Z0-lead}
\bW\widehat\bX_\bA\bQ(\bX_0^\top\bX_0)^{-1}
=\bW\bX_0(\bX_0^\top\bX_0)^{-1}+\bW(\widehat\bX_\bA\bQ-\bX_0)(\bX_0^\top\bX_0)^{-1},
\end{equation}
which extracts the leading term of the lemma. Substituting
\eqref{eq:Z0-cancel} and \eqref{eq:Z0-lead} into \eqref{eq:Z0-split} and
combining the two bracketed summands gives
\[
\widehat\bZ^{(0)}\bQ-\bZ_0=\bW\bX_0(\bX_0^\top\bX_0)^{-1}+\bR^{(0)}_\bZ,
\]
with the remainder given explicitly by
\begin{equation}\label{eq:RZ}
\boxed{
\begin{aligned}
\bR^{(0)}_\bZ
&=\underbrace{\bZ_0\bX_0^\top\bE\bX_0(\bX_0^\top\bX_0)^{-1}\bI_{d_+,d_-}(\bX_0^\top\bX_0)^{-1}}_{=:\,\bT_1}
+\underbrace{\bZ_0\bX_0^\top\bR^{(0)}_\bX(\bX_0^\top\bX_0)^{-1}}_{=:\,\bT_2}
+\underbrace{\bW(\widehat\bX_\bA\bQ-\bX_0)(\bX_0^\top\bX_0)^{-1}}_{=:\,\bT_3}\\
&\quad+\underbrace{(\bZ_0\bX_0^\top+\bW)\widehat\bX_\bA\bQ\bigl\{(\bQ^\top\widehat\bX_\bA^\top\widehat\bX_\bA\bQ)^{-1}-(\bX_0^\top\bX_0)^{-1}\bigr\}}_{=:\,\bT_4}.
\end{aligned}}
\end{equation}
Before bounding these terms separately, their exact regrouping is
\begin{align}\label{eq:Z0-exact-regroup}
 \bR_{\bZ}^{(0)}
 &=-\bZ_0(\widehat\bX_\bA\bQ-\bX_0)^\top\widehat\bX_\bA\bQ
 (\bQ^\top\widehat\bX_\bA^\top\widehat\bX_\bA\bQ)^{-1}\notag\\
 &\quad+\bW\left\{\widehat\bX_\bA\bQ
 (\bQ^\top\widehat\bX_\bA^\top\widehat\bX_\bA\bQ)^{-1}
 -\bX_0(\bX_0^\top\bX_0)^{-1}\right\}.
\end{align}
Indeed, the signal part follows from
\[
 \bX_0^\top\widehat\bX_\bA\bQ
 -\bQ^\top\widehat\bX_\bA^\top\widehat\bX_\bA\bQ
 =-(\widehat\bX_\bA\bQ-\bX_0)^\top\widehat\bX_\bA\bQ.
\]
This regrouping cancels the zeroth-order signal, but not its entire
first-order error. To identify the latter, the derivative at $\bX_0$
of the signal map
$\mathbf U\mapsto\bZ_0\bX_0^\top\mathbf U(\mathbf U^\top\mathbf U)^{-1}$
in a direction $\mathbf V$ is
\[
 \bZ_0\bX_0^\top\mathbf V(\bX_0^\top\bX_0)^{-1}
 -\bZ_0(\bX_0^\top\mathbf V+\mathbf V^\top\bX_0)
 (\bX_0^\top\bX_0)^{-1}
 =-\bZ_0\mathbf V^\top\bX_0(\bX_0^\top\bX_0)^{-1}.
\]
Thus no algebraic identity cancels all signal fluctuations responsible
for the logarithmic high-probability bound.

The four terms are bounded in Steps 2--5.

\emph{Step 2: the term $\bT_1$.}
Fix $l\in[p]$. Write
\[
 \bX_0^\top\bE\bX_0
 =\sum_{j<k}E_{jk}(\bx_{0j}\bx_{0k}^\top+\bx_{0k}\bx_{0j}^\top)
 -\sum_jP_{jj}\bx_{0j}\bx_{0j}^\top.
\]
The off-diagonal coefficients have Frobenius norm $O(\rho_n)$.
Bernstein's inequality in the fixed $d^2$ coordinates gives a bound
$\widetilde O_p\{n\rho_n\sqrt{\rho_n\log n}+\rho_n\log n\}$.
The diagonal correction has norm at most $n\rho_n^2$.
Consequently, using $n\rho_n\gtrsim\log n$,
\[
 \|\bX_0^\top\bE\bX_0(\bX_0^\top\bX_0)^{-1}\|_2
 =\widetilde O_p\{\sqrt{1\vee\rho_n\log n}\}.
\]
The diagonal contribution to $\bT_1$ has row norm at most
$C\|\bZ_0\|_{2\to\infty}/n$ and operator norm at most
$C\|\bZ_0\|_2/n$, by the two inverse factors in its definition.
These are absorbed by the following bounds:
\begin{equation}\label{eq:T1}
\begin{aligned}
 \|\bT_1\|_{2\to\infty}
 &=\widetilde O_p\!\left\{\|\bZ_0\|_{2\to\infty}
 \frac{(\log n)^{1/2}}{n\rho_n^{1/2}}\right\},\\
 \|\bT_1\|_2
 &\le\|\bZ_0\|_2\|\bX_0^\top\bE\bX_0\|_2
 \|(\bX_0^\top\bX_0)^{-1}\|_2^2=\widetilde O_p\!\left\{\|\bZ_0\|_2
 \frac{(\log n)^{1/2}}{n\rho_n^{1/2}}\right\},
\end{aligned}
\end{equation}
where we used $\bI_{d_+,d_-}^\top\bI_{d_+,d_-}=\bI_d$.

\emph{Step 3: the term $\bT_2$.}
Directly from \eqref{eq:Z0-prelim} and the spectral bound
\[
\|\bR^{(0)}_\bX\|_2\le\|\bR^{(0)}_\bX\|_\F
=\widetilde O_p\{(1\vee \rho_n\log n)^{1/2}/(n\rho_n)^{1/2}\}\]
from Lemma \ref{lem:X0}, for any $l\in[p]$,
\begin{align}
\|\be_l^\top\bT_2\|_2
&\le\|\bZ_0\|_{2\to\infty}\|\bX_0\|_2\|\bR^{(0)}_\bX\|_2\|(\bX_0^\top\bX_0)^{-1}\|_2
=\widetilde O_p\!\left\{\|\bZ_0\|_{2\to\infty}\frac{(\log n)^{1/2}}{n\rho_n}\right\},
\label{eq:T2row}\\
\|\bT_2\|_2
&\le\|\bZ_0\|_2\|\bX_0\|_2\|\bR^{(0)}_\bX\|_2\|(\bX_0^\top\bX_0)^{-1}\|_2
=\widetilde O_p\!\left\{\|\bZ_0\|_2\frac{(\log n)^{1/2}}{n\rho_n^{1/2}}\right\}.
\label{eq:T2op}
\end{align}

\emph{Step 4: the term $\bT_3$.}
Here we use that $\bW$ is independent of $\bA$, hence of
$\widehat\bX_\bA$ and of $\bE$, and condition on the latter. Given
$\widehat\bX_\bA$, the $l$th row of
$\bW(\widehat\bX_\bA\bQ-\bX_0)$ is a centered Gaussian vector in $\RR^d$
with covariance
$(\widehat\bX_\bA\bQ-\bX_0)^\top(\widehat\bX_\bA\bQ-\bX_0)$, whose
spectral norm is \emph{exactly} $\|\widehat\bX_\bA\bQ-\bX_0\|_2^2$.
Sub-Gaussian concentration with $d$ fixed, a union bound over $l\in[p]$,
and \Cref{lem:X0} therefore give, for any $m_n=O(n)$,
\begin{equation}\label{eq:T3row}
\|\be_l^\top\bT_3\|_2
=\widetilde O_p\!\left\{\frac{(\log m_n)^{1/2}\|\widehat\bX_\bA\bQ-\bX_0\|_2}{n\rho_n};m_n\right\}
=\widetilde O_p\!\left\{\frac{(\log m_n)^{1/2}}{n\rho_n};m_n\right\},
\end{equation}
since $\|\widehat\bX_\bA\bQ-\bX_0\|_2=\widetilde O_p(1)$ by
\Cref{lem:X0}. In the spectral norm, by
$\|\bW\|_2=\widetilde O_p\{(p+n)^{1/2}\}$
\citep[Cor.\,3.11]{bandeira2016sharp},
\begin{equation}\label{eq:T3op}
\|\bT_3\|_2
\le\|\bW\|_2\|\widehat\bX_\bA\bQ-\bX_0\|_2\|(\bX_0^\top\bX_0)^{-1}\|_2
=\widetilde O_p\!\left\{\frac{(p+n)^{1/2}(1\vee\log n)^{1/2}}{n\rho_n}\right\}.
\end{equation}

\emph{Step 5: the term $\bT_4$.}
We first bound the bracketed inverse difference. By the identity
$\bC^{-1}-\bD^{-1}=\bC^{-1}(\bD-\bC)\bD^{-1}$ with
$\bC=\bQ^\top\widehat\bX_\bA^\top\widehat\bX_\bA\bQ$ and
$\bD=\bX_0^\top\bX_0$, and the expansion
$\bQ^\top\widehat\bX_\bA^\top\widehat\bX_\bA\bQ-\bX_0^\top\bX_0
=(\widehat\bX_\bA\bQ-\bX_0)^\top(\widehat\bX_\bA\bQ-\bX_0)
+(\widehat\bX_\bA\bQ-\bX_0)^\top\bX_0+\bX_0^\top(\widehat\bX_\bA\bQ-\bX_0)$,
\begin{align}
&\bigl\|(\bQ^\top\widehat\bX_\bA^\top\widehat\bX_\bA\bQ)^{-1}-(\bX_0^\top\bX_0)^{-1}\bigr\|_2\nonumber\\
&\quad\le\|(\widehat\bX_\bA^\top\widehat\bX_\bA)^{-1}\|_2\,
\bigl\|\bQ^\top\widehat\bX_\bA^\top\widehat\bX_\bA\bQ-\bX_0^\top\bX_0\bigr\|_2\,
\|(\bX_0^\top\bX_0)^{-1}\|_2\nonumber\\
&\quad\le\|(\widehat\bX_\bA^\top\widehat\bX_\bA)^{-1}\|_2
\Bigl\{\|\widehat\bX_\bA\bQ-\bX_0\|_2^2
+2\bigl\|\bX_0^\top(\widehat\bX_\bA\bQ-\bX_0)\bigr\|_2\Bigr\}
\|(\bX_0^\top\bX_0)^{-1}\|_2\nonumber\\
&\quad\le\|(\widehat\bX_\bA^\top\widehat\bX_\bA)^{-1}\|_2
\Bigl\{\|\widehat\bX_\bA\bQ-\bX_0\|_2^2 +2\bigl\|\bX_0^\top\bE\bX_0(\bX_0^\top\bX_0)^{-1}\bI_{d_+,d_-}\bigr\|_2 +2\|\bX_0\|_2\|\bR^{(0)}_\bX\|_2\Bigr\}
\|(\bX_0^\top\bX_0)^{-1}\|_2\nonumber\\
&\quad=\widetilde O_p\!\left\{\frac{(1 \vee \rho_n\log n)^{1/2}}{(n\rho_n)^2}\right\},
\label{eq:inverse-diff}
\end{align}
using \Cref{lem:X0} and the bounds
\[
\begin{aligned}
 \|(\widehat\bX_\bA^\top\widehat\bX_\bA)^{-1}\|_2
 &\lesssim(n\rho_n)^{-1},\;
 \|\widehat\bX_\bA\bQ-\bX_0\|_2^2
 =\widetilde O_p(1),\\
 \|\bX_0^\top\bE\bX_0(\bX_0^\top\bX_0)^{-1}\|_2
 &=\widetilde O_p\{(1\vee\rho_n\log n)^{1/2}\},\;
 \|\bX_0\|_2\|\bR^{(0)}_\bX\|_2
 =\widetilde O_p\{(1\vee\rho_n\log n)^{1/2}\}.
\end{aligned}
\]

Next we need $\|\bW\widehat\bX_\bA\|_{2\to\infty}$. Conditioning on
$\widehat\bX_\bA$, which is independent of $\bW$, the $l$th row of
$\bW\widehat\bX_\bA$ is centered Gaussian with covariance
$\widehat\bX_\bA^\top\widehat\bX_\bA$; sub-Gaussian concentration and a
union bound over $l\in[p]$, with $p=O(n)$, give
\begin{equation}\label{eq:WXA}
\|\bW\widehat\bX_\bA\|_{2\to\infty}
=\widetilde O_p\bigl\{(n\log n)^{1/2}\|\widehat\bX_\bA\|_{2\to\infty}\bigr\}
=\widetilde O_p\bigl\{(n\rho_n\log n)^{1/2}\bigr\}.
\end{equation}
Combining \eqref{eq:inverse-diff} and \eqref{eq:WXA} with
$\|\widehat\bX_\bA\|_2\lesssim(n\rho_n)^{1/2}$ and \eqref{eq:Z0-prelim},
\begin{align}
\|\bT_4\|_{2\to\infty}
&\le\bigl(\|\bZ_0\|_{2\to\infty}\|\bX_0\|_2\|\widehat\bX_\bA\|_2+\|\bW\widehat\bX_\bA\|_{2\to\infty}\bigr)
\bigl\|(\bQ^\top\widehat\bX_\bA^\top\widehat\bX_\bA\bQ)^{-1}-(\bX_0^\top\bX_0)^{-1}\bigr\|_2\nonumber\\
&=\widetilde O_p\!\left\{\|\bZ_0\|_{2\to\infty}\frac{(1 \vee \rho_n\log n)^{1/2}}{n\rho_n}
+\frac{(\log n)^{1/2}(1 \vee \rho_n\log n)^{1/2}}{(n\rho_n)^{3/2}}\right\},
\label{eq:T4row}\\
\|\bT_4\|_2
&\le\bigl(\|\bZ_0\|_2\|\bX_0\|_2\|\widehat\bX_\bA\|_2+\|\bW\|_2\|\widehat\bX_\bA\|_2\bigr)
\bigl\|(\bQ^\top\widehat\bX_\bA^\top\widehat\bX_\bA\bQ)^{-1}-(\bX_0^\top\bX_0)^{-1}\bigr\|_2\nonumber\\
&=\widetilde O_p\!\left\{\|\bZ_0\|_2\frac{(1 \vee \log n)^{1/2}}{n\rho_n}
+\frac{(p+n)^{1/2}(1 \vee \rho_n\log n)^{1/2}}{(n\rho_n)^{3/2}}\right\}.
\label{eq:T4op}
\end{align}

\emph{Step 6: assembly.}
Adding \eqref{eq:T1}, \eqref{eq:T2row}, \eqref{eq:T3row} and
\eqref{eq:T4row}, and using $\|\bZ_0\|_{2\to\infty}=O(1)$
(\Cref{assump:snr_covariates}) together with
$(\log n)^{1/2}/(n\rho_n)^{1/2}\lesssim1$, every contribution except that
of $\bT_3$ is $\widetilde O_p\{(\log n)^{1/2}/(n\rho_n)\}$,
whence for each fixed $l\in[p]$
\[
\|\be_l^\top\bR^{(0)}_\bZ\|_2
=\widetilde O_p\!\left\{\frac{(\log n)^{1/2}}{n\rho_n}+\frac{(\log m_n)^{1/2}}{n\rho_n};m_n\right\},
\]
which is the stated rowwise bound. Adding \eqref{eq:T1},
\eqref{eq:T2op}, \eqref{eq:T3op} and \eqref{eq:T4op}, and using
$\|\bZ_0\|_2\lesssim\sqrt p$,
\begin{equation}\label{eq:RZop}
\begin{aligned}
\|\bR^{(0)}_\bZ\|_2
&=\widetilde O_p\!\left\{\|\bZ_0\|_2\frac{(1\vee\log n)^{1/2}}{n\rho_n}
+\frac{(p+n)^{1/2}}{n\rho_n}\right\}\\
&=\widetilde O_p\!\left\{\frac{\sqrt{p\log n}+\sqrt{p+n}}{n\rho_n}\right\}.
\end{aligned}
\end{equation}

\emph{Step 7: the leading term.}
It remains to bound $\bW\bX_0(\bX_0^\top\bX_0)^{-1}$. Its $l$th row is a
centered Gaussian vector with covariance
$(\bX_0^\top\bX_0)^{-1}\bX_0^\top\bX_0(\bX_0^\top\bX_0)^{-1}=(\bX_0^\top\bX_0)^{-1}$,
of spectral norm $\lesssim(n\rho_n)^{-1}$; sub-Gaussian concentration and
a union bound over $l\in[p]$ with $p=O(n)$ give
\begin{equation}\label{eq:lead-2inf}
\|\bW\bX_0(\bX_0^\top\bX_0)^{-1}\|_{2\to\infty}
=\widetilde O_p\!\left\{\frac{(\log n)^{1/2}}{(n\rho_n)^{1/2}}\right\},
\end{equation}
while $\|\bW\|_2=\widetilde O_p\{(p+n)^{1/2}\}$
\citep[Cor.\,3.11]{bandeira2016sharp} and
$\|\bX_0(\bX_0^\top\bX_0)^{-1}\|_2=\sigma_d(\bX_0)^{-1}\lesssim(n\rho_n)^{-1/2}$
give
\begin{equation}\label{eq:lead-op}
\|\bW\bX_0(\bX_0^\top\bX_0)^{-1}\|_2
=\widetilde O_p\!\left\{\frac{(p+n)^{1/2}}{(n\rho_n)^{1/2}}\right\}.
\end{equation}
Taking $m_n=n$ in the rowwise remainder bound and applying a
union bound over $l\in[p]$, using $p=O(n)$, yields
\[
\|\bR_{\bZ}^{(0)}\|_{2\to\infty}
=
\widetilde O_p\!\left\{
\frac{(\log n)^{1/2}}{n\rho_n}
\right\},
\]
since $(1\vee\rho_n\log n)^{1/2}\lesssim(\log n)^{1/2}$.
This is dominated by \eqref{eq:lead-2inf}.
For the spectral norm, retain the bound
\[
\|\bR_{\bZ}^{(0)}\|_2
=
\widetilde O_p\!\left\{
\frac{\|\bZ_0\|_2(1\vee\rho_n\log n)^{1/2}
      +(p+n)^{1/2}}{n\rho_n}
\right\}.
\]
Using $\|\bZ_0\|_2\lesssim\sqrt p\le\sqrt{p+n}$ and
$(1\vee\rho_n\log n)^{1/2}/(n\rho_n)^{1/2}\lesssim1$,
this is dominated by \eqref{eq:lead-op}.
The two global assertions now follow from the triangle inequality.
\end{proof}

\section{Linearization and Entrywise Bounds}\label{sec:result-linearization}

This section derives an entrywise bound from the approximate
linearization in \Cref{def:linearization}. The first-iterate
linearization is verified in \Cref{sec:first-step}, after the one-step
stochastic expansion. These results supply the entrywise estimates
and the base case used in \Cref{sec:iterative}.


\begin{result}\label{res:linearization-rate}
Suppose $\widehat\bX=[\widehat\bx_1,\ldots,\widehat\bx_n]^\top$ is
approximately linearized with orthogonal alignment $\bQ\in\mathbb O(d)$
(\Cref{def:linearization}). That is, for each $i\in[n]$,
\[
\bQ^\top\widehat\bx_i-\bx_{0i}
=\sum_{j=1}^n\bds\alpha_{nij}E_{ij}+\sum_{l=1}^p\bds\beta_{nil}W_{li}+\br_{\bx_i},
\]
where
\[
\max_{i,j\in[n]}\|\bds\alpha_{nij}\|_2=O\!\left(\frac{1}{n\rho_n^{1/2}}\right),
\qquad
\max_{i\in[n],\,l\in[p]}\|\bds\beta_{nil}\|_2=O\!\left(\frac{1}{n}\right),
\]
and, for every $\eps>0$,
\[
\max_{i\in[n]}\|\br_{\bx_i}\|_2=\widetilde O_p\!\left(\frac{(\log n)^{2+\eps}}{n\rho_n^{1/2}}\right).
\]
If \Cref{assump:3} holds, in particular $n\rho_n=\Omega\{(\log n)^{4\xi}\}$
for some $\xi>1$, then
\[
\|\widehat\bX\bQ-\bX_0\|_{2\to\infty}
=\widetilde O_p\!\left(\frac{(\log n)^{1/2}}{\sqrt n}\right).
\]
\end{result}

\begin{proof}
For each $i\in[n]$, write
\[
\bL_{Ei}\coloneqq\sum_{j=1}^n\bds\alpha_{nij}E_{ij},
\qquad
\bL_{Wi}\coloneqq\sum_{l=1}^p\bds\beta_{nil}W_{li}.
\]
Then $\bQ^\top\widehat\bx_i-\bx_{0i}=\bL_{Ei}+\bL_{Wi}+\br_{\bx_i}$, and hence
\begin{equation}\label{eq:res-basic-decomp}
\|\widehat\bX\bQ-\bX_0\|_{2\to\infty}
\le\max_{i\in[n]}\|\bL_{Ei}\|_2+\max_{i\in[n]}\|\bL_{Wi}\|_2+\max_{i\in[n]}\|\br_{\bx_i}\|_2 .
\end{equation}
We control the three terms separately.

\emph{1. Network-noise term.} Conditional on $\bX_0$, the variables
$E_{ij}=A_{ij}-P_{ij}$, $j\ne i$, are centered and independent, and satisfy $\Var(E_{ij}\mid\bX_0)=P_{ij}(1-P_{ij})\lesssim\rho_n$.
Fix $i$ and a unit vector $\bu\in\RR^d$. Then $\bu^\top\bL_{Ei}=\sum_{j=1}^n\bu^\top\bds\alpha_{nij}E_{ij}$,
whose conditional variance proxy satisfies
\[
\sum_{j=1}^n\Var(\bu^\top\bds\alpha_{nij}E_{ij}\mid\bX_0)
\lesssim\rho_n\sum_{j=1}^n(\bu^\top\bds\alpha_{nij})^2
\le\rho_n\sum_{j=1}^n\|\bds\alpha_{nij}\|_2^2
\le n\rho_n\max_j\|\bds\alpha_{nij}\|_2^2
=O\!\left(\frac1n\right),
\]
and $|\bu^\top\bds\alpha_{nij}E_{ij}|\le\|\bds\alpha_{nij}\|_2=O\{(n\rho_n^{1/2})^{-1}\}$.
The diagonal contribution to $\bL_{Ei}$ is
$-P_{ii}\bds\alpha_{nii}$ and has norm $O(\sqrt{\rho_n}/n)$.
Bernstein's inequality for the off-diagonal sum, followed by addition
of this deterministic correction, therefore gives
$|\bu^\top\bL_{Ei}|=\widetilde O_p\{\sqrt{\log n/n}+\log n/(n\rho_n^{1/2})\}$
uniformly over $i\in[n]$; since $d$ is fixed, a finite-net argument yields
\[
\max_{i\in[n]}\|\bL_{Ei}\|_2=\widetilde O_p\!\left(\sqrt{\frac{\log n}{n}}+\frac{\log n}{n\rho_n^{1/2}}\right).
\]
Under \Cref{assump:3},
\begin{equation}\label{eq:res-absorb}
\frac{\log n/(n\rho_n^{1/2})}{\sqrt{\log n/n}}=\frac{(\log n)^{1/2}}{(n\rho_n)^{1/2}}=O\{(\log n)^{1/2-2\xi}\}=o(1),
\end{equation}
so
\begin{equation}\label{eq:res-E-bound}
\max_{i\in[n]}\|\bL_{Ei}\|_2=\widetilde O_p\!\left(\sqrt{\frac{\log n}{n}}\right).
\end{equation}

\emph{2. Covariate-noise term.} The entries $W_{li}$ are $\iid\ N(0,1)$,
hence independent sub-Gaussian. For fixed $i$ and unit $\bu\in\RR^d$,
$\bu^\top\bL_{Wi}=\sum_{l=1}^p\bu^\top\bds\beta_{nil}W_{li}$ is
sub-Gaussian with variance proxy bounded by
\[
\sum_{l=1}^p(\bu^\top\bds\beta_{nil})^2
\le\sum_{l=1}^p\|\bds\beta_{nil}\|_2^2
\le p\max_l\|\bds\beta_{nil}\|_2^2
=O\!\left(\frac{p}{n^2}\right)=O\!\left(\frac1n\right)
\]
under $p=O(n)$ (\Cref{assump:snr_covariates}). Sub-Gaussian
concentration, a union bound over $i\in[n]$, and a finite-net argument
in $\RR^d$ then give
\begin{equation}\label{eq:res-W-bound}
\max_{i\in[n]}\|\bL_{Wi}\|_2=\widetilde O_p\!\left(\sqrt{\frac{\log n}{n}}\right).
\end{equation}

\emph{3. Remainder term.} By approximate linearization,
$\max_i\|\br_{\bx_i}\|_2=\widetilde O_p\{(\log n)^{2+\eps}/(n\rho_n^{1/2})\}$.
Moreover,
\[
\frac{(\log n)^{2+\eps}/(n\rho_n^{1/2})}{(\log n)^{1/2}/\sqrt n}
=\frac{(\log n)^{3/2+\eps}}{(n\rho_n)^{1/2}},
\]
and \Cref{assump:3} gives $(n\rho_n)^{1/2}\gtrsim(\log n)^{2\xi}$, so
\[
\frac{(\log n)^{3/2+\eps}}{(n\rho_n)^{1/2}}\lesssim(\log n)^{3/2+\eps-2\xi}.
\]
Since $\xi>1$, choose $\eps>0$ small enough that $3/2+\eps-2\xi<0$.
Therefore
\begin{equation}\label{eq:res-R-bound}
\max_{i\in[n]}\|\br_{\bx_i}\|_2=o_p\!\left(\sqrt{\frac{\log n}{n}}\right).
\end{equation}
Combining \eqref{eq:res-basic-decomp}--\eqref{eq:res-R-bound} gives
$\|\widehat\bX\bQ-\bX_0\|_{2\to\infty}=\widetilde O_p\{(\log n)^{1/2}/\sqrt n\}$.
\end{proof}

\section{Score Expansion and Remainder Bounds}\label{sec:taylor}

\begin{lemma}\label[lemma]{lem:score-remainder}
Under Assumptions~\ref{assump:snr_network} and~\ref{assump:snr_covariates},
if $\tau=\delta\rho_n$ for some constant $\delta\in(0,1/2)$, then,
for any fixed $i\in[n]$ and deterministic nonnegative diverging
sequence $(m_n)_{n=1}^{\infty}$ with $m_n=O(n)$,
\[
 \|\mathbf r_{i1}^{(0)}\|_2
 =\tOp\left\{\frac{\sqrt{\log m_n}}{\sqrt n}
 +\frac{\log m_n+\sqrt{\log n}}
 {n\sqrt{\rho_n}};m_n\right\},
\]
and
\[
 \left\|\mathbf r_{i1}^{(0)}+
 \frac1{n+p}\left\{
 \sum_{j=1}^n\psi'(\mathbf x_{0i}^\top\mathbf y_{0j})
 \mathbf y_{0j}\mathbf y_{0j}^\top
 +\sum_{l=1}^p\mathbf z_{0l}\mathbf z_{0l}^\top
 \right\}(\mathbf Q^\top\widehat{\mathbf x}_i^{(0)}-\mathbf x_{0i})\right\|_2
 =\tOp\left\{\frac{\log m_n+\sqrt{\log n}}
 {n\sqrt{\rho_n}};m_n\right\}.
\]
Furthermore, if Assumption~\ref{assump:3} holds,
$\widehat{\mathbf X}^{(t)}$ is approximately linearized with orthogonal
alignment $\mathbf Q$ in the sense of Definition~\ref{def:linearization},
and the remainder
\(
 \mathbf R_{\mathbf Z}^{(t)}
 =\widehat{\mathbf Z}^{(t)}\mathbf Q-\mathbf Z_0
 -\mathbf W\mathbf X_0(\mathbf X_0^\top\mathbf X_0)^{-1}
\)
satisfies
\[
 \|\mathbf R_{\mathbf Z}^{(t)}\|_{2\to\infty}
 =\tOp\left\{\frac{(\log n)^{2+\epsilon}}{n\rho_n}\right\},
\]
then for any $\epsilon>0$,
\begin{align}
&\|\mathbf r_{i1}^{(t)}\|_2
 =\tOp\left\{\frac{\sqrt{\log n}}{\sqrt n}\right\},\notag\\
&\left\|\mathbf r_{i1}^{(t)} +
 \frac1{n+p}\left\{
 \sum_{j=1}^n\psi'(\mathbf x_{0i}^\top\mathbf y_{0j})
 \mathbf y_{0j}\mathbf y_{0j}^\top
 +\sum_{l=1}^p\mathbf z_{0l}\mathbf z_{0l}^\top
 \right\}(\mathbf Q^\top\widehat{\mathbf x}_i^{(t)}-\mathbf x_{0i})\right\|_2
 =\tOp\left\{\frac{(\log n)^{2+\epsilon}}{n\sqrt{\rho_n}}\right\}.\notag
\end{align}
\end{lemma}

Fix $t\ge0$. We evaluate the functions below at the aligned coordinates
$\bQ^\top\widehat\bx_i^{(t)}$, $\bQ^\top\widehat\by_j^{(t)}$,
and $\bQ^\top\widehat\bz_l^{(t)}$.
For $\bu,\bv,\bz\in\RR^d$, define
\begin{equation}\label{eq:ghphi}
\begin{aligned}
g(\bu,\bv)&=\psi'(\bu^\top\bv)\bv,\\
h_{ij}(\bu,\bv)&=\psi'(\bu^\top\bv)(\bx_{0i}^\top\by_{0j}-\bu^\top\bv)\bv,\\
\phi_{1ij}(\bu,\bv)&=E_{ij}g(\bu,\bv)+h_{ij}(\bu,\bv),\\
\phi_{2il}(\bu,\bz)&=W_{li}\bz+(\bx_{0i}^\top\bz_{0l}-\bu^\top\bz)\bz.
\end{aligned}
\end{equation}
The first-order Taylor expansions are
\begin{align*}
g(\bu,\bv)-g(\bx_{0i},\by_{0j})
&=\psi''(\bx_{0i}^\top\by_{0j})\by_{0j}\by_{0j}^\top(\bu-\bx_{0i})\\
&\quad+\{\psi''(\bx_{0i}^\top\by_{0j})\by_{0j}\bx_{0i}^\top
+\psi'(\bx_{0i}^\top\by_{0j})\bI_d\}(\bv-\by_{0j})
+\br_g(\bu,\bv,\bx_{0i},\by_{0j}),\\
h_{ij}(\bu,\bv)-h_{ij}(\bx_{0i},\by_{0j})
&=-\psi'(\bx_{0i}^\top\by_{0j})\by_{0j}\by_{0j}^\top(\bu-\bx_{0i})\\
&\quad-\psi'(\bx_{0i}^\top\by_{0j})\by_{0j}\bx_{0i}^\top(\bv-\by_{0j})
+\br_{h_{ij}}(\bu,\bv,\bx_{0i},\by_{0j}),\\
\phi_{2il}(\bu,\bz)-\phi_{2il}(\bx_{0i},\bz_{0l})
&=W_{li}(\bz-\bz_{0l})-\bz_{0l}\bx_{0i}^\top(\bz-\bz_{0l})\\
&\quad-\bz_{0l}\bz_{0l}^\top(\bu-\bx_{0i})
+\br_{\phi_{2il}}(\bu,\bz,\bx_{0i},\bz_{0l}).
\end{align*}
The remainder bounds are established in Lemma~\ref{lem:Taylor_bounds}.

\paragraph{Decomposition of the one-step update.}
Summing $\phi_{1ij}$ over $j\in[n]$ and $\phi_{2il}$ over $l\in[p]$
and dividing by $n+p$ yields the normalized score at the aligned coordinates.
Substituting this expression into the one-step update and using
$(\widehat\bG_{in}^{(t)})^{-1}=\bG_{in}^{-1}+\bR_{i2}^{(t)}$
gives, for every $i\in[n]$ and $t\ge0$,
\begin{equation}\label{lem:taylor}
\begin{aligned}
\bQ^\top\widehat\bx_i^{(t+1)}-\bx_{0i}
&=\bG_{in}^{-1}\frac1{n+p}
\frac{\partial\ell_{in}}{\partial\bx_i}(\bx_{0i};\bY_0,\bZ_0)\\
&\quad+\bG_{in}^{-1}\{\bG_{in}(\bQ^\top\widehat\bx_i^{(t)}-\bx_{0i})
+\br_{i1}^{(t)}\}
+\bR_{i2}^{(t)}\br_{i1}^{(t)}+\br_{i3}^{(t)}.
\end{aligned}
\end{equation}
Here $\bR_{i2}^{(t)}=(\widehat\bG_{in}^{(t)})^{-1}-\bG_{in}^{-1}$,
\[
\br_{i3}^{(t)}=\bR_{i2}^{(t)}\left\{
\frac1{n+p}\sum_{j=1}^n\psi'(\bx_{0i}^\top\by_{0j})E_{ij}\by_{0j}
+\frac1{n+p}\sum_{l=1}^pW_{li}\bz_{0l}\right\},
\]
and $\br_{i1}^{(t)}$ is the difference between the normalized scores
at the aligned estimated coordinates and at the oracle coordinates.
Substituting the preceding first-order Taylor expansions into this
score difference and collecting terms yields, for every $t\ge0$,
the following eight terms whose sum is $\br_{i1}^{(t)}$:
\begin{align}
\br_{i1}^{(t)} 
=& \frac1{n+p}\Biggl[
\sum_{j=1}^n
\Bigl\{
\phi_{1ij}(\bQ^\top\widehat\bx_i^{(t)},
          \bQ^\top\widehat\by_j^{(t)})
-\phi_{1ij}(\bx_{0i},\by_{0j})\Bigr\} +\sum_{l=1}^p
\Bigl\{
\phi_{2il}(\bQ^\top\widehat\bx_i^{(t)},
          \bQ^\top\widehat\bz_l^{(t)})
-\phi_{2il}(\bx_{0i},\bz_{0l})
\Bigr\}\Biggr]\notag\\
=&\frac1{n+p}\sum_{j=1}^nE_{ij}\psi''(\bx_{0i}^\top\by_{0j})\by_{0j}\by_{0j}^\top(\bQ^\top\widehat\bx_i^{(t)}-\bx_{0i}) \label{eq:term1}\\
&-\frac1{n+p}\Bigl\{\sum_{j=1}^n\psi'(\bx_{0i}^\top\by_{0j})\by_{0j}\by_{0j}^\top+\sum_{l=1}^p\bz_{0l}\bz_{0l}^\top\Bigr\}(\bQ^\top\widehat\bx_i^{(t)}-\bx_{0i}) \label{eq:term2}\\
&-\frac1{n+p}\sum_{j=1}^n\psi'(\bx_{0i}^\top\by_{0j})\by_{0j}\bx_{0i}^\top(\bQ^\top\widehat\by_j^{(t)}-\by_{0j}) \label{eq:term3}\\
&+\frac1{n+p}\Bigl\{\sum_{j=1}^n\br_{h_{ij}}(\bQ^\top\widehat\bx_i^{(t)},\bQ^\top\widehat\by_j^{(t)},\bx_{0i},\by_{0j}) +\sum_{l=1}^p\br_{\phi_{2il}}(\bQ^\top\widehat\bx_i^{(t)},\bQ^\top\widehat\bz_l^{(t)},\bx_{0i},\bz_{0l})\Bigr\} \label{eq:term4}\\
&+\frac1{n+p}\sum_{j=1}^nE_{ij}\{\psi''(\bx_{0i}^\top\by_{0j})\by_{0j}\bx_{0i}^\top+\psi'(\bx_{0i}^\top\by_{0j})\bI_d\}(\bQ^\top\widehat\by_j^{(t)}-\by_{0j}) \label{eq:term5}\\
&+\frac1{n+p}\sum_{j=1}^nE_{ij}\br_g(\bQ^\top\widehat\bx_i^{(t)},\bQ^\top\widehat\by_j^{(t)},\bx_{0i},\by_{0j}) \label{eq:term6}\\
&+\frac1{n+p}\sum_{l=1}^pW_{li}(\bQ^\top\widehat\bz_l^{(t)}-\bz_{0l}) \label{eq:term7}\\
&-\frac1{n+p}\sum_{l=1}^p\bz_{0l}\bx_{0i}^\top(\bQ^\top\widehat\bz_l^{(t)}-\bz_{0l}). \label{eq:term8}
\end{align}
We bound these terms in the following subsections and then combine
the bounds to control $\br_{i1}^{(t)}$ and
$\br_{i1}^{(t)}+\bG_{in}(\bQ^\top\widehat\bx_i^{(t)}-\bx_{0i})$.

\subsection{Taylor Expansions and Remainder Bounds}
\label{sec:g-lip}

Let $\psi=\psi_1-\psi_2$ be the extended logarithmic surrogate already
defined in the manuscript, with $\tau=\delta\rho_n$ and fixed
$0<\delta<1/2$. Its definition gives
\begin{equation}\label{eq:global}
 \sup_{s\in\RR}|\psi''(s)|\le\frac2{\tau^2},\qquad
 |\psi''(s)-\psi''(t)|\le\frac4{\tau^3}|s-t|\quad(s,t\in\RR).
\end{equation}
In particular, $|\psi'''(s)|\lesssim_\delta\rho_n^{-3}$ almost everywhere.

Indeed, each extended logarithm has a continuous second derivative
bounded in absolute value by $\tau^{-2}$, and its piecewise third
derivative is bounded by $2\tau^{-3}$. Adding these bounds for
$\psi_1-\psi_2$ gives \eqref{eq:global}. The continuity at the joining
points allows the piecewise third-derivative bound to be integrated
across those points as well.

On the range needed below,
\begin{equation}\label{eq:local}
 \sup_{|s|\le K^2\rho_n}|\psi'(s)|
 \lesssim_{\delta,K}\rho_n^{-1}.
\end{equation}
To verify this local bound explicitly, note that
$\psi'(0)=2/\tau+1$. By \eqref{eq:global} and the fundamental theorem
of calculus, for $|s|\le K^2\rho_n$,
\[
 |\psi'(s)|
 \le |\psi'(0)|+\frac{2}{\tau^2}|s|
 \le\left(\frac2\delta+1+\frac{2K^2}{\delta^2}\right)\rho_n^{-1},
\]
where $\rho_n\le1$ and $\tau=\delta\rho_n$ were used in the last step.
Thus, no lower bound on the inner products is needed.

Write $\mathcal X(r)=\{\mathbf{x}\in\RR^d:\left\|\mathbf{x}\right\|_2\le r\}$ and fix $i,j$.
Define
\[
 \mathbf{g}(\mathbf{u},\mathbf{v})=\psi'(\mathbf{u}^\top \mathbf{v})\mathbf{v},\qquad
 \mathbf{h}_{ij}(\mathbf{u},\mathbf{v})=\psi'(\mathbf{u}^\top \mathbf{v})(\mathbf{x}_{0i}^\top\mathbf{y}_{0j}-\mathbf{u}^\top \mathbf{v})\mathbf{v}.
\]
For the covariate score, define
\[
 \boldsymbol{\phi}_{2il}(\mathbf{u},\mathbf{z})=W_{li}\mathbf{z}+(\mathbf{x}_{0i}^\top \mathbf{z}_{0l}-\mathbf{u}^\top \mathbf{z})\mathbf{z}.
\]
For any of these vector-valued functions, its remainder means the
first-order Taylor remainder at the oracle arguments. For example,
\[
 \mathbf{r}_{\mathbf{g}}(\mathbf{u},\mathbf{v},\mathbf{x}_{0i},\mathbf{y}_{0j})=\mathbf{g}(\mathbf{u},\mathbf{v})-\mathbf{g}(\mathbf{x}_{0i},\mathbf{y}_{0j})
 -D_{\mathbf{u}}\mathbf{g}(\mathbf{x}_{0i},\mathbf{y}_{0j})(\mathbf{u}-\mathbf{x}_{0i})-D_{\mathbf{v}}\mathbf{g}(\mathbf{x}_{0i},\mathbf{y}_{0j})(\mathbf{v}-\mathbf{y}_{0j}).
\]
Here $D_{\mathbf{u}}\mathbf{g}$ and $D_{\mathbf{v}}\mathbf{g}$ are the
$d\times d$ Jacobian matrices with respect to the indicated vector
arguments. Their explicit expressions are
\begin{align*}
 D_{\mathbf{u}}\mathbf{g}(\mathbf{x}_{0i},\mathbf{y}_{0j})
 &=\psi''(\mathbf{x}_{0i}^{\top}\mathbf{y}_{0j})
       \mathbf{y}_{0j}\mathbf{y}_{0j}^{\top},\\
 D_{\mathbf{v}}\mathbf{g}(\mathbf{x}_{0i},\mathbf{y}_{0j})
 &=\psi''(\mathbf{x}_{0i}^{\top}\mathbf{y}_{0j})
       \mathbf{y}_{0j}\mathbf{x}_{0i}^{\top}
   +\psi'(\mathbf{x}_{0i}^{\top}\mathbf{y}_{0j})\mathbf I_d.
\end{align*}

\begin{lemma}[Taylor bounds]\label[lemma]{lem:Taylor_bounds}
Under the existing setup, let $K\ge1$ bound the scaled oracle
coordinates and the arguments under consideration:
\[
 \mathbf{x}_{0i},\mathbf{y}_{0j},\mathbf{u},\mathbf{v}\in\mathcal X(K\sqrt{\rho_n}),\qquad
 \mathbf{z}_{0l},\mathbf{z}\in\mathcal X(K).
\]
There are constants depending only on $\delta,K$ (and only on $K$ for
the covariate remainder) such that
\begin{align}
 \left\|\mathbf{g}(\mathbf{u},\mathbf{v})-\mathbf{g}(\mathbf{x}_{0i},\mathbf{y}_{0j})\right\|_2
 &\lesssim_{\delta,K}\rho_n^{-1}
       (\left\|\mathbf{u}-\mathbf{x}_{0i}\right\|_2+\left\|\mathbf{v}-\mathbf{y}_{0j}\right\|_2),\label{eq:gLip}\\
 \left\|\mathbf{h}_{ij}(\mathbf{u},\mathbf{v})-\mathbf{h}_{ij}(\mathbf{x}_{0i},\mathbf{y}_{0j})\right\|_2
 &\lesssim_{\delta,K}\left\|\mathbf{u}-\mathbf{x}_{0i}\right\|_2+\left\|\mathbf{v}-\mathbf{y}_{0j}\right\|_2,\label{eq:hLip}\\
 \left\|\mathbf{r}_{\mathbf{g}}(\mathbf{u},\mathbf{v},\mathbf{x}_{0i},\mathbf{y}_{0j})\right\|_2
 &\lesssim_{\delta,K}\rho_n^{-3/2}
       (\left\|\mathbf{u}-\mathbf{x}_{0i}\right\|_2^2+\left\|\mathbf{v}-\mathbf{y}_{0j}\right\|_2^2),\label{eq:rg}\\
 \left\|\mathbf{r}_{\mathbf{h}_{ij}}(\mathbf{u},\mathbf{v},\mathbf{x}_{0i},\mathbf{y}_{0j})\right\|_2
 &\lesssim_{\delta,K}\rho_n^{-1/2}
       (\left\|\mathbf{u}-\mathbf{x}_{0i}\right\|_2^2+\left\|\mathbf{v}-\mathbf{y}_{0j}\right\|_2^2),\label{eq:rh}\\
 \left\|\mathbf{r}_{\boldsymbol{\phi}_{2il}}(\mathbf{u},\mathbf{z},\mathbf{x}_{0i},\mathbf{z}_{0l})\right\|_2
 &\lesssim_K \sqrt{\rho_n}\left\|\mathbf{z}-\mathbf{z}_{0l}\right\|_2^2
                 +\left\|\mathbf{u}-\mathbf{x}_{0i}\right\|_2\left\|\mathbf{z}-\mathbf{z}_{0l}\right\|_2.\label{eq:rphi}
\end{align}
\end{lemma}

\begin{proof}
Let $\mathbf{a}=\mathbf{u}-\mathbf{x}_{0i}$, $\mathbf{b}=\mathbf{v}-\mathbf{y}_{0j}$, and set
\[
 \mathbf{u}_t=\mathbf{x}_{0i}+t\mathbf{a},\qquad \mathbf{v}_t=\mathbf{y}_{0j}+t\mathbf{b},\qquad s_t=\mathbf{u}_t^\top \mathbf{v}_t,
 \qquad 0\le t\le1.
\]
Convexity of the balls gives $\left\|\mathbf{u}_t\right\|_2,\left\|\mathbf{v}_t\right\|_2\le K\sqrt{\rho_n}$. More explicitly, $\mathbf{u}_t=(1-t)\mathbf{x}_{0i}+t\mathbf{u}$, so
\[
 \left\|\mathbf{u}_t\right\|_2
 \le(1-t)\left\|\mathbf{x}_{0i}\right\|_2
      +t\left\|\mathbf{u}\right\|_2\le K\sqrt{\rho_n};
\]
the same argument applies to $\mathbf{v}_t$.
Since $\mathbf{u}_t'=\mathbf a$ and $\mathbf{v}_t'=\mathbf b$,
\[
 s_t'=\mathbf a^\top\mathbf v_t+\mathbf u_t^\top\mathbf b,
 \qquad s_t''=2\mathbf a^\top\mathbf b.
\]
Writing $L=\left\|\mathbf{a}\right\|_2+\left\|\mathbf{b}\right\|_2$, we have
\begin{equation}\label{eq:path}
 |s_t|\le K^2\rho_n,\quad
 |s_t'|=|\mathbf{a}^\top \mathbf{v}_t+\mathbf{u}_t^\top \mathbf{b}|\le K\sqrt{\rho_n}L,\quad
 |s_t''|=2|\mathbf{a}^\top \mathbf{b}|\le L^2.
\end{equation}
These inequalities follow from Cauchy--Schwarz: for example,
\[
 |s_t'|\le\left\|\mathbf a\right\|_2\left\|\mathbf v_t\right\|_2
       +\left\|\mathbf u_t\right\|_2\left\|\mathbf b\right\|_2,
 \qquad
 2|\mathbf a^\top\mathbf b|
 \le2\left\|\mathbf a\right\|_2\left\|\mathbf b\right\|_2\le L^2.
\]
Combining \eqref{eq:global}--\eqref{eq:local} gives, along this path,
\[
 |\psi'(s_t)|\lesssim_{\delta,K}\rho_n^{-1},
 \qquad
 |\psi''(s_t)|\lesssim_\delta\rho_n^{-2},
 \qquad \text{for all }t\in[0,1],
\]
and
\[
 |\psi'''(s_t)|\lesssim_\delta\rho_n^{-3}
 \qquad
 \text{for all }t\in[0,1]\text{ such that }
 s_t\notin\{\tau,1-\tau\}.
\]

\paragraph{The function $\mathbf{g}$.}
The first derivative along the segment is
\[
 \frac{d}{dt}\mathbf{g}(\mathbf{u}_t,\mathbf{v}_t)=\psi''(s_t)s_t'\mathbf{v}_t+\psi'(s_t)\mathbf{b}.
\]
In detail, the chain rule differentiates $\psi'(s_t)$ and the product
rule differentiates $\mathbf v_t$. The resulting norm bound is
\begin{align*}
 \left\|\frac{d}{dt}\mathbf g(\mathbf u_t,\mathbf v_t)\right\|_2
 &\le |\psi''(s_t)|\,|s_t'|\left\|\mathbf v_t\right\|_2
          +|\psi'(s_t)|\left\|\mathbf b\right\|_2\\
 &\lesssim_{\delta,K}
 \rho_n^{-2}(\sqrt{\rho_n}L)\sqrt{\rho_n}
       +\rho_n^{-1}\left\|\mathbf b\right\|_2
 \lesssim_{\delta,K}\rho_n^{-1}L.
\end{align*}
The fundamental theorem of calculus then gives
\[
 \left\|\mathbf g(\mathbf u,\mathbf v)
             -\mathbf g(\mathbf x_{0i},\mathbf y_{0j})\right\|_2
 \le\int_0^1
 \left\|\frac{d}{dt}\mathbf g(\mathbf u_t,\mathbf v_t)\right\|_2\,dt.
\]
Its norm is at most $C_{\delta,K}\rho_n^{-1}L$; integrating from zero
to one proves \eqref{eq:gLip}. Differentiating again almost everywhere,
\[
 \frac{d^2}{dt^2}\mathbf{g}(\mathbf{u}_t,\mathbf{v}_t)
 =\psi'''(s_t)(s_t')^2\mathbf{v}_t+\psi''(s_t)s_t''\mathbf{v}_t
       +2\psi''(s_t)s_t'\mathbf{b}.
\]
By \eqref{eq:path}, the three terms have norms bounded, respectively, by
constants times
\[
 \rho_n^{-3}(\rho_n L^2)\sqrt{\rho_n},\qquad
 \rho_n^{-2}L^2\sqrt{\rho_n},\qquad
 \rho_n^{-2}(\sqrt{\rho_n}L)L.
\]
All are $O_{\delta,K}(\rho_n^{-3/2}L^2)$. The integral Taylor formula gives
\[
 \mathbf{r}_{\mathbf{g}}=\int_0^1(1-t)\frac{d^2}{dt^2}\mathbf{g}(\mathbf{u}_t,\mathbf{v}_t)\,dt.
\]
To see why this is the required multivariate remainder, apply the
one-dimensional integral Taylor identity to the vector-valued path
$t\mapsto\mathbf g(\mathbf u_t,\mathbf v_t)$. Its derivative at zero is
\[
 D_{\mathbf u}\mathbf g(\mathbf x_{0i},\mathbf y_{0j})\mathbf a
 +D_{\mathbf v}\mathbf g(\mathbf x_{0i},\mathbf y_{0j})\mathbf b,
\]
which is exactly the first-order term subtracted in the definition.
Consequently,
\[
 \left\|\mathbf r_{\mathbf g}\right\|_2
 \le C_{\delta,K}\rho_n^{-3/2}L^2\int_0^1(1-t)\,dt
 =\frac{C_{\delta,K}}2\rho_n^{-3/2}L^2.
\]
Since $L^2\le2(\left\|\mathbf{a}\right\|_2^2+\left\|\mathbf{b}\right\|_2^2)$, this proves \eqref{eq:rg}.

\paragraph{The function $\mathbf{h}_{ij}$.}
Define $f(s)=\psi'(s)(\mathbf{x}_{0i}^\top\mathbf{y}_{0j}-s)$. Along the segment,
$|\mathbf{x}_{0i}^\top\mathbf{y}_{0j}-s_t|\le2K^2\rho_n$, and hence using $|\mathbf x_{0i}^{\top}\mathbf y_{0j}|\le K^2\rho_n$ and $|s_t|\le K^2\rho_n$,
\begin{align*}
 |f(s_t)|&\lesssim_{\delta,K}1,\quad
 f'(s)=\psi''(s)(\mathbf{x}_{0i}^\top\mathbf{y}_{0j}-s)-\psi'(s),
 |f'(s_t)|\lesssim_{\delta,K}\rho_n^{-1},\\
 f''(s)&=\psi'''(s)(\mathbf{x}_{0i}^\top\mathbf{y}_{0j}-s)-2\psi''(s),\quad
 |f''(s_t)|\lesssim_{\delta,K}\rho_n^{-2}\quad\text{a.e.}
\end{align*}
The powers of $\rho_n$ in these estimates can be checked directly:
\begin{align*}
 |f(s_t)|&\lesssim_{\delta,K}\rho_n^{-1}\rho_n=1,\quad
 |f'(s_t)|\lesssim_{\delta,K}
       \rho_n^{-2}\rho_n+\rho_n^{-1}\lesssim\rho_n^{-1},\quad
 |f''(s_t)|\lesssim_{\delta,K}
       \rho_n^{-3}\rho_n+2\rho_n^{-2}\lesssim\rho_n^{-2}.
\end{align*}
Because $\mathbf{h}_{ij}(\mathbf{u}_t,\mathbf{v}_t)=f(s_t)\mathbf{v}_t$,
\[
 \frac{d}{dt}\mathbf{h}_{ij}(\mathbf{u}_t,\mathbf{v}_t)=f'(s_t)s_t'\mathbf{v}_t+f(s_t)\mathbf{b}.
\]
In particular,
\begin{align*}
 \left\|\frac{d}{dt}\mathbf h_{ij}(\mathbf u_t,\mathbf v_t)\right\|_2
 &\le |f'(s_t)|\,|s_t'|\left\|\mathbf v_t\right\|_2
       +|f(s_t)|\left\|\mathbf b\right\|_2
 \lesssim_{\delta,K}
       \rho_n^{-1}(\sqrt{\rho_n}L)\sqrt{\rho_n}
                   +\left\|\mathbf b\right\|_2
 \lesssim_{\delta,K}L.
\end{align*}
Its norm is at most $C_{\delta,K}L$, proving \eqref{eq:hLip} by integration.
Also,
\[
 \frac{d^2}{dt^2}\mathbf{h}_{ij}(\mathbf{u}_t,\mathbf{v}_t)
 =f''(s_t)(s_t')^2\mathbf{v}_t+f'(s_t)s_t''\mathbf{v}_t+2f'(s_t)s_t'\mathbf{b}.
\]
The norms of these three terms are bounded by constants times
\[
 \rho_n^{-2}(\rho_n L^2)\sqrt{\rho_n},\qquad
 \rho_n^{-1}L^2\sqrt{\rho_n},\qquad
 \rho_n^{-1}(\sqrt{\rho_n}L)L.
\]
Thus the second derivative has norm at most
$C_{\delta,K}\rho_n^{-1/2}L^2$. The integral Taylor formula proves
\eqref{eq:rh}.
Explicitly, $f(\mathbf x_{0i}^{\top}\mathbf y_{0j})=0$ and
$f'(\mathbf x_{0i}^{\top}\mathbf y_{0j})
=-\psi'(\mathbf x_{0i}^{\top}\mathbf y_{0j})$. Hence the derivative at
$t=0$ is
\[
 -\psi'(\mathbf x_{0i}^{\top}\mathbf y_{0j})
 \mathbf y_{0j}\mathbf y_{0j}^{\top}\mathbf a
 -\psi'(\mathbf x_{0i}^{\top}\mathbf y_{0j})
 \mathbf y_{0j}\mathbf x_{0i}^{\top}\mathbf b,
\]
matching the first-order terms of $\mathbf h_{ij}$ in the manuscript.
Subtracting these terms and integrating the second derivative yields
\begin{align*}
 \left\|\mathbf r_{\mathbf h_{ij}}\right\|_2
 &\le\int_0^1(1-t)
 \left\|\frac{d^2}{dt^2}\mathbf h_{ij}(\mathbf u_t,\mathbf v_t)\right\|_2\,dt
 \le\frac{C_{\delta,K}}2\rho_n^{-1/2}L^2
 \le C_{\delta,K}\rho_n^{-1/2}
      (\left\|\mathbf a\right\|_2^2+\left\|\mathbf b\right\|_2^2).
\end{align*}

\paragraph{The covariate score.}

Set $\mathbf{c}=\mathbf{z}-\mathbf{z}_{0l}$. Expanding the polynomial score gives the following identity. Indeed, first expand its scalar residual:
\[
 \mathbf x_{0i}^{\top}\mathbf z_{0l}-\mathbf u^{\top}\mathbf z
 =-\mathbf a^{\top}\mathbf z_{0l}
  -\mathbf x_{0i}^{\top}\mathbf c-\mathbf a^{\top}\mathbf c.
\]
Multiplying by $\mathbf z=\mathbf z_{0l}+\mathbf c$, and subtracting
$\boldsymbol\phi_{2il}(\mathbf x_{0i},\mathbf z_{0l})=W_{li}\mathbf z_{0l}$,
separates the linear terms from the products of increments:
\[
 \boldsymbol{\phi}_{2il}(\mathbf{u},\mathbf{z})-\boldsymbol{\phi}_{2il}(\mathbf{x}_{0i},\mathbf{z}_{0l})
 =W_{li}\mathbf{c}-\mathbf{z}_{0l}\mathbf{z}_{0l}^\top \mathbf{a}-\mathbf{z}_{0l}\mathbf{x}_{0i}^\top \mathbf{c}+\mathbf{r}_{\boldsymbol{\phi}_{2il}},
\]
where, exactly,
\[
 \mathbf{r}_{\boldsymbol{\phi}_{2il}}=-(\mathbf{a}^\top \mathbf{c})\mathbf{z}_{0l}-(\mathbf{a}^\top \mathbf{z}_{0l})\mathbf{c}-(\mathbf{u}^\top \mathbf{c})\mathbf{c}.
\]
In the last term above, we combined
$-(\mathbf x_{0i}^{\top}\mathbf c)\mathbf c$
and $-(\mathbf a^{\top}\mathbf c)\mathbf c$, using
$\mathbf u=\mathbf x_{0i}+\mathbf a$.
The noise term $W_{li}\mathbf z$ is linear in $\mathbf z$, so it
contributes no Taylor remainder. By Cauchy--Schwarz,
\begin{align*}
 \left\|\mathbf r_{\boldsymbol\phi_{2il}}\right\|_2
 &\le |\mathbf a^{\top}\mathbf c|\left\|\mathbf z_{0l}\right\|_2
       +|\mathbf a^{\top}\mathbf z_{0l}|\left\|\mathbf c\right\|_2
       +|\mathbf u^{\top}\mathbf c|\left\|\mathbf c\right\|_2\\
 &\le 2\left\|\mathbf z_{0l}\right\|_2
        \left\|\mathbf a\right\|_2\left\|\mathbf c\right\|_2
       +\left\|\mathbf u\right\|_2\left\|\mathbf c\right\|_2^2.
\end{align*}
Therefore
\[
 \left\|\mathbf{r}_{\boldsymbol{\phi}_{2il}}\right\|_2
 \le2K\left\|\mathbf{a}\right\|_2\left\|\mathbf{c}\right\|_2+K\sqrt{\rho_n}\left\|\mathbf{c}\right\|_2^2,
\]
which is \eqref{eq:rphi}.
\end{proof}


\subsection{Term \texorpdfstring{\eqref{eq:term1}}{(\ref{eq:term1})}}
\label{sec:term1}

\begin{lemma}\label[lemma]{lem:term1}
Under Assumptions~\ref{assump:snr_network} and \ref{assump:snr_covariates}, if $\tau=\delta\rho_n$ for some constant
$\delta\in(0,1/2)$, then
\begin{align*}
&\left\|\frac1{n+p}\sum_{j=1}^n
 E_{ij}\psi''(\mathbf x_{0i}^{\top}\mathbf y_{0j})
 \mathbf y_{0j}\mathbf y_{0j}^{\top}
 (\mathbf Q^{\top}\widehat{\mathbf x}_i^{(t)}-\mathbf x_{0i})\right\|_2\\
&\qquad=
 \left\|\mathbf Q^{\top}\widehat{\mathbf x}_i^{(t)}-\mathbf x_{0i}\right\|_2
 \widetilde O_p\!\left\{
 \frac{(\log m_n)^{1/2}}{(n\rho_n)^{1/2}};\,m_n\right\},
\end{align*}
and for any deterministic nonnegative diverging sequence $(m_n)_{n=1}^{\infty}$
with $m_n=O(n)$,
\begin{align*}
&\left\|\frac1{n+p}\sum_{j=1}^n
 E_{ij}\psi''(\mathbf x_{0i}^{\top}\mathbf y_{0j})
 \mathbf y_{0j}\mathbf y_{0j}^{\top}
 (\mathbf Q^{\top}\widehat{\mathbf x}_i^{(0)}-\mathbf x_{0i})\right\|_2
 =\widetilde O_p\!\left\{
 \frac{\log m_n}{n\rho_n^{1/2}};\,m_n\right\}.
\end{align*}
Furthermore, if
$\|\widehat{\mathbf X}^{(t)}\mathbf Q-\mathbf X_0\|_{2\to\infty}
=\widetilde O_p\{(\log n)^{1/2}/\sqrt n\}$, then
\begin{align*}
&\left\|\frac1{n+p}\sum_{j=1}^n
 E_{ij}\psi''(\mathbf x_{0i}^{\top}\mathbf y_{0j})
 \mathbf y_{0j}\mathbf y_{0j}^{\top}
 (\mathbf Q^{\top}\widehat{\mathbf x}_i^{(t)}-\mathbf x_{0i})\right\|_2
 =\widetilde O_p\!\left\{\frac{\log n}{n\rho_n^{1/2}}\right\}.
\end{align*}
\end{lemma}

\begin{proof}
We first bound the random matrix multiplying the estimation error, then
combine this bound with the initialization error rate or the error rate
of the current iterate. Throughout, we treat the oracle latent
positions as fixed; hence, the weights appearing below are deterministic.
We apply the Bernstein inequality in Lemma S2.1 of the supplementary
material to \cite{xie2024entrywise}. Fix $i\in[n]$.
For this fixed row, the off-diagonal variables $E_{ij}$, $j\ne i$, are independent and satisfy
\[
 \mathbb E E_{ij}=0,\qquad |E_{ij}|\le1,\qquad
 \operatorname{Var}(E_{ij})
 =(\mathbf x_{0i}^{\top}\mathbf y_{0j})
 (1-\mathbf x_{0i}^{\top}\mathbf y_{0j})\le\rho_n.
\]
Define the deterministic matrices
\[
 \mathbf D_{nij}
 =\psi''(\mathbf x_{0i}^{\top}\mathbf y_{0j})
       \mathbf y_{0j}\mathbf y_{0j}^{\top}.
\]
The derivative bound $|\psi''(s)|\le2/\tau^2$ and Assumption~\ref{assump:snr_network} give
\begin{align*}
 \|\mathbf D_{nij}\|_F
 &=|\psi''(\mathbf x_{0i}^{\top}\mathbf y_{0j})|
       \|\mathbf y_{0j}\mathbf y_{0j}^{\top}\|_F =|\psi''(\mathbf x_{0i}^{\top}\mathbf y_{0j})|
       \|\mathbf y_{0j}\|_2^2
 \le\frac{2}{\delta^2\rho_n^2}\rho_n
 =\frac{2}{\delta^2\rho_n}.
\end{align*}
The omitted diagonal matrix is $-P_{ii}\mathbf D_{nii}/(n+p)$,
whose norm is at most $C_\delta/n$. Applying Lemma S2.1 of the
supplementary material to \cite{xie2024entrywise} to the vectorized
off-diagonal weighted sum, and adding this correction, while using
$n+p\ge n$, gives
\[
 \left\|\frac1{n+p}\sum_{j=1}^n E_{ij}\mathbf D_{nij}\right\|_2
 =\widetilde O_p\!\left\{
 \frac{\log m_n}{n\rho_n}
 +\sqrt{\frac{\log m_n}{n\rho_n}};\,m_n\right\}.
\]
By Assumption~\ref{assump:snr_network} and $m_n=O(n)$, the first term is absorbed into the second. Hence,
\begin{equation}\label{eq:term1-matrix-rate}
 \left\|\frac1{n+p}\sum_{j=1}^nE_{ij}
 \psi''(\mathbf x_{0i}^{\top}\mathbf y_{0j})
 \mathbf y_{0j}\mathbf y_{0j}^{\top}\right\|_2
 =\widetilde O_p\!\left\{
 \frac{(\log m_n)^{1/2}}{(n\rho_n)^{1/2}};\,m_n\right\}.
\end{equation}
By the multiplicative property of matrix norms, we therefore have
\begin{align*}
&\frac1{n+p}\left\|\sum_{j=1}^n
 E_{ij}\psi''(\mathbf x_{0i}^{\top}\mathbf y_{0j})
 \mathbf y_{0j}\mathbf y_{0j}^{\top}
 (\mathbf Q^{\top}\widehat{\mathbf x}_i^{(t)}-\mathbf x_{0i})\right\|_2\\
&\quad\le\frac1{n+p}
 \left\|\mathbf Q^{\top}\widehat{\mathbf x}_i^{(t)}-\mathbf x_{0i}\right\|_2
 \left\|\sum_{j=1}^n E_{ij}
 \psi''(\mathbf x_{0i}^{\top}\mathbf y_{0j})
 \mathbf y_{0j}\mathbf y_{0j}^{\top}\right\|_2
=
 \left\|\mathbf Q^{\top}\widehat{\mathbf x}_i^{(t)}-\mathbf x_{0i}\right\|_2
 \widetilde O_p\!\left\{
 \frac{(\log m_n)^{1/2}}{(n\rho_n)^{1/2}};\,m_n\right\}.
\end{align*}
The norm inequality holds for every realization, so independence
between the estimation error and the random matrix is not required.

For $t=0$, Lemma~\ref{lem:X0} gives
\[
\|\mathbf Q^{\top}\widehat{\mathbf x}_i^{(0)}-\mathbf x_{0i}\|_2
=\widetilde O_p\left\{
\frac{(\log m_n)^{1/2}}{\sqrt n};\,m_n\right\}.
\]
Intersecting the two probability events and using
\[
 \frac{(\log m_n)^{1/2}}{(n\rho_n)^{1/2}}
 \frac{(\log m_n)^{1/2}}{\sqrt n}
 =\frac{\log m_n}{n\rho_n^{1/2}},
\]
one obtains
\begin{align*}
&\left\|\frac1{n+p}\sum_{j=1}^n
 E_{ij}\psi''(\mathbf x_{0i}^{\top}\mathbf y_{0j})
 \mathbf y_{0j}\mathbf y_{0j}^{\top}
 (\mathbf Q^{\top}\widehat{\mathbf x}_i^{(0)}-\mathbf x_{0i})\right\|_2
 =\widetilde O_p\!\left\{\frac{\log m_n}{n\rho_n^{1/2}};\,m_n\right\}.
\end{align*}
For the final assertion, take $m_n=n$ in
\eqref{eq:term1-matrix-rate}. If
$\|\widehat{\mathbf X}^{(t)}\mathbf Q-\mathbf X_0\|_{2\to\infty}
=\widetilde O_p\{(\log n)^{1/2}/\sqrt n\}$, then
\begin{align*}
&\left\|\frac1{n+p}\sum_{j=1}^n
 E_{ij}\psi''(\mathbf x_{0i}^{\top}\mathbf y_{0j})
 \mathbf y_{0j}\mathbf y_{0j}^{\top}
 (\mathbf Q^{\top}\widehat{\mathbf x}_i^{(t)}-\mathbf x_{0i})\right\|_2
 =\widetilde O_p\!\left\{\frac{\log n}{n\rho_n^{1/2}}\right\}.
\end{align*}
Here we used
\[
 \|\mathbf Q^{\top}\widehat{\mathbf x}_i^{(t)}-\mathbf x_{0i}\|_2
 \le\|\widehat{\mathbf X}^{(t)}\mathbf Q-\mathbf X_0\|_{2\to\infty},
 \qquad
 \sqrt{\frac{\log n}{n\rho_n}}\sqrt{\frac{\log n}{n}}
 =\frac{\log n}{n\rho_n^{1/2}}.
\]
This proves all three assertions.
\end{proof}

\subsection{Term \texorpdfstring{\eqref{eq:term2}}{(\ref{eq:term2})}}
\begin{lemma}\label[lemma]{lem:term2}
Under Assumptions~\ref{assump:snr_network} and~\ref{assump:snr_covariates}, if $\tau=\delta\rho_n$ for some constant
$\delta\in(0,1/2)$, then
\begin{equation}\label{eq:44-general}
\left\|\frac1{n+p}\left\{
 \sum_{j=1}^n\psi'(\mathbf x_{0i}^\top\mathbf y_{0j})
 \mathbf y_{0j}\mathbf y_{0j}^\top
 +\sum_{l=1}^p\mathbf z_{0l}\mathbf z_{0l}^\top
 \right\}(\mathbf Q^\top\widehat{\mathbf x}_i^{(t)}-\mathbf x_{0i})\right\|_2
 =O\!\left(\|\mathbf Q^\top\widehat{\mathbf x}_i^{(t)}-\mathbf x_{0i}\|_2\right),
\end{equation}
and for any deterministic nonnegative diverging sequence $(m_n)_{n=1}^{\infty}$
with $m_n=O(n)$,
\begin{equation}\label{eq:44-init}
\left\|\frac1{n+p}\left\{
 \sum_{j=1}^n\psi'(\mathbf x_{0i}^\top\mathbf y_{0j})
 \mathbf y_{0j}\mathbf y_{0j}^\top
 +\sum_{l=1}^p\mathbf z_{0l}\mathbf z_{0l}^\top
 \right\}(\mathbf Q^\top\widehat{\mathbf x}_i^{(0)}-\mathbf x_{0i})\right\|_2
 =\widetilde O_p\!\left\{\frac{(\log m_n)^{1/2}}{\sqrt n};\,m_n\right\}.
\end{equation}
Furthermore, if
$\|\widehat{\mathbf X}^{(t)}\mathbf Q-\mathbf X_0\|_{2\to\infty}
=\widetilde O_p\{(\log n)^{1/2}/\sqrt n\}$, then
\begin{equation}\label{eq:44-iterate}
\left\|\frac1{n+p}\left\{
 \sum_{j=1}^n\psi'(\mathbf x_{0i}^\top\mathbf y_{0j})
 \mathbf y_{0j}\mathbf y_{0j}^\top
 +\sum_{l=1}^p\mathbf z_{0l}\mathbf z_{0l}^\top
 \right\}(\mathbf Q^\top\widehat{\mathbf x}_i^{(t)}-\mathbf x_{0i})\right\|_2
 =\widetilde O_p\!\left\{\frac{(\log n)^{1/2}}{\sqrt n}\right\}.
\end{equation}
\end{lemma}

\begin{proof}
Retain the oracle matrix $\mathbf G_{in}$ and write
$ \mathbf e_i^{(t)}:=\mathbf Q^\top\widehat{\mathbf x}_i^{(t)}
                         -\mathbf x_{0i}$.
We bound the two components of $\mathbf G_{in}$ separately.
Since $\mathbf y_{0j}=\mathbf I_{d_+,d_-}\mathbf x_{0j}$ and
$\mathbf I_{d_+,d_-}$ is orthogonal, Assumption~\ref{assump:snr_network} implies
 $\|\mathbf y_{0j}\|_2^2=\|\mathbf x_{0j}\|_2^2\le\rho_n$.
Using $\|\mathbf v\mathbf v^\top\|_2=\|\mathbf v\|_2^2$ and
$|\psi'(s)|\le 2/(\delta\rho_n)$, we obtain
\begin{align*}
 \left\|\frac1{n+p}\sum_{j=1}^n
 \psi'(\mathbf x_{0i}^\top\mathbf y_{0j})
       \mathbf y_{0j}\mathbf y_{0j}^\top\right\|_2
 &\le\frac1{n+p}\sum_{j=1}^n
 |\psi'(\mathbf x_{0i}^\top\mathbf y_{0j})|\,
       \|\mathbf y_{0j}\|_2^2
 \le\frac{n}{n+p}\frac{2}{\delta\rho_n}\rho_n
 =\frac{2n}{\delta(n+p)}.
\end{align*}
By Assumption~\ref{assump:snr_covariates}, there is a constant $C_Z$
such that $\max_l\|\mathbf z_{0l}\|_2\le C_Z$ for all sufficiently
large $n$. Hence,
\[
 \left\|\frac1{n+p}\sum_{l=1}^p
       \mathbf z_{0l}\mathbf z_{0l}^\top\right\|_2
 \le\frac1{n+p}\sum_{l=1}^p\|\mathbf z_{0l}\|_2^2
 \le\frac{pC_Z^2}{n+p}.
\]
Consequently, uniformly over $i\in[n]$,
\begin{equation}\label{eq:44-matrix}
 \|\mathbf G_{in}\|_2
 \le\frac{2n/\delta+pC_Z^2}{n+p}
 \le\max\{2/\delta,C_Z^2\}=:C.
\end{equation}
Submultiplicativity gives \eqref{eq:44-general}.

For initialization, Lemma~\ref{lem:X0} gives
\[
 \|\mathbf e_i^{(0)}\|_2
 =\widetilde O_p\!\left\{
 \frac{\sqrt{\log m_n}}{\sqrt n};\,m_n\right\}.
\]
Combining this with \eqref{eq:44-general} proves \eqref{eq:44-init}.
Finally,
\[
 \|\mathbf G_{in}\mathbf e_i^{(t)}\|_2
 \le C\max_{i\in[n]}\|\mathbf e_i^{(t)}\|_2
 =C\|\widehat{\mathbf X}^{(t)}\mathbf Q-\mathbf X_0\|_{2\to\infty},
\]
which proves \eqref{eq:44-iterate} under the stated iterate-rate condition.
\end{proof}

\subsection{Term \texorpdfstring{\eqref{eq:term3}}{(\ref{eq:term3})}}
\begin{lemma}\label[lemma]{lem:term3}
Suppose $\widehat{\mathbf X}$ is approximately linearized with orthogonal
alignment $\mathbf Q$ and remainder $\mathbf R_X$.
Let $\widehat{\mathbf Y}=\widehat{\mathbf X}\mathbf I_{d_+,d_-}
=[\widehat{\mathbf y}_1,\ldots,\widehat{\mathbf y}_n]^\top$.
Under Assumptions~\ref{assump:snr_network} and~\ref{assump:snr_covariates}, if $\tau=\delta\rho_n$ for some constant
$\delta\in(0,1/2)$, then
\begin{align*}
 &\left\|\frac1{n+p}\sum_{j=1}^n
 \psi'(\mathbf x_{0i}^\top\mathbf y_{0j})
 \mathbf y_{0j}\mathbf x_{0i}^\top
 (\mathbf Q^\top\widehat{\mathbf y}_j-\mathbf y_{0j})\right\|_2
 =\widetilde O_p\!\left\{
 \frac{(\log m_n)^{1/2}}{n\rho_n^{1/2}};\,m_n\right\}
 +O\!\left(\frac{\|\mathbf R_X\|_F}{\sqrt n}\right),
\end{align*}
and for any deterministic nonnegative diverging sequence $(m_n)_{n=1}^\infty$
with $m_n=O(n)$,
\begin{align*}
 &\left\|\frac1{n+p}\sum_{j=1}^n
 \psi'(\mathbf x_{0i}^\top\mathbf y_{0j})
 \mathbf y_{0j}\mathbf x_{0i}^\top
 (\mathbf Q^\top\widehat{\mathbf y}_j^{(0)}-\mathbf y_{0j})\right\|_2
 =\widetilde O_p\!\left\{
 \frac{(\log m_n)^{1/2}+(1\vee\rho_n\log n)^{1/2}}
 {n\rho_n^{1/2}};\,m_n\right\}.
\end{align*}
Furthermore, if Assumption~\ref{assump:3} holds and $\widehat{\mathbf X}^{(t)}$
is approximately linearized with orthogonal alignment, then, for any
$\epsilon>0$,
\[
 \left\|\frac1{n+p}\sum_{j=1}^n
 \psi'(\mathbf x_{0i}^\top\mathbf y_{0j})
 \mathbf y_{0j}\mathbf x_{0i}^\top
 (\mathbf Q^\top\widehat{\mathbf y}_j^{(t)}-\mathbf y_{0j})\right\|_2
 =\widetilde O_p\!\left\{
 \frac{(\log n)^{2+\epsilon}}{n\rho_n^{1/2}}\right\}.
\]
\end{lemma}

\begin{proof}
The following argument uses
$\mathbf Q\mathbf I_{d_+,d_-}=\mathbf I_{d_+,d_-}\mathbf Q$.
This holds for the block-diagonal initialization alignment; for a general
iterate it must be ensured by the alignment convention. Orthogonality
alone does not imply this identity. Subject to this condition,
\[
 \mathbf Q^\top\widehat{\mathbf y}_j-\mathbf y_{0j}
 =\mathbf I_{d_+,d_-}(\mathbf Q^\top\widehat{\mathbf x}_j-\mathbf x_{0j}).
\]
Fix $i\in[n]$ and write
 $\mathbf B_{nij}:=\psi'(\mathbf x_{0i}^\top\mathbf y_{0j})
                   \mathbf y_{0j}\mathbf x_{0i}^\top$.
Since $|\psi'(s)|\le2/(\delta\rho_n)$ and
$\|\mathbf y_{0j}\|_2,\|\mathbf x_{0i}\|_2\le\sqrt{\rho_n}$,
\[
 \max_{i,j}\|\mathbf B_{nij}\|_2\le2/\delta.
\]
Approximate linearization gives
\[
 \mathbf Q^\top\widehat{\mathbf x}_j-\mathbf x_{0j}
 =\sum_{a=1}^n\boldsymbol\alpha_{nja}E_{ja}
 +\sum_{l=1}^p\boldsymbol\beta_{njl}W_{lj}+\mathbf r_{x_j},
\]
where the coefficients are deterministic and
\[
 \max_{j,a}\|\boldsymbol\alpha_{nja}\|_2
 =O\!\left(\frac1{n\sqrt{\rho_n}}\right),\qquad
 \max_{j,l}\|\boldsymbol\beta_{njl}\|_2=O(1/n).
\]
Thus the vector of interest decomposes as
\begin{align*}
 \mathbf T_i
 &:=\frac1{n+p}\sum_j\mathbf B_{nij}
       (\mathbf Q^\top\widehat{\mathbf y}_j-\mathbf y_{0j})
 =\mathbf T_{Ei}+\mathbf T_{Wi}+\mathbf T_{Ri},\\
 \mathbf T_{Ei}&:=\frac1{n+p}\sum_{j,a}
 \mathbf B_{nij}\mathbf I_{d_+,d_-}\boldsymbol\alpha_{nja}E_{ja},\\
 \mathbf T_{Wi}&:=\frac1{n+p}\sum_{j,l}
 \mathbf B_{nij}\mathbf I_{d_+,d_-}\boldsymbol\beta_{njl}W_{lj},\\
 \mathbf T_{Ri}&:=\frac1{n+p}\sum_j
 \mathbf B_{nij}\mathbf I_{d_+,d_-}\mathbf r_{x_j}.
\end{align*}

\paragraph{Network-noise term.}
Using $E_{ja}=E_{aj}$, collect the independent off-diagonal entries:
\[
 \mathbf T_{Ei}^{\mathrm{off}}=\sum_{j<a}\mathbf c_{i,ja}E_{ja},\qquad
 \mathbf c_{i,ja}:=\frac{
 \mathbf B_{nij}\mathbf I_{d_+,d_-}\boldsymbol\alpha_{nja}
 +\mathbf B_{nia}\mathbf I_{d_+,d_-}\boldsymbol\alpha_{naj}}{n+p}.
\]
Then $\max_{j<a}\|\mathbf c_{i,ja}\|_2=O(1/(n^2\sqrt{\rho_n}))$.
Let $\mathbf V_i$ have rows $\mathbf c_{i,ja}^\top$, $j<a$.
Since it has $N=n(n-1)/2$ rows,
\[
 \|\mathbf V_i\|_{2\to\infty}=O\!\left(\frac1{n^2\sqrt{\rho_n}}\right),
 \qquad
 \|\mathbf V_i\|_F\le\sqrt N\|\mathbf V_i\|_{2\to\infty}
 =O\!\left(\frac1{n\sqrt{\rho_n}}\right).
\]
The off-diagonal variables are centered, independent, bounded in absolute
value by one, and have variance at most $\rho_n$.
Lemma S2.1 of the supplementary material to \cite{xie2024entrywise}
therefore gives, for $u\ge1$,
\[
 \mathbb P\!\left\{\|\mathbf T_{Ei}^{\mathrm{off}}\|_2>
 3u^2\|\mathbf V_i\|_{2\to\infty}
 +\sqrt6\,\sqrt{\rho_n}\,u\|\mathbf V_i\|_F\right\}
 \le28e^{-u^2}.
\]
Taking $u=\sqrt{(c+1)\log m_n}$ for any prescribed $c>0$ yields
\[
 \|\mathbf T_{Ei}^{\mathrm{off}}\|_2
 =\widetilde O_p\!\left\{
 \frac{\sqrt{\log m_n}}n+
 \frac{\log m_n}{n^2\sqrt{\rho_n}};\,m_n\right\}.
\]
Under Assumption~\ref{assump:snr_network} and $m_n=O(n)$,
\[
 \frac{\log m_n/(n^2\sqrt{\rho_n})}{\sqrt{\log m_n}/n}
 =\sqrt{\frac{\log m_n}{n^2\rho_n}}=O(n^{-1/2}),
\]
so the second term is absorbed into the first.
For the zero-diagonal model, $E_{jj}=-P_{jj}$ and
$|P_{jj}|\le\|\mathbf x_{0j}\|_2^2\le\rho_n$. Consequently,
\[
 \|\mathbf T_{Ei}^{\mathrm{diag}}\|_2
 \le\frac1{n+p}\sum_j
 \|\mathbf B_{nij}\|_2\|\boldsymbol\alpha_{njj}\|_2|P_{jj}|
 =O\!\left(\frac1n\,n\,\frac{\rho_n}{n\sqrt{\rho_n}}\right)
 =O(\sqrt{\rho_n}/n).
\]
This is absorbed into $\sqrt{\log m_n}/n$, and hence
\begin{equation}\label{eq:45-network}
 \|\mathbf T_{Ei}\|_2
 =\widetilde O_p\!\left\{\frac{\sqrt{\log m_n}}n;\,m_n\right\}.
\end{equation}

\paragraph{Covariate-noise term.}
Each coefficient of $W_{lj}$ has norm $O(n^{-2})$, and $p=O(n)$ gives
\[
 \sum_{j,l}\left\|\frac{
 \mathbf B_{nij}\mathbf I_{d_+,d_-}\boldsymbol\beta_{njl}}{n+p}\right\|_2^2
 =O(np/n^4)=O(n^{-2}).
\]
Since the $W_{lj}$ are independent standard Gaussian variables,
Gaussian concentration in fixed dimension gives
\begin{equation}\label{eq:45-covariate}
 \|\mathbf T_{Wi}\|_2
 =\widetilde O_p\!\left\{\frac{\sqrt{\log m_n}}n;\,m_n\right\}.
\end{equation}

\paragraph{Remainder term.}
Cauchy--Schwarz yields
\[
 \|\mathbf T_{Ri}\|_2
 \le\frac C{n+p}\sum_j\|\mathbf r_{x_j}\|_2
 \le\frac{C\sqrt n}{n+p}\|\mathbf R_X\|_F
 \le C\frac{\|\mathbf R_X\|_F}{\sqrt n}.
\]
Combining the three bounds and using $\rho_n\le1$ proves
\begin{align*}
 \|\mathbf T_i\|_2
 &=\widetilde O_p\!\left\{\frac{\sqrt{\log m_n}}n;\,m_n\right\}
   +O\!\left(\frac{\|\mathbf R_X\|_F}{\sqrt n}\right)\\
 &=\widetilde O_p\!\left\{\frac{\sqrt{\log m_n}}{n\sqrt{\rho_n}};\,m_n\right\}
   +O\!\left(\frac{\|\mathbf R_X\|_F}{\sqrt n}\right).
\end{align*}

\paragraph{Initialization and iterates.}
Lemma~\ref{lem:X0} gives
\[
 \|\mathbf R_X^{(0)}\|_F
 =\widetilde O_p\!\left\{
 \frac{\sqrt{1\vee\rho_n\log n}}{\sqrt{n\rho_n}}\right\}.
\]
Since $m_n=O(n)$, the corresponding $n$-based probability bound also
implies the $m_n$-based bound. Substitution yields
\[
 \|\mathbf T_i^{(0)}\|_2
 =\widetilde O_p\!\left\{
 \frac{\sqrt{\log m_n}+\sqrt{1\vee\rho_n\log n}}
 {n\sqrt{\rho_n}};\,m_n\right\}.
\]
For the final assertion, applying a union bound to the
remainder estimates in approximate linearization,
whose constants are uniform over $j\in[n]$, gives
\[
 \|\mathbf R_X^{(t)}\|_{2\to\infty}
 =\max_{j\in[n]}\|\mathbf r_{x_j}^{(t)}\|_2
 =\widetilde O_p\!\left\{
 \frac{(\log n)^{2+\epsilon}}{n\sqrt{\rho_n}}\right\}.
\]
Because $\|\mathbf R_X^{(t)}\|_F/\sqrt n
\le\|\mathbf R_X^{(t)}\|_{2\to\infty}$, taking $m_n=n$ in the
first bound proves
\[
 \|\mathbf T_i^{(t)}\|_2
 =\widetilde O_p\!\left\{
 \frac{(\log n)^{2+\epsilon}}{n\sqrt{\rho_n}}\right\}.
\]
\end{proof}

\subsection{Term \texorpdfstring{\eqref{eq:term4}}{(\ref{eq:term4})}}
\begin{lemma}\label[lemma]{lem:term4}
Under Assumptions~\ref{assump:snr_network} and~\ref{assump:snr_covariates}, if $\tau=\delta\rho_n$ for some constant
$\delta\in(0,1/2)$, then for any fixed $i\in[n]$ and deterministic
nonnegative diverging sequence $(m_n)$ with $m_n=O(n)$,
\begin{align*}
 &\left\|\frac1{n+p}\left\{
 \sum_{j=1}^n\mathbf r_{h_{ij}}(\mathbf Q^\top\widehat{\mathbf x}_i^{(0)},
 \mathbf Q^\top\widehat{\mathbf y}_j^{(0)},\mathbf x_{0i},\mathbf y_{0j})
 +\sum_{l=1}^p\mathbf r_{\phi_{2il}}(\mathbf Q^\top\widehat{\mathbf x}_i^{(0)},
 \mathbf Q^\top\widehat{\mathbf z}_l^{(0)},\mathbf x_{0i},\mathbf z_{0l})
 \right\}\right\|_2\\
 &\qquad=\widetilde O_p\!\left\{\frac{\log m_n}{n\rho_n^{1/2}};\,m_n\right\}.
\end{align*}
Furthermore, if
\[
\begin{aligned}
 \|\widehat{\mathbf X}^{(t)}\mathbf Q-\mathbf X_0\|_{2\to\infty}
 &=\widetilde O_p\{(\log n)^{1/2}/\sqrt n\},\\
 \|\widehat{\mathbf Z}^{(t)}\mathbf Q-\mathbf Z_0\|_{2\to\infty}
 &=\widetilde O_p\{(\log n)^{1/2}/(n\rho_n)^{1/2}\},
\end{aligned}
\]
then
\begin{align*}
 &\left\|\frac1{n+p}\left\{
 \sum_{j=1}^n\mathbf r_{h_{ij}}(\mathbf Q^\top\widehat{\mathbf x}_i^{(t)},
 \mathbf Q^\top\widehat{\mathbf y}_j^{(t)},\mathbf x_{0i},\mathbf y_{0j})
 +\sum_{l=1}^p\mathbf r_{\phi_{2il}}(\mathbf Q^\top\widehat{\mathbf x}_i^{(t)},
 \mathbf Q^\top\widehat{\mathbf z}_l^{(t)},\mathbf x_{0i},\mathbf z_{0l})
 \right\}\right\|_2\\
 &\qquad=\widetilde O_p\!\left\{\frac{\log n}{n\rho_n^{1/2}}\right\}.
\end{align*}
\end{lemma}
\begin{proof}
We use the signature-preserving alignment
$\mathbf Q\mathbf I_{d_+,d_-}=\mathbf I_{d_+,d_-}\mathbf Q$.
For brevity, write $\mathbf r_{h_{ij}}^{(t)}$ and
$\mathbf r_{\phi_{2il}}^{(t)}$ for the remainders evaluated at the arguments
in the statement. Set
\[
 \boldsymbol\Delta_X^{(t)}=\widehat{\mathbf X}^{(t)}\mathbf Q-\mathbf X_0,
 \qquad
 \boldsymbol\Delta_Z^{(t)}=\widehat{\mathbf Z}^{(t)}\mathbf Q-\mathbf Z_0.
\]
The uniform initialization rates in Lemmas~\ref{lem:X0} and~\ref{lem:Z0}, or the stated
iterate rates, together with $n\rho_n\gtrsim\log n$, ensure that the
estimated network coordinates have norms at most $K\sqrt{\rho_n}$
and the estimated covariate coordinates have norms at most $K$, on an
event with arbitrarily high polynomial probability. Lemma~\ref{lem:Taylor_bounds} therefore
applies on this event.

\emph{Initialization.}
Write $\mathbf a_i^{(0)}=\mathbf Q^\top\widehat{\mathbf x}_i^{(0)}-\mathbf x_{0i}$.
By Lemma~\ref{lem:Taylor_bounds} and Cauchy--Schwarz,
\begin{align*}
 \left\|\frac1{n+p}\sum_j\mathbf r_{h_{ij}}^{(0)}\right\|_2
 &\le C\left\{\frac{\|\mathbf a_i^{(0)}\|_2^2}{\sqrt{\rho_n}}
 +\frac{\|\boldsymbol\Delta_X^{(0)}\|_F^2}{n\sqrt{\rho_n}}\right\},\\
 \left\|\frac1{n+p}\sum_l\mathbf r_{\phi_{2il}}^{(0)}\right\|_2
 &\le C\left\{\frac{\sqrt{\rho_n}}n\|\boldsymbol\Delta_Z^{(0)}\|_F^2
 +\frac{\sqrt p}n\|\mathbf a_i^{(0)}\|_2\|\boldsymbol\Delta_Z^{(0)}\|_F\right\}.
\end{align*}
Since $d$ is fixed, $\|\mathbf M\|_F\le\sqrt d\|\mathbf M\|_2$
for the error matrices. Lemmas~\ref{lem:X0} and~\ref{lem:Z0} give
\[
 \|\mathbf a_i^{(0)}\|_2=\widetilde O_p\!\left\{\sqrt{\frac{\log m_n}{n}};m_n\right\},
 \quad \|\boldsymbol\Delta_X^{(0)}\|_F=\widetilde O_p(1),
 \quad \|\boldsymbol\Delta_Z^{(0)}\|_F=\widetilde O_p(\rho_n^{-1/2}).
\]
Here an $n$-based probability bound also gives an $m_n$-based bound,
since $m_n=O(n)$. Substituting these rates and using $p=O(n)$ gives
\begin{align*}
 \left\|\frac1{n+p}\sum_j\mathbf r_{h_{ij}}^{(0)}\right\|_2
 &=\widetilde O_p\!\left\{\frac{\log m_n}{n\sqrt{\rho_n}};m_n\right\},\\
 \left\|\frac1{n+p}\sum_l\mathbf r_{\phi_{2il}}^{(0)}\right\|_2
 &=\widetilde O_p\!\left\{\frac{\sqrt{\log m_n}}{n\sqrt{\rho_n}};m_n\right\}.
\end{align*}
The triangle inequality proves the initialization assertion.

\emph{Iterates.}

Define the estimation errors
\[
 \mathbf a_i^{(t)}:=\mathbf Q^\top\widehat{\mathbf x}_i^{(t)}-\mathbf x_{0i},
 \qquad
 \mathbf b_j^{(t)}:=\mathbf Q^\top\widehat{\mathbf y}_j^{(t)}-\mathbf y_{0j},
\]
and set
\[
 D_t:=\|\widehat{\mathbf X}^{(t)}\mathbf Q-\mathbf X_0\|_{2\to\infty}
 =\max_{k\in[n]}\|\mathbf a_k^{(t)}\|_2.
\]
Under the alignment condition
$\mathbf Q\mathbf I_{d_+,d_-}=\mathbf I_{d_+,d_-}\mathbf Q$,
we have $\mathbf b_j^{(t)}=\mathbf I_{d_+,d_-}\mathbf a_j^{(t)}$ and hence
\[
 \|\mathbf a_i^{(t)}\|_2\le D_t,
 \qquad \|\mathbf b_j^{(t)}\|_2=\|\mathbf a_j^{(t)}\|_2\le D_t.
\]
By Lemma~\ref{lem:Taylor_bounds}, on its applicable high-probability event,
\begin{align*}
 \left\|\frac1{n+p}\sum_{j=1}^n\mathbf r_{h_{ij}}^{(t)}\right\|_2
 &\le\frac1{n+p}\sum_{j=1}^n\|\mathbf r_{h_{ij}}^{(t)}\|_2
 \le\frac{C_0}{(n+p)\sqrt{\rho_n}}
       \sum_{j=1}^n(D_t^2+D_t^2)=\frac{2C_0n}{(n+p)\sqrt{\rho_n}}D_t^2
       \le\frac{2C_0}{\sqrt{\rho_n}}D_t^2.
\end{align*}
In particular, if
\[
 D_t=\widetilde O_p\!\left(\sqrt{\frac{\log n}{n}}\right),
\]
then $D_t^2=\widetilde O_p(\log n/n)$, and therefore
\[
 \left\|\frac1{n+p}\sum_{j=1}^n\mathbf r_{h_{ij}}^{(t)}\right\|_2
 =\widetilde O_p\!\left(\frac{\log n}{n\sqrt{\rho_n}}\right).
\]
For the covariate term, put
$F_t=\|\boldsymbol\Delta_Z^{(t)}\|_{2\to\infty}$.
By Lemma~\ref{lem:Taylor_bounds},
\[
 \left\|\frac1{n+p}\sum_l\mathbf r_{\phi_{2il}}^{(t)}\right\|_2
 \le\frac{Cp}{n+p}\left(\sqrt{\rho_n}F_t^2+D_tF_t\right)
 =\widetilde O_p\!\left(\frac{\log n}{n\sqrt{\rho_n}}\right).
\]
Combining the network and covariate bounds proves the final assertion.
\end{proof}

\subsection{Term \texorpdfstring{\eqref{eq:term5}}{(\ref{eq:term5})}}\label{sec:term5}
\paragraph{Leave-one-out notation.}
All probabilities below are conditional on the oracle coordinates.
For $m\in[n]$, define
\[
 A_{jk}^{(m)}=
 \begin{cases}
 A_{jk},&j\ne m\text{ and }k\ne m,\\
 \mathbb E A_{jk},&j=m\text{ or }k=m.
 \end{cases}
\]
Conditional on the oracle coordinates, $\mathbf A^{(m)}$ is
independent of $(E_{m1},\ldots,E_{mn})$.
Suppose
\[
 \mathbf A^{(m)}=\sum_{k=1}^n\widehat\lambda_k^{(m)}
 \widehat{\mathbf u}_k^{(m)}(\widehat{\mathbf u}_k^{(m)})^\top,
 \qquad \widehat\lambda_1^{(m)}\ge\cdots\ge\widehat\lambda_n^{(m)}.
\]
Define
\[
 \mathbf U_{\mathbf A+}^{(m)}=[\widehat{\mathbf u}_1^{(m)},\ldots,\widehat{\mathbf u}_{d_+}^{(m)}],
 \qquad
 \mathbf U_{\mathbf A-}^{(m)}=[\widehat{\mathbf u}_{n-d_-+1}^{(m)},\ldots,\widehat{\mathbf u}_n^{(m)}],
 \qquad
 \mathbf U_{\mathbf A}^{(m)}=[\mathbf U_{\mathbf A+}^{(m)},\mathbf U_{\mathbf A-}^{(m)}].
\]
Use the same notation without $(m)$ for $\mathbf A$.
Let $\mathbf U_{\mathbf P}=[\mathbf U_{\mathbf P+},\mathbf U_{\mathbf P-}]$ contain the positive
and negative signal eigenvectors of $\mathbf P$, and let
$\mathbf S_{\mathbf A},\mathbf S_{\mathbf P}$ be the corresponding signed eigenvalue matrices.
Set
\begin{align*}
 \mathbf H&=\operatorname{diag}(\mathbf U_{\mathbf A+}^\top\mathbf U_{\mathbf P+},
                              \mathbf U_{\mathbf A-}^\top\mathbf U_{\mathbf P-}),\\
 \mathbf H^{(m)}&=\operatorname{diag}((\mathbf U_{\mathbf A+}^{(m)})^\top\mathbf U_{\mathbf P+},
                              (\mathbf U_{\mathbf A-}^{(m)})^\top\mathbf U_{\mathbf P-}).
\end{align*}
Here $\operatorname{sgn}(\mathbf H)$ denotes the orthogonal polar factor.

We use the initialization estimates in Lemma~\ref{lem:X0} and
\citep[Lemma~S2.3]{xie2024entrywise}, together with the Davis--Kahan
theorem. For a fixed $i$, put
\[
 \mathbf L_i=\mathbf U_{\mathbf A}^{(i)}\mathbf H^{(i)}-\mathbf U_{\mathbf P},
 \qquad
 \mathbf K_i=\mathbf U_{\mathbf A}\mathbf H-\mathbf U_{\mathbf A}^{(i)}\mathbf H^{(i)}.
\]
The required estimates are
\begin{align}
&\|\mathbf L_i\|_2=\tOp((n\rho_n)^{-1/2}),\quad
 \|\mathbf L_i\|_{2\to\infty}=\tOp(n^{-1/2}),\label{eq:47-loo}\\
& \|\mathbf K_i\|_2=\tOp\!\left\{
 \frac{\sqrt{\log m_n}}{n\sqrt{\rho_n}};m_n\right\},\quad
 \|\mathbf U_{\mathbf A}\|_{2\to\infty}=\tOp(n^{-1/2}),\label{eq:47-align}\\
& \|\operatorname{sgn}(\mathbf H)-\mathbf H\|_2
 =\tOp((n\rho_n)^{-1}),\label{eq:47-polar}\\
 &\big\||\mathbf S_{\mathbf A}|^{1/2}\operatorname{sgn}(\mathbf H)
 -\operatorname{sgn}(\mathbf H)|\mathbf S_{\mathbf P}|^{1/2}\big\|_2
 =\tOp\!\left\{
 \frac{\sqrt{1\vee\rho_n\log m_n}}{\sqrt{n\rho_n}};m_n\right\}.
 \label{eq:47-eigenvalues}
\end{align}
These are spectral inputs; they are not consequences of independence
alone. For a zero-diagonal adjacency matrix with
$\mathbf P=\mathbf X_0\mathbf I_{d_+,d_-}\mathbf X_0^\top$, these inputs are
understood for that model, including its deterministic diagonal perturbation.

\begin{lemma}\label[lemma]{lem:term5}
Let $(\mathbf B_{nij})_{i,j\in[n]}$ be a collection of deterministic $d\times d$ matrices
such that $\sup_{i,j}\|\mathbf B_{nij}\|_2\lesssim\rho_n^{-1}$.
Under Assumptions~\ref{assump:snr_network} and~\ref{assump:snr_covariates},
for any deterministic nonnegative diverging sequence $(m_n)_{n=1}^{\infty}$
with $m_n=O(n)$,
\[
 \left\|\frac1{n+p}\left\{\sum_{j=1}^n E_{ij}\mathbf B_{nij}
 (\mathbf Q^\top\widehat{\mathbf y}_j^{(0)}-\mathbf y_{0j})\right\}\right\|_2
 =\tOp\!\left\{\frac{\log m_n}{n\rho_n^{1/2}};m_n\right\}.
\]
Furthermore, if Assumption~\ref{assump:3} holds and
$\widehat{\mathbf X}^{(t)}$ is approximately linearized with orthogonal
alignment $\mathbf Q$, then for any $\epsilon>0$,
\[
 \left\|\frac1{n+p}\sum_{j=1}^n E_{ij}\mathbf B_{nij}
 (\mathbf Q^\top\widehat{\mathbf y}_j^{(t)}-\mathbf y_{0j})\right\|_2
 =\tOp\!\left\{\frac{(\log n)^{2+\epsilon}}{n\rho_n^{1/2}}\right\}.
\]
\end{lemma}

\begin{proof}
The signature-preserving alignment gives
\[
 \mathbf Q^\top\widehat{\mathbf y}_j^{(t)}-\mathbf y_{0j}
 =\mathbf I_{d_+,d_-}(\mathbf Q^\top\widehat{\mathbf x}_j^{(t)}-\mathbf x_{0j}).
\]
We use the elementary noise bounds
\begin{equation}\label{eq:47-rownoise}
 \sum_j|E_{ij}|=\tOp(n\rho_n),\qquad
 \left(\sum_jE_{ij}^2\right)^{1/2}=\tOp(\sqrt{n\rho_n}),
\end{equation}
which follow from Bernstein's inequality and $n\rho_n\gtrsim\log n$.
Also $\||\mathbf S_{\mathbf P}|^{1/2}\|_2\lesssim\sqrt{n\rho_n}$.
The dimension $d$ is fixed throughout.

\emph{Initialization.}
Work in oracle spectral coordinates, so that
\[
 \mathbf X_0=\mathbf U_{\mathbf P}|\mathbf S_{\mathbf P}|^{1/2},
 \qquad
 \mathbf Q=\operatorname{sgn}(\mathbf H).
\]
Then $\widehat{\mathbf X}^{(0)}\mathbf Q-\mathbf X_0
=\mathbf D_1+\mathbf D_2+\mathbf D_3+\mathbf D_4$, where
\begin{align*}
 \mathbf D_1&=\mathbf U_{\mathbf A}\bigl\{|\mathbf S_{\mathbf A}|^{1/2}\operatorname{sgn}(\mathbf H)
 -\operatorname{sgn}(\mathbf H)|\mathbf S_{\mathbf P}|^{1/2}\bigr\},\\
 \mathbf D_2&=\mathbf U_{\mathbf A}\{\operatorname{sgn}(\mathbf H)-\mathbf H\}|\mathbf S_{\mathbf P}|^{1/2},\\
 \mathbf D_3&=\mathbf K_i|\mathbf S_{\mathbf P}|^{1/2},\qquad
 \mathbf D_4=\mathbf L_i|\mathbf S_{\mathbf P}|^{1/2}.
\end{align*}
Denote their contributions by
\[
 \mathbf T_{i,r}:=\frac1{n+p}\sum_j
 E_{ij}\mathbf B_{nij}\mathbf I_{d_+,d_-}\mathbf D_r^\top\mathbf e_j,
 \qquad r=1,2,3,4.
\]
By \eqref{eq:47-align}--\eqref{eq:47-rownoise},
\begin{align*}
 \|\mathbf T_{i,1}\|_2
 &\le\frac{\sum_j|E_{ij}|}{n+p}\max_j\|\mathbf B_{nij}\|_2
 \|\mathbf U_{\mathbf A}\|_{2\to\infty}
 \big\||\mathbf S_{\mathbf A}|^{1/2}\operatorname{sgn}(\mathbf H)
 -\operatorname{sgn}(\mathbf H)|\mathbf S_{\mathbf P}|^{1/2}\big\|_2
 \\&
 =\tOp\!\left\{\frac{\sqrt{1\vee\rho_n\log m_n}}
 {n\sqrt{\rho_n}};m_n\right\},\\
 \|\mathbf T_{i,2}\|_2
 &\le\frac{\sum_j|E_{ij}|}{n+p}\max_j\|\mathbf B_{nij}\|_2
 \|\mathbf U_{\mathbf A}\|_{2\to\infty}
 \|\operatorname{sgn}(\mathbf H)-\mathbf H\|_2\||\mathbf S_{\mathbf P}|^{1/2}\|_2\\
 &=\tOp\!\left(\frac1{n\sqrt{\rho_n}}\right).
\end{align*}
Cauchy--Schwarz and $\|\mathbf M\|_F\le\sqrt d\|\mathbf M\|_2$ give
\begin{align*}
 \|\mathbf T_{i,3}\|_2
 &\le\frac{(\sum_jE_{ij}^2)^{1/2}}{n+p}
 \max_j\|\mathbf B_{nij}\|_2\|\mathbf D_3\|_F\le\frac{(\sum_jE_{ij}^2)^{1/2}}{n+p}
 \max_j\|\mathbf B_{nij}\|_2\sqrt d\|\mathbf K_i\|_2\||\mathbf S_{\mathbf P}|^{1/2}\|_2\\
 &=\tOp\!\left\{\frac{\sqrt{\log m_n}}{n\sqrt{\rho_n}};m_n\right\}.
\end{align*}
For the last term, condition on $\mathbf A^{(i)}$ and put
\[
 \mathbf v_j=\mathbf B_{nij}\mathbf I_{d_+,d_-}|\mathbf S_{\mathbf P}|^{1/2}\mathbf L_i^\top\mathbf e_j.
\]
By \eqref{eq:47-loo},
\[
 \max_j\|\mathbf v_j\|_2=\tOp(\rho_n^{-1/2}),\qquad
 \left(\sum_j\|\mathbf v_j\|_2^2\right)^{1/2}=\tOp(\rho_n^{-1}).
\]
Conditional on $\mathbf A^{(i)}$, the weights are fixed and the
$E_{ij}$, $j\ne i$, are centered and independent.
Lemma S2.1 of the supplement to \cite{xie2024entrywise}, followed by
integration over the leave-one-out event, gives
\[
 \left\|\frac1{n+p}\sum_{j\ne i}E_{ij}\mathbf v_j\right\|_2
 =\tOp\!\left\{\frac1n\left(
 \frac{\log m_n}{\sqrt{\rho_n}}+
 \frac{\sqrt{\rho_n\log m_n}}{\rho_n}\right);m_n\right\}.
\]
The deterministic diagonal term satisfies
$\|E_{ii}\mathbf v_i/(n+p)\|_2=\tOp(\sqrt{\rho_n}/n)$,
since $|E_{ii}|=|P_{ii}|\le\rho_n$. Thus
$\|\mathbf T_{i,4}\|_2=\tOp\{\log m_n/(n\sqrt{\rho_n});m_n\}$.
All other terms are absorbed into this rate. An $n$-based probability
bound is also an $m_n$-based bound because $m_n=O(n)$.
This proves the initialization assertion.

\emph{General iterates.}
Approximate linearization yields
\[
 \mathbf Q^\top\widehat{\mathbf x}_j^{(t)}-\mathbf x_{0j}
 =\sum_a\boldsymbol\alpha_{nja}E_{ja}
 +\sum_l\boldsymbol\beta_{njl}W_{lj}+\mathbf r_{x_j}^{(t)},
\]
with deterministic coefficients satisfying
$\max_{j,a}\|\boldsymbol\alpha_{nja}\|_2\lesssim(n\sqrt{\rho_n})^{-1}$
and $\max_{j,l}\|\boldsymbol\beta_{njl}\|_2\lesssim n^{-1}$.
The vector in the assertion is the sum of
\begin{align*}
 \mathbf T_{EE,i}&=\frac1{n+p}\sum_{j,a}
 E_{ij}E_{ja}\mathbf B_{nij}\mathbf I_{d_+,d_-}\boldsymbol\alpha_{nja},\\
 \mathbf T_{EW,i}&=\frac1{n+p}\sum_{j,l}
 E_{ij}W_{lj}\mathbf B_{nij}\mathbf I_{d_+,d_-}\boldsymbol\beta_{njl},\\
 \mathbf T_{R,i}&=\frac1{n+p}\sum_j
 E_{ij}\mathbf B_{nij}\mathbf I_{d_+,d_-}\mathbf r_{x_j}^{(t)}.
\end{align*}

For the $EE$ term, define
\[
 \gamma_{nija}^{(k)}
 :=\frac1{n+p}\mathbf e_k^\top
 \mathbf B_{nij}\mathbf I_{d_+,d_-}\boldsymbol\alpha_{nja},
 \qquad
 \max_{i,j,a}|\gamma_{nija}^{(k)}|
 \lesssim\frac1{n^2\rho_n^{3/2}}.
\]
Let $\xi_0>1$ be the exponent in Assumption~\ref{assump:3}
and choose $1<\eta<\min\{\xi_0,1+\epsilon/2\}$.
By Result A.3 of \cite{xie2023eigenvector}, applied
coordinatewise to the off-diagonal errors,
\[
 \left|
 \sum_{\substack{j\ne i\\a\ne j}}
 \gamma_{nija}^{(k)}E_{ij}E_{ja}
 \right|
 =\tOp\!\left\{
 (n\rho_n)(\log n)^{2\eta}
 \max_{i,j,a}|\gamma_{nija}^{(k)}|
 \right\}
 =\tOp\!\left\{
 \frac{(\log n)^{2\eta}}{n\sqrt{\rho_n}}
 \right\}.
\]
Here deterministic rescaling of the coefficients is understood
if needed. The terms with $j=i$ or $a=j$ satisfy, by
$|E_{jj}|\le\rho_n$ and \eqref{eq:47-rownoise},
\[
 \left|
 \sum_{\substack{j,a:\\j=i\ \mathrm{or}\ a=j}}
 \gamma_{nija}^{(k)}E_{ij}E_{ja}
 \right|
 \lesssim
 \frac{\rho_n}{n^2\rho_n^{3/2}}\sum_j|E_{ij}|
 =\tOp(\sqrt{\rho_n}/n).
\]
Since $d$ is fixed and $2\eta<2+\epsilon$,
\begin{equation}\label{eq:47-EE}
 \|\mathbf T_{EE,i}\|_2
 =\tOp\!\left\{
 \frac{(\log n)^{2+\epsilon}}{n\sqrt{\rho_n}}
 \right\}.
\end{equation}

For the $EW$ term, condition on $\mathbf E$.
The $W_{lj}$ are independent standard Gaussian variables, independent
of $\mathbf E$. Since $p=O(n)$, their coefficient squared norms satisfy
\[
 \sum_{j,l}\left\|\frac{E_{ij}\mathbf B_{nij}\mathbf I_{d_+,d_-}\boldsymbol\beta_{njl}}
 {n+p}\right\|_2^2
 \lesssim\frac{p}{n^4\rho_n^2}\sum_jE_{ij}^2
 =\tOp\!\left(\frac1{n^2\rho_n}\right).
\]
Conditional Gaussian concentration in fixed dimension therefore gives
\begin{equation}\label{eq:47-EW}
 \|\mathbf T_{EW,i}\|_2
 =\tOp\!\left\{\frac{\sqrt{\log n}}{n\sqrt{\rho_n}}\right\}.
\end{equation}

For the remainder, weighted Cauchy--Schwarz gives
\begin{align*}
 \|\mathbf T_{R,i}\|_2
 &\le\frac1{n+p}\sum_j|E_{ij}|
       \|\mathbf B_{nij}\mathbf I_{d_+,d_-}\mathbf r_{x_j}^{(t)}\|_2
       \le\frac1{n+p}\left(\sum_j|E_{ij}|\right)^{1/2}
 \left(\sum_j|E_{ij}|\|\mathbf B_{nij}\mathbf I_{d_+,d_-}\mathbf r_{x_j}^{(t)}\|_2^2\right)^{1/2}\\
 &\le\frac{C}{(n+p)\rho_n}\left(\sum_j|E_{ij}|\right)
       \|\mathbf R_X^{(t)}\|_{2\to\infty}
       =\tOp\!\left\{\frac{(\log n)^{2+\epsilon}}{n\sqrt{\rho_n}}\right\}.
\end{align*}
The last step uses \eqref{eq:47-rownoise} and the remainder bound in the definition of approximate linearization. It does not require independence between the remainder and the noise, and retains the factor $\|\mathbf B_{nij}\|_2\lesssim\rho_n^{-1}$.
Combining this bound with \eqref{eq:47-EE} and \eqref{eq:47-EW}
proves the final assertion.
\end{proof}

\subsection{Term \texorpdfstring{\eqref{eq:term6}}{(\ref{eq:term6})}}\label{sec:term6}

The proof combines the initialization estimate in
Lemma~\ref{lem:X0}, the leave-one-out spectral estimates used
in the proof of Lemma~\ref{lem:term5}, and the Taylor bound in
Lemma~\ref{lem:Taylor_bounds}.
Retaining the notation of the preceding lemmas, write
\[
 \boldsymbol\Delta_j^{(t)}
 =\mathbf Q^\top\widehat{\mathbf x}_j^{(t)}-\mathbf x_{0j},
 \qquad
 \mathbf D^{(t)}
 =\widehat{\mathbf X}^{(t)}\mathbf Q-\mathbf X_0.
\]


\begin{lemma}\label[lemma]{lem:term6}
Under Assumptions~\ref{assump:snr_network} and~\ref{assump:snr_covariates}, let $\tau=\delta\rho_n$ with fixed
$\delta\in(0,1/2)$. For fixed $i$ and every deterministic diverging sequence
$m_n=O(n)$,
\[
 \left\|\frac1{n+p}\sum_{j=1}^n E_{ij}
 \mathbf r_g(\mathbf Q^\top\widehat{\mathbf x}_i^{(0)},
 \mathbf Q^\top\widehat{\mathbf y}_j^{(0)},\mathbf x_{0i},\mathbf y_{0j})\right\|_2
 =\tOp\left\{\frac{\log m_n}{n\sqrt{\rho_n}};m_n\right\}.
\]
Furthermore, if $\|\mathbf D^{(t)}\|_{2\to\infty}
=\tOp\{\sqrt{\log n/n}\}$, then
\[
 \left\|\frac1{n+p}\sum_{j=1}^n E_{ij}
 \mathbf r_g(\mathbf Q^\top\widehat{\mathbf x}_i^{(t)},
 \mathbf Q^\top\widehat{\mathbf y}_j^{(t)},\mathbf x_{0i},\mathbf y_{0j})\right\|_2
 =\tOp\left\{\frac{\log n}{n\sqrt{\rho_n}}\right\}.
\]
\end{lemma}
\begin{proof}
For fixed $i$, define the noise-weighted Taylor remainder vector
\[
 \mathbf T_i^{(t)}:=\frac1{n+p}\sum_{j=1}^n E_{ij}
 \mathbf r_g\!\left(
 \mathbf Q^\top\widehat{\mathbf x}_i^{(t)},
 \mathbf Q^\top\widehat{\mathbf y}_j^{(t)},
 \mathbf x_{0i},\mathbf y_{0j}\right)\in\mathbb R^d.
\]
We bound $\|\mathbf T_i^{(0)}\|_2$ for initialization and
$\|\mathbf T_i^{(t)}\|_2$ under the stated row-error condition for later iterates.

Fix $c>0$ throughout; constants may depend on $c$ and $\delta$.
The initialization localization in Lemma~\ref{lem:X0}, or the assumed row bound
for the later iterate, places the aligned arguments in a ball of radius
$K_c\sqrt{\rho_n}$ except with probability $O(n^{-c})$.
For later iterates, Assumption~\ref{assump:snr_network} gives
$\sqrt{\log n/n}\lesssim\sqrt{\rho_n}$. On the row-error event,
\[
 \max_j\|\mathbf Q^\top\widehat{\mathbf x}_j^{(t)}\|_2
 \le\|\mathbf X_0\|_{2\to\infty}+\|\mathbf D^{(t)}\|_{2\to\infty}
 \le\sqrt{\rho_n}+C_c\sqrt{\log n/n}
 \le K_c\sqrt{\rho_n}.
\]
The same bound holds for the aligned $\mathbf y$ arguments because
$\|\I\mathbf v\|_2=\|\mathbf v\|_2$.
Lemma~\ref{lem:Taylor_bounds} therefore gives
\begin{equation}\label{eq:X0-lineart}
 \|\mathbf T_i^{(t)}\|_2\le
 \frac{C_{c,\delta}}{n\rho_n^{3/2}}
 \left\{s_i\|\boldsymbol\Delta_i^{(t)}\|_2^2+
 \sum_j|E_{ij}|\|\boldsymbol\Delta_j^{(t)}\|_2^2\right\},
 \qquad s_i=\sum_j|E_{ij}|,
\end{equation}
where $\mathbf T_i^{(t)}$ denotes the vector in the statement. We used
$n+p\ge n$ and the triangle inequality; no independence of the errors
and $\mathbf E$ is needed in this step.
For $j\ne i$, $\mathbb E|E_{ij}|=2P_{ij}(1-P_{ij})\le2\rho_n$;
the diagonal contributes $|E_{ii}|=|P_{ii}|\le\rho_n$. Since
$|E_{ij}|\le1$, and $\operatorname{Var}(|E_{ij}|)\le\rho_n$,
Bernstein's inequality and a union bound imply
\begin{equation}\label{eq:noise}
 \max_i s_i=\tOp(n\rho_n).
\end{equation}
Here $n\rho_n\gtrsim\log n$ is supplied by Assumption~\ref{assump:snr_network}.

\paragraph{Initialization: the fixed row.}
Lemma~\ref{lem:X0} supplies
$\|\boldsymbol\Delta_i^{(0)}\|_2=\tOp\{\sqrt{\log m_n/n};m_n\}$.
Thus the first contribution in \eqref{eq:X0-lineart} is
$\tOp\{\log m_n/(n\sqrt{\rho_n});m_n\}$.
An $n$-based failure probability is also $O(m_n^{-c})$ because $m_n=O(n)$.

\paragraph{Initialization: the weighted errors of the other rows.}

The full embedding $\widehat{\mathbf X}^{(0)}$ depends on all entries
of $\mathbf A$, including the noise row $(E_{ij})_{j=1}^n$.
Consequently, its squared row errors cannot be treated as fixed weights
independent of this noise row when applying Bernstein's inequality.
To separate this dependence, use the leave-one-out matrix
$\mathbf A^{(i)}$ defined in the proof of Lemma~\ref{lem:term5}.
The random entries retained in $\mathbf A^{(i)}$ are independent of
$(E_{ij})_{j=1}^n$. Hence all spectral quantities constructed from
$\mathbf A^{(i)}$ are independent of that row, including $E_{ii}$.
We split the full embedding error into the difference between the full
and leave-one-out spectral quantities and the error of the leave-one-out
quantities relative to the oracle. The former is controlled by
norm inequalities; the latter admits conditional concentration.

As in the proof of Lemma~\ref{lem:term5}, work in oracle spectral
coordinates, so that
\[
 \mathbf X_0=\mathbf U_{\mathbf P}|\mathbf S_{\mathbf P}|^{1/2}.
\]
Let
\[
 \mathbf H=\operatorname{diag}(\mathbf U_{\mathbf A+}^\top\mathbf U_{\mathbf P+},
 \mathbf U_{\mathbf A-}^\top\mathbf U_{\mathbf P-}),
\]
and define $\mathbf H^{(i)}$ analogously. As in Lemma~\ref{lem:term5}, take
$\mathbf Q=\operatorname{sgn}(\mathbf H)$.
Here $\operatorname{sgn}$ is the orthogonal polar factor.
Exactly as in the proof of Lemma~\ref{lem:term5},
\begin{align*}
 \mathbf D^{(0)}&=\mathbf D_1+\mathbf D_2+\mathbf D_3+\mathbf D_4,\\
 \mathbf D_1&=\mathbf U_{\mathbf A}\bigl\{|\mathbf S_{\mathbf A}|^{1/2}\operatorname{sgn}(\mathbf H)-\operatorname{sgn}(\mathbf H)|\mathbf S_{\mathbf P}|^{1/2}\bigr\},\\
 \mathbf D_2&=\mathbf U_{\mathbf A}(\operatorname{sgn}(\mathbf H)-\mathbf H)|\mathbf S_{\mathbf P}|^{1/2},\\
 \mathbf D_3&=\bigl(\mathbf U_{\mathbf A}\mathbf H-\mathbf U_{\mathbf A}^{(i)}\mathbf H^{(i)}\bigr)|\mathbf S_{\mathbf P}|^{1/2},\\
 \mathbf D_4&=\bigl(\mathbf U_{\mathbf A}^{(i)}\mathbf H^{(i)}-\mathbf U_{\mathbf P}\bigr)|\mathbf S_{\mathbf P}|^{1/2}.
\end{align*}
We use the following spectral inputs from that argument:
\begin{align*}
& \||\mathbf S_{\mathbf P}|^{1/2}\|_2\lesssim\sqrt{n\rho_n},\quad
  \|\mathbf U_{\mathbf A}\|_{2\to\infty}=\tOp(n^{-1/2}),\\
& \|\bigl\{|\mathbf S_{\mathbf A}|^{1/2}\operatorname{sgn}(\mathbf H)-\operatorname{sgn}(\mathbf H)|\mathbf S_{\mathbf P}|^{1/2}\bigr\}\|_2=\tOp\left\{
 \frac{\sqrt{1\vee\rho_n\log m_n}}{\sqrt{n\rho_n}};m_n\right\},\quad
  \|\operatorname{sgn}(\mathbf H)-\mathbf H\|_2=\tOp((n\rho_n)^{-1}),\\
& \|\bigl(\mathbf U_{\mathbf A}\mathbf H-\mathbf U_{\mathbf A}^{(i)}\mathbf H^{(i)}\bigr)\|_2=\tOp\left\{\frac{\sqrt{\log m_n}}{n\sqrt{\rho_n}};m_n\right\},\quad
  \|\bigl(\mathbf U_{\mathbf A}^{(i)}\mathbf H^{(i)}-\mathbf U_{\mathbf P}\bigr)\|_2=\tOp((n\rho_n)^{-1/2}),\\
& \|\bigl(\mathbf U_{\mathbf A}^{(i)}\mathbf H^{(i)}-\mathbf U_{\mathbf P}\bigr)\|_{2\to\infty}=\tOp(n^{-1/2}).
\end{align*}
For $\ell_n=\log m_n$ (eventually $\ell_n\ge1$), define
\[
 S_r=\frac1{n\rho_n^{3/2}}\sum_j|E_{ij}|
       \|\mathbf e_j^\top\mathbf D_r\|_2^2.
\]
The inequality $\|\sum_{r=1}^4\mathbf v_r\|_2^2\le4\sum_r\|\mathbf v_r\|_2^2$
shows that the second contribution in \eqref{eq:X0-lineart} is at most
$4C_{c,\delta}\sum_rS_r$. The factor four must be retained.
The spectral estimates and \eqref{eq:noise} yield
\begin{align*}
 S_1&\le\frac{s_i}{n\rho_n^{3/2}}
       \|\mathbf U_{\mathbf A}\|_{2\to\infty}^2\|\bigl\{|\mathbf S_{\mathbf A}|^{1/2}\operatorname{sgn}(\mathbf H)-\operatorname{sgn}(\mathbf H)|\mathbf S_{\mathbf P}|^{1/2}\bigr\}\|_2^2
 =\tOp\left\{\frac{1\vee\rho_n\ell_n}{n^2\rho_n^{3/2}};m_n\right\},\\
 S_2&\le\frac{s_i}{n\rho_n^{3/2}}
       \|\mathbf U_{\mathbf A}\|_{2\to\infty}^2
       \|\operatorname{sgn}(\mathbf H)-\mathbf H\|_2^2\||\mathbf S_{\mathbf P}|^{1/2}\|_2^2
 =\tOp\{(n^2\rho_n^{3/2})^{-1}\},\\
 S_3&\le\frac{\|\bigl(\mathbf U_{\mathbf A}\mathbf H-\mathbf U_{\mathbf A}^{(i)}\mathbf H^{(i)}\bigr)|\mathbf S_{\mathbf P}|^{1/2}\|_F^2}{n\rho_n^{3/2}}
 \le\frac{d\|\bigl(\mathbf U_{\mathbf A}\mathbf H-\mathbf U_{\mathbf A}^{(i)}\mathbf H^{(i)}\bigr)\|_2^2\||\mathbf S_{\mathbf P}|^{1/2}\|_2^2}{n\rho_n^{3/2}}
 =\tOp\left\{\frac{\ell_n}{n^2\rho_n^{3/2}};m_n\right\}.
\end{align*}
In $S_3$ we used $|E_{ij}|\le1$ and fixed $d$.
All three rates are $O(\ell_n/(n\sqrt{\rho_n}))$ since $n\rho_n\gtrsim1$.

\paragraph{Conditional concentration for $S_4$.}
The remaining component $\mathbf D_4$ depends only on the leave-one-out
matrix and the oracle quantities. Define
\[
 a_j=\left\|\mathbf e_j^\top\bigl(\mathbf U_{\mathbf A}^{(i)}\mathbf H^{(i)}-\mathbf U_{\mathbf P}\bigr)
                  |\mathbf S_{\mathbf P}|^{1/2}\right\|_2^2.
\]
Fix the prescribed tail exponent $c>0$. The spectral estimates for
$\bigl(\mathbf U_{\mathbf A}^{(i)}\mathbf H^{(i)}-\mathbf U_{\mathbf P}\bigr)$ give an event $\mathcal G_i$, measurable with respect to
$\mathbf A^{(i)}$, such that $\mathbb P(\mathcal G_i^c)=O(n^{-c})$ and,
on $\mathcal G_i$,
\[
 \|\bigl(\mathbf U_{\mathbf A}^{(i)}\mathbf H^{(i)}-\mathbf U_{\mathbf P}\bigr)\|_2\le C_c(n\rho_n)^{-1/2},\qquad
 \|\bigl(\mathbf U_{\mathbf A}^{(i)}\mathbf H^{(i)}-\mathbf U_{\mathbf P}\bigr)\|_{2\to\infty}\le C_cn^{-1/2}.
\]
Using the deterministic bound
$\big\||\mathbf S_{\mathbf P}|^{1/2}\big\|_2\lesssim\sqrt{n\rho_n}$,
we obtain on this event
\begin{align*}
 \max_j a_j
 &\le\|\bigl(\mathbf U_{\mathbf A}^{(i)}\mathbf H^{(i)}-\mathbf U_{\mathbf P}\bigr)\|_{2\to\infty}^2
       \big\||\mathbf S_{\mathbf P}|^{1/2}\big\|_2^2
 \lesssim_c n^{-1}(n\rho_n)=\rho_n,\\
 \sum_j a_j
 &=\big\|\bigl(\mathbf U_{\mathbf A}^{(i)}\mathbf H^{(i)}-\mathbf U_{\mathbf P}\bigr)|\mathbf S_{\mathbf P}|^{1/2}\big\|_F^2
 \le d\|\bigl(\mathbf U_{\mathbf A}^{(i)}\mathbf H^{(i)}-\mathbf U_{\mathbf P}\bigr)\|_2^2
       \big\||\mathbf S_{\mathbf P}|^{1/2}\big\|_2^2
 \lesssim_c d(n\rho_n)^{-1}(n\rho_n)\lesssim_c1,\\
 \sum_j a_j^2&\le(\max_j a_j)\sum_j a_j\lesssim_c\rho_n.
\end{align*}
Here $d$ is fixed. These are bounds on $\mathcal G_i$, rather than
deterministic bounds valid for every realization.

Conditional on $\mathbf A^{(i)}$, the weights $a_j$ are fixed, while the
off-diagonal noise variables retain their independent Bernoulli-error distributions; the diagonal remains deterministic.
Let $V_j=|E_{ij}|-\mathbb E|E_{ij}|$. In particular, $V_i=0$. For $j\ne i$,
\[
 \mathbb E(V_j\mid\mathbf A^{(i)})=0,\qquad |V_j|\le1,
 \qquad
 \mathbb E(V_j^2\mid\mathbf A^{(i)})
 =\operatorname{Var}(|E_{ij}|)
 \le\mathbb E E_{ij}^2=P_{ij}(1-P_{ij})\le\rho_n.
\]
Thus $a_jV_j$ are conditionally independent and centered, their absolute
values are at most $\max_j a_j$, and their total conditional variance is
at most $\rho_n\sum_j a_j^2$. Bernstein's inequality gives,
for every $u>0$,
\[
 \mathbb P\left(\left|\sum_jV_ja_j\right|>
 \sqrt{2u\rho_n\sum_j a_j^2}+\frac{2u}{3}\max_j a_j
 \,\middle|\,\mathbf A^{(i)}\right)\le2e^{-u}.
\]
Take $u=(c+2)\ell_n$. On $\mathcal G_i$, the threshold satisfies
\[
 \sqrt{2u\rho_n\sum_j a_j^2}+\frac{2u}{3}\max_j a_j
 \le C_c\rho_n(\sqrt{\ell_n}+\ell_n).
\]
To recover the uncentered sum, write
\[
 \sum_j|E_{ij}|a_j
 =\sum_jV_ja_j+
   \mathbb E\!\left(\sum_j|E_{ij}|a_j\,\middle|\,\mathbf A^{(i)}\right).
\]
The conditional mean in this identity is bounded on $\mathcal G_i$ by
\[
 \mathbb E\!\left(\sum_j|E_{ij}|a_j\,\middle|\,\mathbf A^{(i)}\right)
 =\sum_{j\ne i}2P_{ij}(1-P_{ij})a_j+|P_{ii}|a_i
 \le2\rho_n\sum_j a_j\lesssim_c\rho_n.
\]
After enlarging $C_c$, it follows that, for every realization in
$\mathcal G_i$,
\[
 \mathbb P\left(\sum_j|E_{ij}|a_j>
 C_c\rho_n(1+\sqrt{\ell_n}+\ell_n)
 \,\middle|\,\mathbf A^{(i)}\right)
 \le2e^{-(c+2)\ell_n}=2m_n^{-(c+2)}.
\]

To pass from the conditional bound to an unconditional one, let
$\mathcal B_i$ be the event inside this last probability. Since
$\mathcal G_i$ is measurable with respect to $\mathbf A^{(i)}$, the law
of total expectation gives
\begin{align*}
 \mathbb P(\mathcal B_i)
 &\le\mathbb P(\mathcal B_i\cap\mathcal G_i)
      +\mathbb P(\mathcal G_i^c)=\mathbb E\!\left[
      \mathbf1_{\mathcal G_i}
      \mathbb P(\mathcal B_i\mid\mathbf A^{(i)})\right]
      +\mathbb P(\mathcal G_i^c)\le2m_n^{-(c+2)}+O(n^{-c})=O(m_n^{-c}),
\end{align*}
where the last step uses $m_n=O(n)$. Thus the two sources of failure
are the failure of the leave-one-out spectral event and the conditional
Bernstein deviation on that event. No independence between
$\mathcal B_i$ and $\mathcal G_i$ is needed.
Dividing the resulting threshold by $n\rho_n^{3/2}$ proves
\[
 S_4=\frac{\sum_j|E_{ij}|a_j}{n\rho_n^{3/2}}
 =\tOp\left\{\frac{1+\sqrt{\ell_n}+\ell_n}{n\sqrt{\rho_n}};m_n\right\}
 =\tOp\left\{\frac{\ell_n}{n\sqrt{\rho_n}};m_n\right\}.
\]
Indeed, $\ell_n=\log m_n\to\infty$, so eventually
$1+\sqrt{\ell_n}+\ell_n\le3\ell_n$.
Only the leave-one-out weights $a_j$ were treated as independent of the
noise row. The full embedding errors and the first three components
were controlled by norm bounds without asserting such independence.
Combining the four bounds with the fixed-row contribution proves the
initialization assertion.

\paragraph{Later iterates.}
Both squared errors in \eqref{eq:X0-lineart} are at most
$\|\mathbf D^{(t)}\|_{2\to\infty}^2$, so
\[
 \|\mathbf T_i^{(t)}\|_2\le
 \frac{2C_{c,\delta}s_i \|\mathbf D^{(t)}\|_{2\to\infty}^2}{n\rho_n^{3/2}}
 =\tOp\left\{\frac{\log n}{n\sqrt{\rho_n}}\right\}.
\]
This uses \eqref{eq:noise} and $\|\mathbf D^{(t)}\|_{2\to\infty}^2=\tOp(\log n/n)$.
There is no need for approximate linearization or independence of the
iterate and the noise. Finite intersections of the probability events
preserve the stated tail orders, by choosing the input exponents larger
than $c$. This completes the proof.
\end{proof}

\subsection{Terms \texorpdfstring{\eqref{eq:term7}}{(\ref{eq:term7})} and \texorpdfstring{\eqref{eq:term8}}{(\ref{eq:term8})}}
\label{sec:term78}
\begin{lemma}\label[lemma]{lem:term78}
Under Assumptions~\ref{assump:snr_network} and~\ref{assump:snr_covariates},
if $\tau=\delta\rho_n$ for some constant $\delta\in(0,1/2)$, then for
any deterministic nonnegative diverging sequence $(m_n)_{n=1}^{\infty}$
with $m_n=O(n)$,
\begin{align*}
 &\left\|\frac1{n+p}\sum_{l=1}^pW_{li}
 (\mathbf Q^\top\widehat{\mathbf z}_l^{(0)}-\mathbf z_{0l})^\top\right\|_2
 +\left\|\frac1{n+p}\sum_{l=1}^p\mathbf z_{0l}\mathbf x_{0i}^\top
 \{\mathbf e_l^\top(\widehat{\mathbf Z}^{(0)}\mathbf Q-\mathbf Z_0)\}^\top\right\|_2
 \\
 &
 \quad
 =\tOp\left\{\frac{\sqrt{\log m_n}+\sqrt{\log n}}{n\sqrt{\rho_n}};m_n\right\}.
\end{align*}
Furthermore, suppose Assumption~\ref{assump:3} holds. If
$\|\widehat{\mathbf Z}^{(t)}\mathbf Q-\mathbf Z_0\|_{2\to\infty}
=\tOp\{\sqrt{\log n/(n\rho_n)}\}$ and
\[
 \widehat{\mathbf Z}^{(t)}\mathbf Q-\mathbf Z_0
 =\mathbf W\mathbf X_0(\mathbf X_0^\top\mathbf X_0)^{-1}
  +\mathbf R_{\mathbf Z}^{(t)},\qquad
 \|\mathbf R_{\mathbf Z}^{(t)}\|_{2\to\infty}
 =\tOp\left\{\frac{(\log n)^{2+\epsilon}}{n\rho_n}\right\}
\]
for any $\epsilon>0$, then
\begin{align*}
 &\left\|\frac1{n+p}\sum_{l=1}^pW_{li}
 (\mathbf Q^\top\widehat{\mathbf z}_l^{(t)}-\mathbf z_{0l})^\top\right\|_2
 +\left\|\frac1{n+p}\sum_{l=1}^p\mathbf z_{0l}\mathbf x_{0i}^\top
 \{\mathbf e_l^\top(\widehat{\mathbf Z}^{(t)}\mathbf Q-\mathbf Z_0)\}^\top\right\|_2\\
 &\hspace{2cm}=\tOp\left\{\frac{(\log n)^{2+\epsilon}}{n\sqrt{\rho_n}}\right\}.
\end{align*}
\end{lemma}

\begin{proof}
Fix $i$ and retain the notation of the preceding lemmas. Define
$\mathbf W(i)=[W_{lj}(i)]_{p\times n}$ by
\[
 W_{lj}(i)=\begin{cases}W_{lj},&j\ne i,\\0,&j=i,\end{cases}
 \qquad
 \widehat{\mathbf Z}^{(0)}(i)
 =\{\mathbf Z_0\mathbf X_0^\top+\mathbf W(i)\}
 \widehat{\mathbf X}^{(0)}
 \{(\widehat{\mathbf X}^{(0)})^\top\widehat{\mathbf X}^{(0)}\}^{-1}.
\]
Because $\widehat{\mathbf X}^{(0)}$ and $\mathbf Q$ depend only on
$\mathbf A$, the error $\widehat{\mathbf Z}^{(0)}(i)\mathbf Q-\mathbf Z_0$
is independent of $(W_{1i},\ldots,W_{pi})$.
By Lemma~\ref{lem:X0},
\[
 \|\widehat{\mathbf X}^{(0)}\|_{2\to\infty}=\tOp(\sqrt{\rho_n}),\qquad
 \|\{(\widehat{\mathbf X}^{(0)})^\top\widehat{\mathbf X}^{(0)}\}^{-1}\|_2
 =\tOp\{(n\rho_n)^{-1}\}.
\]
Corollary~3.11 of \cite{bandeira2016sharp}, applied to the Gaussian
column $\mathbf W\mathbf e_i$ and the rectangular Gaussian matrix
$\mathbf W(i)$, gives
\begin{equation}\label{eq:49-gaussian-norms}
 \|\mathbf W\mathbf e_i\|_2=\tOp\{\sqrt{p+\log m_n};m_n\},\qquad
 \|\mathbf W(i)\|_2=\tOp(\sqrt{p+n})=\tOp(\sqrt n).
\end{equation}
To verify both bounds, recall that for a rectangular Gaussian matrix
with entry standard deviations $b_{lj}$, the parameters in
Corollary~3.11 are
\[
 \sigma_1=\max_l\Bigl(\sum_j b_{lj}^2\Bigr)^{1/2},\qquad
 \sigma_2=\max_j\Bigl(\sum_l b_{lj}^2\Bigr)^{1/2},\qquad
 \sigma_*=\max_{l,j}b_{lj}.
\]
Fix the corollary's parameter $\varepsilon=1/2$.
For the $p\times1$ column $\mathbf W\mathbf e_i$, these parameters
are $\sigma_1=1$, $\sigma_2=\sqrt p$, and $\sigma_*=1$.
The dimension-dependent logarithmic term is zero because
$\log(p\wedge1)=0$. Hence Corollary~3.11 gives
\[
 \mathbb P\!\left(\|\mathbf W\mathbf e_i\|_2>
 \tfrac32(1+\sqrt p)+r\right)\le e^{-r^2/2},\qquad r\ge0.
\]
For any prescribed $c>0$, take
$r=\sqrt{2(c+1)\log m_n}$. The failure probability is
$m_n^{-(c+1)}$, and
$1+\sqrt p+\sqrt{\log m_n}\lesssim_c\sqrt{p+\log m_n}$
for all sufficiently large $n$. This proves the first bound in
\eqref{eq:49-gaussian-norms}.

For $\mathbf W(i)$, remove its zero column; this does not change the
operator norm. The remaining $p\times(n-1)$ matrix has independent
standard Gaussian entries, so
$\sigma_1=\sqrt{n-1}$, $\sigma_2=\sqrt p$, and $\sigma_*=1$.
The same corollary therefore gives
\[
 \mathbb P\!\left(\|\mathbf W(i)\|_2>
 C\{\sqrt{n-1}+\sqrt p+\sqrt{\log(p\wedge(n-1))}\}+r\right)
 \le e^{-r^2/2}.
\]
Take $r=\sqrt{2(c+1)\log n}$. Since $p=O(n)$, the threshold is
$O_c(\sqrt{p+n})=O_c(\sqrt n)$, while the failure probability is
$n^{-(c+1)}$. This proves the second bound in
\eqref{eq:49-gaussian-norms}.
Throughout, an $n$-based polynomial failure bound is also
an $m_n$-based bound since $m_n=O(n)$.

\paragraph{Term \eqref{eq:term7} at initialization.}
Since $\mathbf W-\mathbf W(i)=(\mathbf W\mathbf e_i)\mathbf e_i^\top$,
\begin{align*}
 \|(\widehat{\mathbf Z}^{(0)}-\widehat{\mathbf Z}^{(0)}(i))\mathbf Q\|_F
 &=\|\mathbf W\mathbf e_i\|_2
 \left\|\mathbf e_i^\top\widehat{\mathbf X}^{(0)}
 \{(\widehat{\mathbf X}^{(0)})^\top\widehat{\mathbf X}^{(0)}\}^{-1}\mathbf Q\right\|_2\\
 &=\tOp\left\{\frac{\sqrt{p+\log m_n}}{n\sqrt{\rho_n}};m_n\right\}
 =\tOp\{(n\rho_n)^{-1/2};m_n\}.
\end{align*}
The Frobenius identity holds because the difference is rank one.
Lemma~\ref{lem:Z0} and fixed $d$ give
\[
 \|\widehat{\mathbf Z}^{(0)}\mathbf Q-\mathbf Z_0\|_F
 =\tOp\left\{\frac{\sqrt{p+n}}{\sqrt{n\rho_n}}\right\}
 =\tOp(\rho_n^{-1/2}).
\]
The triangle inequality therefore also gives
$\|\widehat{\mathbf Z}^{(0)}(i)\mathbf Q-\mathbf Z_0\|_F
=\tOp\{\rho_n^{-1/2};m_n\}$.

Conditional on $(\mathbf A,\mathbf W(i))$, the vector
$\mathbf e_i^\top\mathbf W^\top
(\widehat{\mathbf Z}^{(0)}(i)\mathbf Q-\mathbf Z_0)$ is centered Gaussian.
Sub-Gaussian concentration gives, for $u\ge1$,
\begin{align*}
 &\mathbb P\left(
 \|\mathbf e_i^\top\mathbf W^\top
 (\widehat{\mathbf Z}^{(0)}(i)\mathbf Q-\mathbf Z_0)\|_2
 >C\sqrt u\|\widehat{\mathbf Z}^{(0)}(i)\mathbf Q-\mathbf Z_0\|_F
 \,\middle|\,\mathbf A,\mathbf W(i)\right)\le2d e^{-u}.
\end{align*}
Choose $u=C_c\log m_n$. Averaging this conditional bound and adding
the failure probability of the Frobenius bound yields the corresponding
$\tOp\{\sqrt{\log m_n}/\sqrt{\rho_n};m_n\}$ bound.
Thus, using Corollary~3.11 and sub-Gaussian concentration as above,
\begin{align*}
 &\left\|\frac1{n+p}\mathbf e_i^\top\mathbf W^\top
 (\widehat{\mathbf Z}^{(0)}\mathbf Q-\mathbf Z_0)\right\|_2\\
 &\quad\le\frac{\|\mathbf W\mathbf e_i\|_2}{n+p}
 \|(\widehat{\mathbf Z}^{(0)}-\widehat{\mathbf Z}^{(0)}(i))\mathbf Q\|_F
 +\left\|\frac1{n+p}\mathbf e_i^\top\mathbf W^\top
 (\widehat{\mathbf Z}^{(0)}(i)\mathbf Q-\mathbf Z_0)\right\|_2\\
 &\quad=\tOp\left\{\frac1{n\sqrt{\rho_n}};m_n\right\}
 +\tOp\left\{\frac{\sqrt{\log m_n}}{n\sqrt{\rho_n}};m_n\right\}
 =\tOp\left\{\frac{\sqrt{\log m_n}}{n\sqrt{\rho_n}};m_n\right\}.
\end{align*}
Only the leave-one-out error is treated as independent of the removed
column; the full estimation error is not.

\paragraph{Term \eqref{eq:term8} at initialization.}
By the expansion in Lemma~\ref{lem:Z0} and the global remainder
bound \eqref{eq:RZop}, using $\|\mathbf R_{\mathbf Z}^{(0)}\|_F\le\sqrt d\|\mathbf R_{\mathbf Z}^{(0)}\|_2$ with fixed $d$,
\[
 \widehat{\mathbf Z}^{(0)}\mathbf Q-\mathbf Z_0
 =\mathbf W\mathbf X_0(\mathbf X_0^\top\mathbf X_0)^{-1}
 +\mathbf R_{\mathbf Z}^{(0)},\qquad
 \|\mathbf R_{\mathbf Z}^{(0)}\|_F
 =\tOp\left\{\frac{\sqrt{p+n}\sqrt{\log n}}{n\rho_n}\right\}.
\]
Substituting this expansion gives
\begin{align*}
 &\left\|\frac1{n+p}\sum_{l=1}^p\mathbf z_{0l}\mathbf x_{0i}^\top
 \{\mathbf e_l^\top(\widehat{\mathbf Z}^{(0)}\mathbf Q-\mathbf Z_0)\}^\top\right\|_2\\
 &\quad\le\left\|\frac1{n+p}\sum_{l=1}^p\sum_{j=1}^n
 W_{lj}\mathbf z_{0l}\mathbf x_{0i}^\top
 (\mathbf X_0^\top\mathbf X_0)^{-1}\mathbf x_{0j}\right\|_2
 +\frac{\|\mathbf R_{\mathbf Z}^{(0)}\|_F}{n+p}
 \left(\sum_l\|\mathbf z_{0l}\|_2^2\|\mathbf x_{0i}\|_2^2\right)^{1/2}.
\end{align*}
The Gaussian sum has total second moment at most
\[
 \frac{\sum_l\|\mathbf z_{0l}\|_2^2}{(n+p)^2}
 \mathbf x_{0i}^\top(\mathbf X_0^\top\mathbf X_0)^{-1}\mathbf x_{0i}
 \lesssim\frac{p}{n(n+p)^2},
\]
so sub-Gaussian concentration bounds its norm by
$\tOp\{\sqrt{p\log m_n}/((n+p)\sqrt n);m_n\}$.
The remainder contribution is
\[
 \tOp\left\{\frac{\sqrt{p+n}\sqrt{\log n}\sqrt{p\rho_n}}{(n+p)n\rho_n}\right\}
 =\tOp\{\sqrt{\log n}/(n\sqrt{\rho_n})\}.
\]
Combining the Gaussian leading term with this contribution gives
$\tOp\{(\sqrt{\log m_n}+\sqrt{\log n})/(n\sqrt{\rho_n});m_n\}$.
The $n$-based remainder event also has failure probability
$O(m_n^{-c})$ because $m_n=O(n)$.
This proves the initialization assertion.

\paragraph{Later iterates.}
For the noise-weighted term, apply the assumed expansion and split its
leading part using the same $\mathbf W(i)$:
\begin{align*}
 &\mathbf e_i^\top\mathbf W^\top\mathbf W\mathbf X_0
 (\mathbf X_0^\top\mathbf X_0)^{-1}
 =\|\mathbf W\mathbf e_i\|_2^2\mathbf x_{0i}^\top
 (\mathbf X_0^\top\mathbf X_0)^{-1}
 +(\mathbf W\mathbf e_i)^\top\mathbf W(i)\mathbf X_0
 (\mathbf X_0^\top\mathbf X_0)^{-1}.
\end{align*}
By \eqref{eq:49-gaussian-norms} with $m_n=n$, the first term divided
by $n+p$ is $\tOp\{(n\sqrt{\rho_n})^{-1}\}$.
Conditional on $\mathbf W(i)$, sub-Gaussian concentration bounds the
second term divided by $n+p$ by
\[
 \frac{C_c\sqrt{\log n}}{n+p}
 \|\mathbf W(i)\mathbf X_0(\mathbf X_0^\top\mathbf X_0)^{-1}\|_F
 \lesssim_c\frac{\sqrt{\log n}}{n\sqrt{\rho_n}},
\]
where Corollary~3.11 gives $\|\mathbf W(i)\|_2=\tOp(\sqrt n)$ and
$\|\mathbf X_0(\mathbf X_0^\top\mathbf X_0)^{-1}\|_F
\lesssim(n\rho_n)^{-1/2}$ since $d$ is fixed.
For the noise-weighted remainder, fix $\epsilon>0$ and use its
assumed row bound with exponent $\epsilon/2$. Assumption~\ref{assump:3}
gives $\rho_n^{-1/2}=O\{(\log n)^{\epsilon/2}\}$. Hence
\begin{align*}
 \left\|\frac{\mathbf e_i^\top\mathbf W^\top\mathbf R_{\mathbf Z}^{(t)}}{n+p}\right\|_2
 &\le\frac{\|\mathbf W\mathbf e_i\|_2\sqrt p}{n+p}
       \|\mathbf R_{\mathbf Z}^{(t)}\|_{2\to\infty}
       =\tOp\left\{\frac{(\log n)^{2+\epsilon/2}}{n\rho_n}\right\}
 =\tOp\left\{\frac{(\log n)^{2+\epsilon}}{n\sqrt{\rho_n}}\right\}.
\end{align*}
For the signal-weighted term, the Gaussian-sum calculation used at
initialization bounds the leading part by $\tOp(\sqrt{\log n}/n)$.
The remainder satisfies
\[
 \left\|\frac{\mathbf Z_0^\top\mathbf R_{\mathbf Z}^{(t)}\mathbf x_{0i}}{n+p}\right\|_2
 \le\frac{\|\mathbf Z_0\|_2\sqrt p\|\mathbf x_{0i}\|_2}{n+p}
       \|\mathbf R_{\mathbf Z}^{(t)}\|_{2\to\infty}
 =\tOp\{(\log n)^{2+\epsilon}/(n\sqrt{\rho_n})\}.
\]
Combining these bounds proves the second assertion. The remainder
bounds are deterministic norm inequalities and require no independence
between $\mathbf R_{\mathbf Z}^{(t)}$ and $\mathbf W$.
\end{proof}

\subsection{Control of the inverse Fisher matrix}
\label{sec:Ri2}
\begin{lemma}\label[lemma]{lem:Ri2}
Under Assumptions~\ref{assump:snr_network} and~\ref{assump:snr_covariates},
if $\tau=\delta\rho_n$ for some constant $\delta\in(0,1/2)$, then
\[
 \|\mathbf R_{i2}^{(0)}\|_2
 =\tOp\left\{\frac{\sqrt{\log m_n}}{\sqrt{n\rho_n}};m_n\right\},
\]
for any deterministic nonnegative diverging sequence $(m_n)_{n=1}^{\infty}$
with $m_n=O(n)$. Furthermore, if Assumption~\ref{assump:3} holds and
\[
 \|\widehat{\mathbf X}^{(t)}\mathbf Q-\mathbf X_0\|_{2\to\infty}
 =\tOp\{\sqrt{\log n/n}\},\qquad
 \|\widehat{\mathbf Z}^{(t)}\mathbf Q-\mathbf Z_0\|_{2\to\infty}
 =\tOp\{\sqrt{\log n/(n\rho_n)}\},
\]
then
\[
 \|\mathbf R_{i2}^{(t)}\|_2
 =\tOp\left\{\frac{\sqrt{\log n}}{\sqrt{n\rho_n}}\right\}.
\]
\end{lemma}

\begin{proof}
Recall the definitions
\begin{align*}
 \mathbf G_{in}
 &=\frac1{n+p}\left\{
 \sum_{j=1}^n\psi'(\mathbf x_{0i}^\top\mathbf y_{0j})
                   \mathbf y_{0j}\mathbf y_{0j}^\top
 +\sum_{l=1}^p\mathbf z_{0l}\mathbf z_{0l}^\top\right\},\\
 \widehat{\mathbf G}_{in}^{(t)}
 &=\frac1{n+p}\left\{
 \sum_{j=1}^n\psi'((\widehat{\mathbf x}_i^{(t)})^\top
                            \widehat{\mathbf y}_j^{(t)})
 (\mathbf Q^\top\widehat{\mathbf y}_j^{(t)})
 (\mathbf Q^\top\widehat{\mathbf y}_j^{(t)})^\top
 +\sum_{l=1}^p(\mathbf Q^\top\widehat{\mathbf z}_l^{(t)})
                     (\mathbf Q^\top\widehat{\mathbf z}_l^{(t)})^\top\right\},\\
 \mathbf R_{i2}^{(t)}&=(\widehat{\mathbf G}_{in}^{(t)})^{-1}-\mathbf G_{in}^{-1}.
\end{align*}
We first bound the difference of the two Fisher matrices, then control
both inverse norms. All constants below may depend on the prescribed
tail exponent $c>0$ and $\delta$.

\paragraph{The network contribution.}
The derivative bounds from Lemma~\ref{lem:Taylor_bounds} give
\[
 |\psi'(s)-\psi'(r)|\le C_\delta\rho_n^{-2}|s-r|.
\]
For every $c>0$, on an event of probability at least $1-O(n^{-c})$,
\[
 \mathbf Q^\top\widehat{\mathbf x}_i^{(t)},\quad
 \mathbf Q^\top\widehat{\mathbf y}_j^{(t)}
 \in\mathcal X(K_c\rho_n^{1/2}),\qquad i,j\in[n].
\]
For $t=0$ this follows from Lemma~\ref{lem:X0}; for later iterates it
follows from the assumed row bound and Assumption~\ref{assump:snr_network}.
Consequently,
\[
 \left|(\widehat{\mathbf x}_i^{(t)})^\top\widehat{\mathbf y}_j^{(t)}\right|
 \le K_c^2\rho_n,\qquad
 \sup_{|s|\le K_c^2\rho_n}|\psi'(s)|\le C_{c,\delta}\rho_n^{-1}.
\]
Orthogonality of $\mathbf Q$ and the identity
\begin{align*}
 (\widehat{\mathbf x}_i^{(t)})^\top\widehat{\mathbf y}_j^{(t)}
           -\mathbf x_{0i}^\top\mathbf y_{0j}
 =(\mathbf Q^\top\widehat{\mathbf x}_i^{(t)}-\mathbf x_{0i})^\top
                         \mathbf Q^\top\widehat{\mathbf y}_j^{(t)}
       +\mathbf x_{0i}^\top
                         (\mathbf Q^\top\widehat{\mathbf y}_j^{(t)}-\mathbf y_{0j})
\end{align*}
therefore imply
\begin{align*}
 &\frac1{n+p}\left\|\sum_j\left\{
 \psi'((\widehat{\mathbf x}_i^{(t)})^\top\widehat{\mathbf y}_j^{(t)})
 (\mathbf Q^\top\widehat{\mathbf y}_j^{(t)})
 (\mathbf Q^\top\widehat{\mathbf y}_j^{(t)})^\top
 -\psi'(\mathbf x_{0i}^\top\mathbf y_{0j})\mathbf y_{0j}\mathbf y_{0j}^\top
 \right\}\right\|_2\\
 &\quad\le\frac1{n+p}\sum_j
 \left|\psi'((\widehat{\mathbf x}_i^{(t)})^\top\widehat{\mathbf y}_j^{(t)})
       -\psi'(\mathbf x_{0i}^\top\mathbf y_{0j})\right|
 \|\widehat{\mathbf y}_j^{(t)}\|_2^2\\
 &\qquad+\frac1{n+p}\sum_j
 |\psi'(\mathbf x_{0i}^\top\mathbf y_{0j})|
 \|\mathbf Q^\top\widehat{\mathbf y}_j^{(t)}-\mathbf y_{0j}\|_2
 (\|\widehat{\mathbf y}_j^{(t)}\|_2+\|\mathbf y_{0j}\|_2)\\
 &\quad\le\frac{C_{c,\delta}}{(n+p)\rho_n^2}\sum_j
 \left\{\|\mathbf Q^\top\widehat{\mathbf x}_i^{(t)}-\mathbf x_{0i}\|_2
                           \|\widehat{\mathbf y}_j^{(t)}\|_2
 +\|\mathbf x_{0i}\|_2
                  \|\mathbf Q^\top\widehat{\mathbf y}_j^{(t)}-\mathbf y_{0j}\|_2
 \right\}\|\widehat{\mathbf y}_j^{(t)}\|_2^2\\
 &\qquad+\frac{C_{c,\delta}}{(n+p)\rho_n}\sum_j
 \|\mathbf Q^\top\widehat{\mathbf y}_j^{(t)}-\mathbf y_{0j}\|_2
 (\|\widehat{\mathbf y}_j^{(t)}\|_2+\|\mathbf y_{0j}\|_2)\\
 &\quad\le\frac{C_{c,\delta}n}{(n+p)\rho_n^2}
 \|\mathbf Q^\top\widehat{\mathbf x}_i^{(t)}-\mathbf x_{0i}\|_2
 \max_j\|\widehat{\mathbf y}_j^{(t)}\|_2^3\\
 &\qquad+\frac{C_{c,\delta}\|\mathbf x_{0i}\|_2}{(n+p)\rho_n^2}
 \left(\sum_j\|\widehat{\mathbf y}_j^{(t)}\|_2^4\right)^{1/2}
 \|\widehat{\mathbf X}^{(t)}\mathbf Q-\mathbf X_0\|_F\\
 &\qquad+\frac{C_{c,\delta}}{(n+p)\rho_n}
 \left\{\sum_j(\|\widehat{\mathbf y}_j^{(t)}\|_2+\|\mathbf y_{0j}\|_2)^2\right\}^{1/2}
 \|\widehat{\mathbf X}^{(t)}\mathbf Q-\mathbf X_0\|_F\\
 &\quad\le\frac{C_{c,\delta}}{\sqrt{\rho_n}}
 \left\{\|\mathbf Q^\top\widehat{\mathbf x}_i^{(t)}-\mathbf x_{0i}\|_2
 +\frac1{\sqrt n}\|\widehat{\mathbf X}^{(t)}\mathbf Q-\mathbf X_0\|_F\right\}.
\end{align*}
The first inequality is obtained by adding and subtracting
$\psi'(\mathbf x_{0i}^\top\mathbf y_{0j})
(\mathbf Q^\top\widehat{\mathbf y}_j^{(t)})
(\mathbf Q^\top\widehat{\mathbf y}_j^{(t)})^\top$
inside each summand and using the triangle inequality together with
$\|\mathbf u\mathbf u^\top-\mathbf v\mathbf v^\top\|_2
\le\|\mathbf u-\mathbf v\|_2(\|\mathbf u\|_2+\|\mathbf v\|_2)$.
For the last inequality we used localization, $n+p\ge n$,
Cauchy--Schwarz, and the signature-preserving alignment identity
\[
 \|\mathbf Q^\top\widehat{\mathbf y}_j^{(t)}-\mathbf y_{0j}\|_2
 =\|\mathbf Q^\top\widehat{\mathbf x}_j^{(t)}-\mathbf x_{0j}\|_2.
\]
When $t=0$, Lemma~\ref{lem:X0} gives
\[
 \|\mathbf Q^\top\widehat{\mathbf x}_i^{(0)}-\mathbf x_{0i}\|_2
 =\tOp\{\sqrt{\log m_n/n};m_n\},\qquad
 \|\widehat{\mathbf X}^{(0)}\mathbf Q-\mathbf X_0\|_F=\tOp(1).
\]
Hence the network contribution is
$\tOp\{\sqrt{\log m_n/(n\rho_n)};m_n\}$.
For later iterates, both terms in braces are at most
$\|\widehat{\mathbf X}^{(t)}\mathbf Q-\mathbf X_0\|_{2\to\infty}$;
the contribution is consequently $\tOp\{\sqrt{\log n/(n\rho_n)}\}$.

\paragraph{The covariate contribution.}
On the covariate localization event, the same outer-product inequality
and Cauchy--Schwarz give
\begin{align*}
 \frac1{n+p}\left\|\sum_l
 \{(\mathbf Q^\top\widehat{\mathbf z}_l^{(t)})
       (\mathbf Q^\top\widehat{\mathbf z}_l^{(t)})^\top
                 -\mathbf z_{0l}\mathbf z_{0l}^\top\}\right\|_2
 &\le\frac1{n+p}\sum_l
 \|\mathbf Q^\top\widehat{\mathbf z}_l^{(t)}-\mathbf z_{0l}\|_2
 (\|\widehat{\mathbf z}_l^{(t)}\|_2+\|\mathbf z_{0l}\|_2)\\
 &\le\frac1{n+p}
 \left\{\sum_l(\|\widehat{\mathbf z}_l^{(t)}\|_2+\|\mathbf z_{0l}\|_2)^2\right\}^{1/2}
 \|\widehat{\mathbf Z}^{(t)}\mathbf Q-\mathbf Z_0\|_F\\
 &\le\frac{C_c\sqrt p}{n+p}
 \|\widehat{\mathbf Z}^{(t)}\mathbf Q-\mathbf Z_0\|_F.
\end{align*}
At initialization, Lemma~\ref{lem:Z0} supplies localization and
$\|\widehat{\mathbf Z}^{(0)}\mathbf Q-\mathbf Z_0\|_F
=\tOp\{\sqrt{p+n}/\sqrt{n\rho_n}\}$.
Since $p=O(n)$, the contribution is $\tOp\{(n\rho_n)^{-1/2}\}$.
For later iterates, use
$\|\widehat{\mathbf Z}^{(t)}\mathbf Q-\mathbf Z_0\|_F
\le\sqrt p\|\widehat{\mathbf Z}^{(t)}\mathbf Q-\mathbf Z_0\|_{2\to\infty}$
to obtain $\tOp\{\sqrt{\log n/(n\rho_n)}\}$.
The localization used above follows from the stated row bounds and
the oracle row bounds: Assumption~\ref{assump:snr_network} gives
$\sqrt{\log n/n}\lesssim\sqrt{\rho_n}$ and
$\sqrt{\log n/(n\rho_n)}=O(1)$.
\paragraph{Combining the two contributions.}
The triangle inequality applied to the network and covariate parts
of the Fisher matrices gives, on the localization event,
\begin{align*}
 \|\widehat{\mathbf G}_{in}^{(t)}-\mathbf G_{in}\|_2
 &\le\frac{C_{c,\delta}}{\sqrt{\rho_n}}
 \left\{\|\mathbf Q^\top\widehat{\mathbf x}_i^{(t)}-\mathbf x_{0i}\|_2
 +\frac{\|\widehat{\mathbf X}^{(t)}\mathbf Q-\mathbf X_0\|_F}{\sqrt n}\right\}
 +\frac{C_c\sqrt p}{n+p}
 \|\widehat{\mathbf Z}^{(t)}\mathbf Q-\mathbf Z_0\|_F.
\end{align*}
For $t=0$, substitute the fixed-row and global bounds from
Lemmas~\ref{lem:X0} and~\ref{lem:Z0} separately:
\begin{align*}
 \|\widehat{\mathbf G}_{in}^{(0)}-\mathbf G_{in}\|_2
 &=\tOp\left\{
 \frac1{\sqrt{\rho_n}}\sqrt{\frac{\log m_n}{n}}
 +\frac1{\sqrt{n\rho_n}}
 +\frac{\sqrt p}{n+p}\frac{\sqrt{p+n}}{\sqrt{n\rho_n}};m_n\right\}
 =\tOp\left\{
 \sqrt{\frac{\log m_n}{n\rho_n}}
 +\frac1{\sqrt{n\rho_n}};m_n\right\}.
\end{align*}
Here $\sqrt p\sqrt{p+n}/(n+p)=\sqrt{p/(n+p)}\le1$.
The global bounds have $n$-based tails, which are also $m_n$-based
because $m_n=O(n)$. Since $m_n\to\infty$, eventually $\log m_n\ge1$;
hence the second rate is absorbed into the first, giving
\begin{equation}\label{eq:410-diff0}
 \|\widehat{\mathbf G}_{in}^{(0)}-\mathbf G_{in}\|_2
 =\tOp\left\{\sqrt{\frac{\log m_n}{n\rho_n}};m_n\right\}.
\end{equation}
Keeping the fixed-row error separate from the global error is what
preserves $\log m_n$ instead of replacing it by $\log n$.

For later iterates, the elementary row-to-Frobenius bounds are
\[
 \|\widehat{\mathbf X}^{(t)}\mathbf Q-\mathbf X_0\|_F
 \le\sqrt n\|\widehat{\mathbf X}^{(t)}\mathbf Q-\mathbf X_0\|_{2\to\infty},
 \qquad
 \|\widehat{\mathbf Z}^{(t)}\mathbf Q-\mathbf Z_0\|_F
 \le\sqrt p\|\widehat{\mathbf Z}^{(t)}\mathbf Q-\mathbf Z_0\|_{2\to\infty}.
\]
Also, the fixed-row network error is at most its maximum row error.
Thus the same deterministic inequality implies
\begin{align*}
 \|\widehat{\mathbf G}_{in}^{(t)}-\mathbf G_{in}\|_2
 &\le\frac{2C_{c,\delta}}{\sqrt{\rho_n}}
 \|\widehat{\mathbf X}^{(t)}\mathbf Q-\mathbf X_0\|_{2\to\infty}
 +\frac{C_cp}{n+p}
 \|\widehat{\mathbf Z}^{(t)}\mathbf Q-\mathbf Z_0\|_{2\to\infty}
 \\&
 =\tOp\left\{
 \frac1{\sqrt{\rho_n}}\sqrt{\frac{\log n}{n}}
 +\frac p{n+p}\sqrt{\frac{\log n}{n\rho_n}}\right\}.
\end{align*}
Using $p/(n+p)\le1$, we obtain
\begin{equation}\label{eq:410-difft}
 \|\widehat{\mathbf G}_{in}^{(t)}-\mathbf G_{in}\|_2
 =\tOp\left\{\sqrt{\frac{\log n}{n\rho_n}}\right\}.
\end{equation}

\paragraph{The inverse norms.}
The definition of the extended logarithmic surrogate gives
$\psi'(s)>0$ for all $s$ and, for every fixed $K$, constants
$0<c_{K,\delta}\le C_{K,\delta}<\infty$ such that
\[
 \frac{c_{K,\delta}}{\rho_n}\le\psi'(s)
 \le\frac{C_{K,\delta}}{\rho_n}
 \qquad (|s|\le K^2\rho_n).
\]
In particular, the oracle inner products lie in $[0,\rho_n]$,
so the lower bound on $\psi'$ and the oracle singular-value
condition in Assumption~\ref{assump:snr_network} imply
\[
 \mathbf G_{in}\succeq\frac{c_\delta}{(n+p)\rho_n}\mathbf Y_0^\top\mathbf Y_0,
 \qquad
 \lambda_d(\mathbf G_{in})
 \ge\frac{c_\delta\sigma_d(\mathbf X_0)^2}{(n+p)\rho_n}
 \gtrsim1.
\]
Thus $\|\mathbf G_{in}^{-1}\|_2\lesssim1$.
At initialization, localization and positivity similarly imply
\[
 \widehat{\mathbf G}_{in}^{(0)}
 \succeq\frac{c_{c,\delta}}{(n+p)\rho_n}
       \mathbf Q^\top(\widehat{\mathbf Y}^{(0)})^\top
                      \widehat{\mathbf Y}^{(0)}\mathbf Q.
\]
The initialization singular-value bound in Lemma~\ref{lem:X0} yields
$\sigma_d(\widehat{\mathbf Y}^{(0)})
=\sigma_d(\widehat{\mathbf X}^{(0)})\gtrsim_c\sqrt{n\rho_n}$.
Consequently $\|(\widehat{\mathbf G}_{in}^{(0)})^{-1}\|_2=\tOp(1)$.
This argument does not require the bound in \eqref{eq:410-diff0} to
converge to zero under Assumptions~\ref{assump:snr_network}
and~\ref{assump:snr_covariates}.

For later iterates, Assumption~\ref{assump:3} gives
$\sqrt{\log n/(n\rho_n)}\to0$. On the event in
\eqref{eq:410-difft}, Weyl's inequality therefore implies, for all
sufficiently large $n$,
\[
 \lambda_d(\widehat{\mathbf G}_{in}^{(t)})
 \ge\lambda_d(\mathbf G_{in})
      -\|\widehat{\mathbf G}_{in}^{(t)}-\mathbf G_{in}\|_2
 \ge\tfrac12\lambda_d(\mathbf G_{in})\gtrsim1.
\]
Hence $\|(\widehat{\mathbf G}_{in}^{(t)})^{-1}\|_2=\tOp(1)$.

\paragraph{The inverse difference.}
Finally, the exact identity
\[
 \mathbf R_{i2}^{(t)}
 =(\widehat{\mathbf G}_{in}^{(t)})^{-1}
   (\mathbf G_{in}-\widehat{\mathbf G}_{in}^{(t)})\mathbf G_{in}^{-1}
\]
implies
\[
 \|\mathbf R_{i2}^{(t)}\|_2
 \le\|(\widehat{\mathbf G}_{in}^{(t)})^{-1}\|_2
       \|\widehat{\mathbf G}_{in}^{(t)}-\mathbf G_{in}\|_2
       \|\mathbf G_{in}^{-1}\|_2.
\]
Substituting \eqref{eq:410-diff0} or \eqref{eq:410-difft} and the
inverse bounds proves the respective assertions. Finite intersections
of the input events preserve the stated polynomial tail bounds.
\end{proof}

\subsection{Assembly of the Score Remainder Bounds} 
\begin{proof}[Proof of Lemma~\ref{lem:score-remainder}]
Fix $i\in[n]$. Retain the notation of the preceding analysis and the
Taylor decomposition \eqref{eq:term1}--\eqref{eq:term8}, with the
iteration index $t$ restored in every estimated coordinate.
The term in \eqref{eq:term2} is exactly
\[
 -\mathbf G_{in}(\mathbf Q^\top\widehat{\mathbf x}_i^{(t)}-\mathbf x_{0i}),
 \qquad
 \mathbf G_{in}=\frac1{n+p}\left\{
 \sum_{j=1}^n\psi'(\mathbf x_{0i}^\top\mathbf y_{0j})
 \mathbf y_{0j}\mathbf y_{0j}^\top
 +\sum_{l=1}^p\mathbf z_{0l}\mathbf z_{0l}^\top\right\}.
\]
Consequently, adding $\mathbf G_{in}
(\mathbf Q^\top\widehat{\mathbf x}_i^{(t)}-\mathbf x_{0i})$
to $\mathbf r_{i1}^{(t)}$ cancels \eqref{eq:term2}. The resulting
vector is the sum of the other seven terms. No independence between
these terms is needed: we use their individual bounds and the triangle
inequality.

\paragraph{Initialization.}
Let $m_n\to\infty$ be deterministic with $m_n=O(n)$.
Under Assumptions~\ref{assump:snr_network} and~\ref{assump:snr_covariates},
the initialization bounds in Lemmas~\ref{lem:term1},
\ref{lem:term3}, \ref{lem:term4}, \ref{lem:term5},
\ref{lem:term6}, and~\ref{lem:term78} give the following bounds
for the Euclidean norms of the corresponding terms:
\[
\begin{array}{c|c}
\text{Terms} & \text{Bound, with tail parameter }m_n\\ \hline
\eqref{eq:term1},\ \eqref{eq:term4},\ \eqref{eq:term5},\ \eqref{eq:term6}
 &\displaystyle\tOp\left\{\frac{\log m_n}{n\sqrt{\rho_n}};m_n\right\}\\[6pt]
\eqref{eq:term3}
 &\displaystyle\tOp\left\{\frac{\sqrt{\log m_n}
 +\sqrt{1\vee\rho_n\log n}}{n\sqrt{\rho_n}};m_n\right\}\\[6pt]
\eqref{eq:term7},\ \eqref{eq:term8}
 &\displaystyle\tOp\left\{\frac{\sqrt{\log m_n}+\sqrt{\log n}}{n\sqrt{\rho_n}};m_n\right\}.
\end{array}
\]
Since $\log m_n\to\infty$, we have
$\sqrt{\log m_n}\le\log m_n$ for sufficiently large $n$.
Also $\sqrt{1\vee\rho_n\log n}\lesssim\sqrt{\log n}$
for sufficiently large $n$, since $\rho_n\le1$. We retain the additional
$\sqrt{\log n}$ from the global initial covariate remainder.
The exact cancellation above and the triangle inequality therefore yield
\begin{equation}\label{eq:41-cancel-init}
 \left\|\mathbf r_{i1}^{(0)}+
 \mathbf G_{in}(\mathbf Q^\top\widehat{\mathbf x}_i^{(0)}-\mathbf x_{0i})\right\|_2
 =\tOp\left\{\frac{\log m_n+\sqrt{\log n}}
 {n\sqrt{\rho_n}};m_n\right\}.
\end{equation}
For clarity, the probability statement follows by fixing $c>0$ and
intersecting the finitely many events used in these termwise bounds.
Their failure probabilities sum to $O(m_n^{-c})$; on their intersection,
the norm of the sum is bounded by the sum of their norm bounds.

Lemma~\ref{lem:term2} gives
\[
 \|\mathbf G_{in}(\mathbf Q^\top\widehat{\mathbf x}_i^{(0)}-\mathbf x_{0i})\|_2
 =\tOp\left\{\sqrt{\frac{\log m_n}{n}};m_n\right\}.
\]
Subtracting this vector from the vector in \eqref{eq:41-cancel-init}
and applying the triangle inequality proves
\[
 \|\mathbf r_{i1}^{(0)}\|_2
 =\tOp\left\{\sqrt{\frac{\log m_n}{n}}
 +\frac{\log m_n+\sqrt{\log n}}
 {n\sqrt{\rho_n}};m_n\right\}.
\]
These are the two initialization assertions.

\paragraph{Later iterates.}
Now assume Assumption~\ref{assump:3} and the approximate linearization
and covariate remainder hypotheses in the statement. The uniform
linearization rate established in \Cref{res:linearization-rate} gives
\[
 \|\widehat{\mathbf X}^{(t)}\mathbf Q-\mathbf X_0\|_{2\to\infty}
 =\tOp\left\{\sqrt{\frac{\log n}{n}}\right\}.
\]
The covariate expansion in the statement, together with Gaussian
concentration and a union bound over $p=O(n)$ rows, gives
\[
 \|\widehat{\mathbf Z}^{(t)}\mathbf Q-\mathbf Z_0\|_{2\to\infty}
 =\tOp\left\{\sqrt{\frac{\log n}{n\rho_n}}\right\}.
\]
Indeed, each row of $\mathbf W\mathbf X_0
(\mathbf X_0^\top\mathbf X_0)^{-1}$ is centered Gaussian with covariance
$(\mathbf X_0^\top\mathbf X_0)^{-1}$, whose norm is $O((n\rho_n)^{-1})$.
The remainder is smaller than the displayed rate: choose a positive
exponent $\eta$ for which $3/2+\eta<2\xi$ and use
\[
 \frac{(\log n)^{2+\eta}}{n\rho_n}
 \bigg/\sqrt{\frac{\log n}{n\rho_n}}
 =\frac{(\log n)^{3/2+\eta}}{\sqrt{n\rho_n}}
 \longrightarrow0.
\]
These rowwise rates supply the hypotheses of the later-iterate
termwise bounds.

Fix $\epsilon>0$. Lemmas~\ref{lem:term1}, \ref{lem:term4},
and~\ref{lem:term6} bound their respective terms by
$\tOp\{\log n/(n\sqrt{\rho_n})\}$.
Lemmas~\ref{lem:term3}, \ref{lem:term5}, and~\ref{lem:term78}
bound the remaining terms, excluding \eqref{eq:term2}, by
$\tOp\{(\log n)^{2+\epsilon}/(n\sqrt{\rho_n})\}$.
Canceling \eqref{eq:term2} and intersecting these finitely many events
as above proves
\begin{equation}\label{eq:41-cancel-iterate}
 \left\|\mathbf r_{i1}^{(t)}+
 \mathbf G_{in}(\mathbf Q^\top\widehat{\mathbf x}_i^{(t)}-\mathbf x_{0i})\right\|_2
 =\tOp\left\{\frac{(\log n)^{2+\epsilon}}{n\sqrt{\rho_n}}\right\}.
\end{equation}
Finally, Lemma~\ref{lem:term2} and the network rowwise rate imply
\[
 \|\mathbf G_{in}(\mathbf Q^\top\widehat{\mathbf x}_i^{(t)}-\mathbf x_{0i})\|_2
 =\tOp\left\{\sqrt{\frac{\log n}{n}}\right\}.
\]
To obtain the asserted rate for $\mathbf r_{i1}^{(t)}$, apply
\eqref{eq:41-cancel-iterate} with any $\eta>0$ satisfying
$3/2+\eta<2\xi$. Assumption~\ref{assump:3} then gives
\[
 \frac{(\log n)^{2+\eta}}{n\sqrt{\rho_n}}
 \bigg/\sqrt{\frac{\log n}{n}}
 =\frac{(\log n)^{3/2+\eta}}{\sqrt{n\rho_n}}
 \longrightarrow0.
\]
The triangle inequality thus yields
$\|\mathbf r_{i1}^{(t)}\|_2
=\tOp\{\sqrt{\log n/n}\}$, completing the proof.
\end{proof}

\section{One-Step Stochastic Expansion}\label{sec:onestep}


\subsection{Proof of Theorem \ref{thm:OnestepFisher}}
\label{sub:proof_of_main_thm_onestepfisher}

\begin{proof}[Proof of Stochastic Expansion]
Fix $i\in[n]$ and a deterministic sequence $m_n\to\infty$ with
$m_n=O(n)$. All probability bounds below use tail parameter $m_n$,
unless stated otherwise.

\paragraph{The oracle score.}
At the oracle coordinates, the normalized score is
\begin{equation}\label{eq:21-population-score}
 \frac1{n+p}\frac{\partial\ell_{in}}{\partial\mathbf x_i}
 (\mathbf x_{0i};\mathbf Y_0,\mathbf Z_0)
 =\frac1{n+p}\left\{
 \sum_{j=1}^nE_{ij}\psi'(\mathbf x_{0i}^\top\mathbf y_{0j})\mathbf y_{0j}
 +\sum_{l=1}^pW_{li}\mathbf z_{0l}\right\}.
\end{equation}
We first establish its concentration rate. For the fixed row $i$,
the off-diagonal variables $E_{ij}$, $j\ne i$, are independent and centered, with
$|E_{ij}|\le1$ and $\mathbb E E_{ij}^2\le\rho_n$.
Moreover, Assumption~\ref{assump:snr_network} and $\tau=\delta\rho_n$
imply
\[
 \|\mathbf y_{0j}\|_2\le\sqrt{\rho_n},\qquad
 |\psi'(\mathbf x_{0i}^\top\mathbf y_{0j})|\le C_\delta\rho_n^{-1}.
\]
We apply the first inequality of \citep[Lemma~S2.1]{xie2024entrywise} to the off-diagonal sum, setting the $i$th random variable to zero and the others to $E_{ij}$, its variance parameters set to $\sigma^2=1$ and
$\rho=\rho_n$, and its deterministic coefficient matrix defined by
\[
 \mathbf V=\frac1{n+p}\begin{bmatrix}
 \psi'(\mathbf x_{0i}^\top\mathbf y_{01})\mathbf y_{01}^\top\\
 \vdots\\
 \psi'(\mathbf x_{0i}^\top\mathbf y_{0n})\mathbf y_{0n}^\top
 \end{bmatrix}\in\mathbb R^{n\times d}.
\]
The preceding bounds imply
\begin{align*}
 \|\mathbf V\|_{2\to\infty}
 &=\frac1{n+p}\max_j
 |\psi'(\mathbf x_{0i}^\top\mathbf y_{0j})|\|\mathbf y_{0j}\|_2
 \le\frac{C_\delta}{n\sqrt{\rho_n}},\\
 \|\mathbf V\|_F^2
 &=\frac1{(n+p)^2}\sum_{j=1}^n
 \{\psi'(\mathbf x_{0i}^\top\mathbf y_{0j})\}^2\|\mathbf y_{0j}\|_2^2
 \le\frac{C_\delta n}{(n+p)^2\rho_n}
 \le\frac{C_\delta}{n\rho_n}.
\end{align*}
Since $\mathbb E E_{ij}=0$ for $j\ne i$, \citep[Lemma~S2.1]{xie2024entrywise} gives, for $s\ge1$,
\[
 \mathbb P\left(
 \left\|\frac1{n+p}\sum_{j\ne i} E_{ij}
 \psi'(\mathbf x_{0i}^\top\mathbf y_{0j})\mathbf y_{0j}\right\|_2
 >3s^2\|\mathbf V\|_{2\to\infty}
   +\sqrt6\sqrt{\rho_n}s\|\mathbf V\|_F
 \right)\le28e^{-s^2}.
\]
Here $s$ is the tail parameter of the concentration inequality,
separate from the Fisher-scoring iteration index. Taking $s=\sqrt u$
and substituting the two coefficient norm bounds yields, for $u\ge1$,
\[
 3u\|\mathbf V\|_{2\to\infty}
 +\sqrt6\sqrt{\rho_n u}\|\mathbf V\|_F
 \le C_\delta\left\{\frac{u}{n\sqrt{\rho_n}}+\sqrt{\frac un}\right\}.
\]
Consequently,
\[
 \mathbb P\left(
 \left\|\frac1{n+p}\sum_{j\ne i}E_{ij}
 \psi'(\mathbf x_{0i}^\top\mathbf y_{0j})\mathbf y_{0j}\right\|_2
 >C_\delta\left\{\sqrt{\frac un}+\frac{u}{n\sqrt{\rho_n}}\right\}
 \right)\le28e^{-u}.
\]
The full sum also contains the deterministic vector
$-P_{ii}\psi'(P_{ii})\mathbf y_{0i}/(n+p)$, whose norm is at most
$C_\delta\sqrt{\rho_n}/n$. For $u\ge1$ this is bounded by
$C_\delta\sqrt{u/n}$, so enlarging the constant gives the same
concentration threshold for the full sum in \eqref{eq:21-population-score}.
The covariate score is a centered Gaussian vector with covariance
$(n+p)^{-2}\mathbf Z_0^\top\mathbf Z_0$. By
Assumption~\ref{assump:snr_covariates}, the norm of this covariance
is at most $Cp/(n+p)^2\le C/n$. Gaussian concentration therefore gives
\[
 \mathbb P\left(
 \left\|\frac1{n+p}\sum_{l=1}^pW_{li}\mathbf z_{0l}\right\|_2
 >C\sqrt{\frac un}\right)\le2d e^{-u},\qquad u\ge1,
\]
where constants may depend on the fixed dimension $d$.
Taking $u=(c+2)\log m_n$ for arbitrary $c>0$ and combining the two
bounds proves
\[
 \left\|\frac1{n+p}\frac{\partial\ell_{in}}{\partial\mathbf x_i}
 (\mathbf x_{0i};\mathbf Y_0,\mathbf Z_0)\right\|_2
 =\tOp\left\{\sqrt{\frac{\log m_n}{n}}
 +\frac{\log m_n}{n\sqrt{\rho_n}};m_n\right\}.
\]
Since $m_n=O(n)$ and $n\rho_n\gtrsim\log n$,
\begin{equation}\label{eq:21-bounded-factor}
 \frac{\log m_n}{n\rho_n}\lesssim1.
\end{equation}
Consequently the second score rate is absorbed into the first:
\begin{equation}\label{eq:21-score-rate}
 \left\|\frac1{n+p}\frac{\partial\ell_{in}}{\partial\mathbf x_i}
 (\mathbf x_{0i};\mathbf Y_0,\mathbf Z_0)\right\|_2
 =\tOp\left\{\sqrt{\frac{\log m_n}{n}};m_n\right\}.
\end{equation}

\paragraph{The three remainder contributions.}
Set $t=0$ in the one-step decomposition \eqref{lem:taylor}.
Using the definition of $\mathbf r_{i3}^{(0)}$, the remainder is exactly
\begin{equation}\label{eq:21-remainder}
\begin{aligned}
 \mathbf r_{\mathbf x_i}^{(1)}
 &=\mathbf G_{in}^{-1}\left\{
 \mathbf G_{in}(\mathbf Q^\top\widehat{\mathbf x}_i^{(0)}-\mathbf x_{0i})
 +\mathbf r_{i1}^{(0)}\right\}\\
 &\quad+\mathbf R_{i2}^{(0)}\mathbf r_{i1}^{(0)}
 +\mathbf R_{i2}^{(0)}\frac1{n+p}
 \frac{\partial\ell_{in}}{\partial\mathbf x_i}
 (\mathbf x_{0i};\mathbf Y_0,\mathbf Z_0).
\end{aligned}
\end{equation}
We bound these three contributions separately.

First, Lemma~\ref{lem:score-remainder} gives
\begin{align*}
 \left\|\mathbf G_{in}(\mathbf Q^\top\widehat{\mathbf x}_i^{(0)}-\mathbf x_{0i})
 +\mathbf r_{i1}^{(0)}\right\|_2
 &=\tOp\left\{\frac{\log m_n+\sqrt{\log n}}
 {n\sqrt{\rho_n}};m_n\right\},\\
 \|\mathbf r_{i1}^{(0)}\|_2
 &=\tOp\left\{\sqrt{\frac{\log m_n}{n}}
 +\frac{\log m_n+\sqrt{\log n}}
 {n\sqrt{\rho_n}};m_n\right\}.
\end{align*}
The sharper first bound uses the exact cancellation of the term
$-\mathbf G_{in}(\mathbf Q^\top\widehat{\mathbf x}_i^{(0)}-\mathbf x_{0i})$
in the eight-term score decomposition. The second bound retains this
term and hence includes the initialization rate.
The oracle inverse bound in the proof of Lemma~\ref{lem:Ri2}
is $\|\mathbf G_{in}^{-1}\|_2\lesssim1$. Therefore
\begin{equation}\label{eq:21-first-contribution}
 \left\|\mathbf G_{in}^{-1}\left\{
 \mathbf G_{in}(\mathbf Q^\top\widehat{\mathbf x}_i^{(0)}-\mathbf x_{0i})
 +\mathbf r_{i1}^{(0)}\right\}\right\|_2
 =\tOp\left\{\frac{\log m_n+\sqrt{\log n}}
 {n\sqrt{\rho_n}};m_n\right\}.
\end{equation}

Second, Lemma~\ref{lem:Ri2} supplies
\[
 \|\mathbf R_{i2}^{(0)}\|_2
 =\tOp\left\{\sqrt{\frac{\log m_n}{n\rho_n}};m_n\right\}.
\]
Multiplying this bound by the second bound for $\mathbf r_{i1}^{(0)}$
yields
\begin{align*}
 \|\mathbf R_{i2}^{(0)}\mathbf r_{i1}^{(0)}\|_2
 &=\tOp\left\{
 \frac{\log m_n}{n\sqrt{\rho_n}}
 +\sqrt{\frac{\log m_n}{n\rho_n}}
 \frac{\log m_n+\sqrt{\log n}}
 {n\sqrt{\rho_n}};m_n\right\}\\
 &=\tOp\left\{\frac{\log m_n+\sqrt{\log n}}
 {n\sqrt{\rho_n}};m_n\right\}.
\end{align*}
For the last equality, the square-root factor is bounded by
\eqref{eq:21-bounded-factor}, and the first numerator $\log m_n$
is at most the asserted numerator. This absorption needs only
$n\rho_n\gtrsim\log n$; it does not require the factor to converge to zero.

Third, combining \eqref{eq:21-score-rate} with the same inverse-difference
bound gives
\[
 \left\|\mathbf R_{i2}^{(0)}\frac1{n+p}
 \frac{\partial\ell_{in}}{\partial\mathbf x_i}
 (\mathbf x_{0i};\mathbf Y_0,\mathbf Z_0)\right\|_2
 =\tOp\left\{\frac{\log m_n}{n\sqrt{\rho_n}};m_n\right\}.
\]
Each product bound is obtained on the intersection of its input events;
independence between the factors is unnecessary. The finitely many
failure probabilities sum to $O(m_n^{-c})$ for any prescribed $c>0$.
The triangle inequality in \eqref{eq:21-remainder} now proves the stated
rowwise remainder rate and hence the stochastic expansion.

\paragraph{The maximum row error.}
Take $m_n=n$. The constants in the preceding bounds can be chosen
uniformly over $i\in[n]$, because the oracle bounds and the
initialization estimates used above are uniform in the row index.
For any prescribed $c>0$, apply the fixed-row estimates with failure
probability $O(n^{-(c+1)})$ and take a union bound over the $n$ rows.
Using $\sup_i\|\mathbf G_{in}^{-1}\|_2\lesssim1$, we obtain
\begin{align*}
 \|\widehat{\mathbf X}^{(1)}\mathbf Q-\mathbf X_0\|_{2\to\infty}
 =\tOp\left\{\sqrt{\frac{\log n}{n}}
 +\frac{\log n+\sqrt{\log n}}{n\sqrt{\rho_n}}\right\}.
\end{align*}
For sufficiently large $n$, $\rho_n\le1$ implies
$\sqrt{\log n}\le\log n$, while
\[
 \frac{\log n}{n\sqrt{\rho_n}}
 =\sqrt{\frac{\log n}{n}}
   \sqrt{\frac{\log n}{n\rho_n}}
 \lesssim\sqrt{\frac{\log n}{n}}.
\]
Thus the remainder is absorbed into the leading score rate, proving
the asserted maximum row error bound.
\end{proof}

\begin{proof}[Proof of the asymptotic normality conclusion]
Fix $i$ and abbreviate $N=n+p$. By the expansion in the theorem, it is
enough to show that $\sqrt N\,\mathbf G_{in}^{-1/2}\mathbf s_{in}$ converges
to $\mathrm N(\mathbf0,\mathbf I_d)$ and that
$\sqrt N\,\|\mathbf G_{in}^{1/2}\mathbf r_{\mathbf x_i}^{(1)}\|_2=o_p(1)$.
The latter follows from $\|\mathbf G_{in}^{1/2}\|_2=O(1)$ (the upper
eigenvalue bound follows from the bounded covariate rows and
$\psi'(P_{ij})\,\|\mathbf y_{0j}\|_2^2=O(1)$) and the theorem's remainder
bound, taking $m_n=\log n$. Indeed,
\[
 \sqrt N\,\|\mathbf r_{\mathbf x_i}^{(1)}\|_2
 =\widetilde O_p\!\left(\frac{\log\log n+\sqrt{\log n}}
 {\sqrt{n\rho_n}}\right)=o_p(1)
\]
when $n\rho_n/\log n\to\infty$.

For the score, write $\mathbf U_{in}=\sqrt N\,\mathbf G_{in}^{-1/2}\mathbf s_{in}$.
The off-diagonal graph terms are independent centered vectors
\[
 \mathbf X_{nj}=N^{-1/2}\mathbf G_{in}^{-1/2}
 \psi'(P_{ij})\mathbf y_{0j}E_{ij},\qquad j\ne i,
\]
and are independent of the Gaussian covariate term. The proof of
\eqref{eq:fs-inversebound} gives $\|\mathbf G_{in}^{-1}\|_2=O(1)$.
Since $P_{ij}\in[\tau,1-\tau]$ and $\tau=\delta\rho_n$,
$\psi'(P_{ij})=\{P_{ij}(1-P_{ij})\}^{-1}$, so
\[
 \max_{j\ne i}\|\mathbf X_{nj}\|_2
 \le \frac{C}{\sqrt{n\rho_n}}\longrightarrow0.
\]
Their total covariance, added to that of the Gaussian covariate term,
is $\mathbf I_d-\mathbf D_{in}$, where
\[
 \mathbf D_{in}=N^{-1}\mathbf G_{in}^{-1/2}
 \psi'(P_{ii})\mathbf y_{0i}\mathbf y_{0i}^{\top}
 \mathbf G_{in}^{-1/2},\qquad \|\mathbf D_{in}\|_2=O(N^{-1}).
\]
Here the identity uses
$\operatorname{Var}(E_{ij})=P_{ij}(1-P_{ij})$ and the definition of
$\mathbf G_{in}$. To verify Lindeberg, fix $\mathbf a\in\mathbb R^d$
and $\varepsilon>0$. The scalar graph summands in direction $\mathbf a$
are $\mathbf a^\top\mathbf X_{nj}$, and the condition to check is
\[
 \sum_{j\ne i}\mathbb E\!\left[(\mathbf a^\top\mathbf X_{nj})^2
 \mathbf 1\{|\mathbf a^\top\mathbf X_{nj}|>\varepsilon\}\right]
 \longrightarrow0.
\]
The deterministic maximum bound gives
$|\mathbf a^\top\mathbf X_{nj}|\le C\|\mathbf a\|_2/\sqrt{n\rho_n}$
for every $j\ne i$. Since $n\rho_n\to\infty$, this is at most
$\varepsilon$ for all sufficiently large $n$, so every indicator in the
sum is zero. This proves Lindeberg for the graph terms; the covariate
term is already Gaussian and is handled separately. We now combine them
in a fixed direction $\mathbf a\in\mathbb R^d$. Let
\[
 T_{n,\mathrm{graph}}=\sum_{j\ne i}\mathbf a^\top\mathbf X_{nj},
 \qquad
 T_{n,\mathrm{cov}}=\frac1{\sqrt N}\mathbf a^\top\mathbf G_{in}^{-1/2}
 \sum_{l=1}^pW_{li}\mathbf z_{0l},
\]
and write their variances as $v_{n,\mathrm{graph}}$ and
$v_{n,\mathrm{cov}}$, respectively. Both variances are nonnegative and
bounded, and their sum is
\[
 v_{n,\mathrm{graph}}+v_{n,\mathrm{cov}}
 =\mathbf a^\top(\mathbf I_d-\mathbf D_{in})\mathbf a
 \longrightarrow\|\mathbf a\|_2^2.
\]
The two variance sequences need not converge separately. Start with any
subsequence of sample sizes. Since both sequences are bounded, it has a
further subsequence along which
$v_{n,\mathrm{graph}}\to v_{\mathrm{graph}}$ and
$v_{n,\mathrm{cov}}\to v_{\mathrm{cov}}$ for some nonnegative limits.
The displayed identity forces
$v_{\mathrm{graph}}+v_{\mathrm{cov}}=\|\mathbf a\|_2^2$.

Along this further subsequence, the scalar graph summands are independent,
satisfy the Lindeberg condition proved above, and have total variance
converging to $v_{\mathrm{graph}}$. The Lindeberg--Feller theorem therefore
gives
$T_{n,\mathrm{graph}}\Rightarrow\mathrm N(0,v_{\mathrm{graph}})$.
Meanwhile $T_{n,\mathrm{cov}}$ is exactly Gaussian with variance
$v_{n,\mathrm{cov}}$, so it converges to
$\mathrm N(0,v_{\mathrm{cov}})$, including the degenerate case
$v_{\mathrm{cov}}=0$. The graph and covariate terms are independent.
Consequently, the characteristic function of their sum is the product of
their characteristic functions, and the sum converges to
\[
 T_{n,\mathrm{graph}}+T_{n,\mathrm{cov}}
 \Rightarrow
 \mathrm N(0,v_{\mathrm{graph}}+v_{\mathrm{cov}})
 =\mathrm N(0,\|\mathbf a\|_2^2).
\]
We began with an arbitrary subsequence and found a further subsequence with
this same limit. Hence the full scalar sequence converges to
$\mathrm N(0,\|\mathbf a\|_2^2)$: if it did not, some subsequence would
stay away from that limit and could not have a further subsequence
converging to it. Since this holds for every fixed $\mathbf a$, the
Cram\'er--Wold theorem proves the asserted vector limit.

Finally, because the hollow diagonal has $E_{ii}=-P_{ii}$ rather than mean
zero, $\mathbf U_{in}$ has a deterministic mean contribution from that
term. Its norm is at most $C\sqrt{\rho_n/N}=o(1)$, so it does not change
the limit. Slutsky's theorem and the negligible remainder prove the claim.
\end{proof}

\paragraph{First-iterate remainder bound.}
The fixed-row one-step remainder bound is
\[
 \|\br_{\bx_i}^{(1)}\|_2
 =\widetilde O_p\left\{
 \frac{\log m_n+\sqrt{\log n}}{n\sqrt{\rho_n}};m_n\right\}.
\]
For starting the induction, take $m_n=n$. Since
$\sqrt{\log n}\le\log n$ eventually, the constants are uniform over
rows, and a union bound with arbitrarily high polynomial probability gives
\[
 \max_{i\in[n]}\|\br_{\bx_i}^{(1)}\|_2
 =\widetilde O_p\left\{\frac{\log n}{n\sqrt{\rho_n}}\right\}.
\]
This satisfies the remainder requirement in
Definition~\ref{def:linearization} and provides the base case
for the induction.

\subsection{Linearization of the first Fisher-scoring iterate}\label{sec:first-step}

\Cref{res:linearization-rate} takes \Cref{def:linearization} as a
hypothesis. We now verify that hypothesis at $t=1$, i.e.\ that
$\widehat\bX^{(1)}$ genuinely admits an approximate linearization with
deterministic coefficients of the required size and a remainder of the
required order. The argument uses only the estimates already
established in \Cref{sec:taylor} --- the score-difference bounds
of \Cref{lem:term1,lem:term2,lem:term3,lem:term4,lem:term5,lem:term6,lem:term78}
and the inverse-information perturbation bound of \Cref{lem:Ri2} ---
rather than assuming the expansion or its remainder bound. The
concentration bound for the score at the truth is proved explicitly in
\Cref{lem:popscore} below. Throughout, all logarithms are written as
$\log n$; the tail parameter of \Cref{sec:notation} is taken to be
$m_n=n$.

\paragraph{Oracle score and information.} Recall from
\Cref{sec:notation} that $\bY_0=\bX_0\bI_{d_+,d_-}$, so that
$P_{ij}=\bx_{0i}^\top\by_{0j}$, and that the normalized oracle
information matrix is
\begin{equation}\label{eq:fs-populationG}
\bG_{in}
=\frac1{n+p}\left\{
\sum_{j=1}^n\psi'(\bx_{0i}^\top\by_{0j})\,\by_{0j}\by_{0j}^\top
+\sum_{l=1}^p\bz_{0l}\bz_{0l}^\top
\right\}.
\end{equation}
Recall the normalized oracle score $\bs_{in}$ defined in
Section~\ref{sec:notation}:
\begin{equation}\label{eq:fs-populations}
\bs_{in}
\coloneqq\frac1{n+p}\frac{\partial\ell_{in}}{\partial\bx_i}(\bx_{0i};\bY_0,\bZ_0)
=\frac1{n+p}\left\{
\sum_{j=1}^nE_{ij}\,\psi'(\bx_{0i}^\top\by_{0j})\,\by_{0j}
+\sum_{l=1}^pW_{li}\,\bz_{0l}
\right\},
\end{equation}
which is exactly the leading term appearing in \Cref{thm:OnestepFisher} and
\Cref{thm:efficiency}.

\begin{lemma}[Concentration of the score at the truth]\label{lem:popscore}
Under \Cref{assump:snr_network,assump:snr_covariates,assump:3}, for each
$i\in[n]$,
\[
\|\bs_{in}\|_2=\widetilde O_p\!\left(\sqrt{\frac{\log n}{n}}\right).
\]
\end{lemma}

\begin{proof}
With $\tau=\delta\rho_n$, the piecewise definitions of $\psi_1,\psi_2$
give, exactly as in the proof of \Cref{lem:Taylor_bounds},
\begin{equation}\label{eq:fs-psiupper}
0<\psi'(s)\le C_\delta\rho_n^{-1},\qquad s\in[0,1].
\end{equation}
Since $\bI_{d_+,d_-}$ is orthogonal,
$\|\by_{0j}\|_2=\|\bx_{0j}\|_2\le\rho_n^{1/2}$ by
\Cref{assump:snr_network}. Hence the deterministic graph-score vectors
\[
\bv_{nij}\coloneqq\frac{\psi'(\bx_{0i}^\top\by_{0j})\,\by_{0j}}{n+p}
\]
satisfy
\begin{equation}\label{eq:fs-coefbound}
\max_{i,j\in[n]}\|\bv_{nij}\|_2\le\frac{C_\delta}{(n+p)\rho_n^{1/2}}
\lesssim\frac1{n\rho_n^{1/2}}.
\end{equation}
Moreover $0\le P_{ij}\le\|\bx_{0i}\|_2\|\by_{0j}\|_2\le\rho_n$, so
$\Var(E_{ij})=P_{ij}(1-P_{ij})\le\rho_n$ for $j\ne i$. For fixed $i$ the off-diagonal variables
$(E_{ij})_{j\ne i}$ are independent, centered, and bounded by $1$, and
\[
\sum_{j=1}^n\|\bv_{nij}\|_2^2\,\Var(E_{ij})
\lesssim n\rho_n\cdot\frac1{n^2\rho_n}=\frac1n .
\]
Bernstein's inequality, applied coordinatewise with $d$ fixed, therefore
yields constants $C,C'>0$ such that, for every $t\ge1$,
\[
\prob\left[\Bigl\|\sum_{j\ne i}\bv_{nij}E_{ij}\Bigr\|_2
>C\left\{\sqrt{\frac tn}+\frac{t}{n\rho_n^{1/2}}\right\}\right]\le C'e^{-t}.
\]
The diagonal score vector $-P_{ii}\bv_{nii}$ has norm at most
$C_\delta\sqrt{\rho_n}/n$ and is absorbed into the same threshold
for $t\ge1$. Thus the full sum has the following rate.
Taking $t$ to be a sufficiently large multiple of $\log n$ gives
\begin{equation}\label{eq:fs-graphscore}
\Bigl\|\frac1{n+p}\sum_{j=1}^nE_{ij}\,\psi'(\bx_{0i}^\top\by_{0j})\,\by_{0j}\Bigr\|_2
=\widetilde O_p\!\left(\sqrt{\frac{\log n}{n}}+\frac{\log n}{n\rho_n^{1/2}}\right).
\end{equation}
The covariate contribution is, for fixed $i$, a centered Gaussian vector
with covariance matrix $(n+p)^{-2}\sum_{l=1}^p\bz_{0l}\bz_{0l}^\top$,
whose spectral norm is at most $p\,\|\bZ_0\|_{2\to\infty}^2/(n+p)^2=O(n^{-1})$
by \Cref{assump:snr_covariates}. Gaussian concentration therefore gives
\begin{equation}\label{eq:fs-covscore}
\Bigl\|\frac1{n+p}\sum_{l=1}^pW_{li}\,\bz_{0l}\Bigr\|_2
=\widetilde O_p\!\left(\sqrt{\frac{\log n}{n}}\right).
\end{equation}
Finally, by \eqref{eq:res-absorb} the second term of
\eqref{eq:fs-graphscore} is dominated by the first under
\Cref{assump:3}. Combining \eqref{eq:fs-graphscore} and
\eqref{eq:fs-covscore} completes the proof.
\end{proof}

\begin{remark}[The first-iterate linearization requirement]\label{rem:fs-regime}
To verify Definition~\ref{def:linearization} for the first iterate,
we require, uniformly in the row index,
\begin{equation}\label{eq:fs-target}
 \max_{i\in[n]}\|\be_i^\top\bR_\bX^{(1)}\|_2
 =\widetilde O_p\left\{\frac{(\log n)^{2+\eps}}{n\sqrt{\rho_n}}\right\},
 \quad\forall\eps>0.
\end{equation}
The first-iterate remainder bound established below has numerator
$\log n$, which satisfies this requirement.
\end{remark}

\begin{proposition}[Linearization of the first iterate]\label[proposition]{prop:first-step}
Under \Cref{assump:snr_network,assump:snr_covariates,assump:3}, with
$\tau=\delta\rho_n$ for a fixed $\delta\in(0,1/2)$, and with the
spectral initialization and alignment $\bQ$ of \Cref{lem:X0}, the first
Fisher-scoring iterate admits the representation
\begin{equation}\label{eq:fs-claim}
\widehat\bX^{(1)}\bQ-\bX_0=\calL(\bE,\bW)+\bR_\bX^{(1)},
\end{equation}
where $\calL$ is the deterministic linear map of the two noise matrices
given rowwise by
\begin{equation}\label{eq:fs-linearmap}
\be_i^\top\calL(\bE,\bW)
=\left(\sum_{j=1}^n\bds\alpha_{nij}E_{ij}+\sum_{l=1}^p\bds\beta_{nil}W_{li}\right)^{\!\top},
\end{equation}
with deterministic coefficients
\begin{equation}\label{eq:fs-coefficients}
\bds\alpha_{nij}=\frac{1}{n+p}\,\bG_{in}^{-1}\psi'(\bx_{0i}^\top\by_{0j})\,\by_{0j},
\qquad
\bds\beta_{nil}=\frac{1}{n+p}\,\bG_{in}^{-1}\bz_{0l}.
\end{equation}
These satisfy
\[
\max_{i,j\in[n]}\|\bds\alpha_{nij}\|_2=O\{(n\rho_n^{1/2})^{-1}\},
\qquad
\max_{i\in[n],\,l\in[p]}\|\bds\beta_{nil}\|_2=O(1/n),
\]
and for each $i\in[n]$ the remainder obeys
\begin{equation}\label{eq:fs-stronger}
\|\be_i^\top\bR_\bX^{(1)}\|_2=\widetilde O_p\!\left(\frac{\log n}{n\rho_n^{1/2}}\right).
\end{equation}
In particular \eqref{eq:fs-target} holds, so $\widehat\bX^{(1)}$ is
approximately linearized in the sense of \Cref{def:linearization}.
\end{proposition}

\begin{proof}
Write $\bds\delta_i^{(0)}\coloneqq\bQ^\top\widehat\bx_i^{(0)}-\bx_{0i}$ for the
aligned initial error, and let
\begin{align}
\widehat\bs_{in}^{(0)}
=\frac1{n+p}\Biggl\{&
\sum_{j=1}^n\left(A_{ij}-(\widehat\bx_i^{(0)})^\top\widehat\by_j^{(0)}\right)
\psi'\!\left((\widehat\bx_i^{(0)})^\top\widehat\by_j^{(0)}\right)\bQ^\top\widehat\by_j^{(0)}
\nonumber\\
&
+\sum_{l=1}^p\left(B_{li}-(\widehat\bx_i^{(0)})^\top\widehat\bz_l^{(0)}\right)
\bQ^\top\widehat\bz_l^{(0)}\Biggr\},\label{eq:fs-estscore}\\
\widehat\bG_{in}^{(0)}
=\frac1{n+p}\Biggl\{&
\sum_{j=1}^n\psi'\!\left((\widehat\bx_i^{(0)})^\top\widehat\by_j^{(0)}\right)
(\bQ^\top\widehat\by_j^{(0)})(\bQ^\top\widehat\by_j^{(0)})^\top +\sum_{l=1}^p(\bQ^\top\widehat\bz_l^{(0)})(\bQ^\top\widehat\bz_l^{(0)})^\top\Biggr\}
\label{eq:fs-estG}
\end{align}
be the aligned normalized score and information at the initializer, the
nuisance matrices being held fixed in the row score. These are exactly
the quantities entering the Fisher-scoring update of \Cref{sec:notation},
so that
\begin{equation}\label{eq:fs-update}
\bQ^\top\widehat\bx_i^{(1)}-\bx_{0i}
=\bds\delta_i^{(0)}+(\widehat\bG_{in}^{(0)})^{-1}\widehat\bs_{in}^{(0)} .
\end{equation}
As in \eqref{lem:taylor}, set
\begin{equation}\label{eq:fs-remnotation}
\br_{i1}^{(0)}=\widehat\bs_{in}^{(0)}-\bs_{in},
\qquad
\bR_{i2}^{(0)}=(\widehat\bG_{in}^{(0)})^{-1}-\bG_{in}^{-1},
\end{equation}
the score difference and the inverse-information difference analyzed in
\Cref{sec:taylor,sec:Ri2}. Substituting
\eqref{eq:fs-remnotation} into \eqref{eq:fs-update} gives the exact
identity
\begin{equation}\label{eq:fs-exact}
\bQ^\top\widehat\bx_i^{(1)}-\bx_{0i}
=\bds\delta_i^{(0)}+(\bG_{in}^{-1}+\bR_{i2}^{(0)})(\bs_{in}+\br_{i1}^{(0)})
=\bG_{in}^{-1}\bs_{in}+\br_{\bx_i}^{(1)},
\end{equation}
where
\begin{equation}\label{eq:fs-exactremainder}
\br_{\bx_i}^{(1)}
=\bG_{in}^{-1}\bigl(\bG_{in}\bds\delta_i^{(0)}+\br_{i1}^{(0)}\bigr)
+\bR_{i2}^{(0)}\br_{i1}^{(0)}
+\bR_{i2}^{(0)}\bs_{in},
\end{equation}
the last term being $\br_{i3}^{(0)}$ in the notation of
\eqref{lem:taylor}. We stress that \eqref{eq:fs-exact}--%
\eqref{eq:fs-exactremainder} are algebraic identities; no expansion has
been assumed.

We now bound the three groups in \eqref{eq:fs-exactremainder}, fixing
$i\in[n]$ and taking $m_n=n$ throughout.

\emph{(i) The cancellation term.} Term \eqref{eq:term2} of the
decomposition is exactly $-\bG_{in}\bds\delta_i^{(0)}$, so
$\bG_{in}\bds\delta_i^{(0)}+\br_{i1}^{(0)}$ is the sum of the remaining
seven terms \eqref{eq:term1} and \eqref{eq:term3}--\eqref{eq:term8}.
By \Cref{lem:term1,lem:term3,lem:term4,lem:term5,lem:term6,lem:term78} and the
triangle inequality,
\begin{equation}\label{eq:fs-cancellation}
\bigl\|\bG_{in}\bds\delta_i^{(0)}+\br_{i1}^{(0)}\bigr\|_2
=\widetilde O_p\!\left(\frac{\log n+(1\vee\rho_n\log n)^{1/2}}{n\rho_n^{1/2}}\right)
=\widetilde O_p\!\left(\frac{\log n}{n\rho_n^{1/2}}\right),
\end{equation}
the simplification using $0<\rho_n\le1$. This is the essential
cancellation: the initial error and the score difference are controlled
\emph{jointly}, not separately.

\emph{(ii) The score difference alone.} By \Cref{lem:term2} and
\Cref{lem:X0}, $\|\bG_{in}\bds\delta_i^{(0)}\|_2=\widetilde O_p\{(\log n)^{1/2}/\sqrt n\}$,
so \eqref{eq:fs-cancellation} and \eqref{eq:res-absorb} give
\begin{equation}\label{eq:fs-r1bound}
\|\br_{i1}^{(0)}\|_2
=\widetilde O_p\!\left(\sqrt{\frac{\log n}{n}}+\frac{\log n}{n\rho_n^{1/2}}\right)
=\widetilde O_p\!\left(\sqrt{\frac{\log n}{n}}\right).
\end{equation}

\emph{(iii) The inverse-information difference.} \Cref{lem:Ri2} at
$t=0$ with $m_n=n$ gives
\begin{equation}\label{eq:fs-inversebound}
\|\bR_{i2}^{(0)}\|_2=\widetilde O_p\!\left(\sqrt{\frac{\log n}{n\rho_n}}\right),
\qquad \|\bG_{in}^{-1}\|_2=O(1).
\end{equation}
The inverse bounds and the update \eqref{eq:fs-update} are understood on
the high-probability invertibility events established in the proof of
\Cref{lem:Ri2}; assigning arbitrary values to the estimator on their
complements does not affect any of the stated rates. For completeness,
the lower bound on $\bG_{in}$ is immediate from
\eqref{eq:fs-populationG}: on the oracle range
$0\le\bx_{0i}^\top\by_{0j}\le\rho_n$ the extended logarithms satisfy
$\psi'(\bx_{0i}^\top\by_{0j})\ge c\rho_n^{-1}$, whence
\[
\bG_{in}\succeq\frac{c}{(n+p)\rho_n}\,\bY_0^\top\bY_0,
\qquad
\lambda_{\min}(\bY_0^\top\bY_0)=\sigma_d(\bX_0)^2\gtrsim n\rho_n,
\]
so $\lambda_d(\bG_{in})=\Omega(1)$ by \Cref{assump:snr_network}.

Combining \Cref{lem:popscore} with
\eqref{eq:fs-cancellation}--\eqref{eq:fs-inversebound} in
\eqref{eq:fs-exactremainder},
\begin{align}
\|\br_{\bx_i}^{(1)}\|_2
&\le\|\bG_{in}^{-1}\|_2\bigl\|\bG_{in}\bds\delta_i^{(0)}+\br_{i1}^{(0)}\bigr\|_2
+\|\bR_{i2}^{(0)}\|_2\bigl(\|\br_{i1}^{(0)}\|_2+\|\bs_{in}\|_2\bigr)\nonumber\\
&=\widetilde O_p\!\left(\frac{\log n}{n\rho_n^{1/2}}
+\sqrt{\frac{\log n}{n\rho_n}}\sqrt{\frac{\log n}{n}}\right)
=\widetilde O_p\!\left(\frac{\log n}{n\rho_n^{1/2}}\right).
\label{eq:fs-finalbound}
\end{align}
No independence among these remainder terms is required: each holds with
probability $1-O(n^{-c})$ for every $c>0$, and the corresponding events
may be intersected before the norm inequalities are applied.

It remains to identify the linear part. By \eqref{eq:fs-populations} and
\eqref{eq:fs-coefficients},
\[
\bG_{in}^{-1}\bs_{in}
=\sum_{j=1}^n\bds\alpha_{nij}E_{ij}+\sum_{l=1}^p\bds\beta_{nil}W_{li},
\]
so \eqref{eq:fs-exact} reads
\begin{equation}\label{eq:fs-rowexpansion}
\bQ^\top\widehat\bx_i^{(1)}-\bx_{0i}
=\sum_{j=1}^n\bds\alpha_{nij}E_{ij}+\sum_{l=1}^p\bds\beta_{nil}W_{li}+\br_{\bx_i}^{(1)}.
\end{equation}
The coefficients \eqref{eq:fs-coefficients} are deterministic, so
\eqref{eq:fs-linearmap} defines a deterministic linear map of
$(\bE,\bW)$; stacking the transposes of \eqref{eq:fs-rowexpansion} over
$i\in[n]$, with
$\bR_\bX^{(1)}=[\br_{\bx_1}^{(1)},\ldots,\br_{\bx_n}^{(1)}]^\top$, gives
\eqref{eq:fs-claim}. The coefficient sizes follow from
\eqref{eq:fs-coefbound}, $\|\bG_{in}^{-1}\|_2=O(1)$, and
$\|\bZ_0\|_{2\to\infty}=O(1)$:
\[
\max_{i,j}\|\bds\alpha_{nij}\|_2
\le\|\bG_{in}^{-1}\|_2\max_{i,j}\|\bv_{nij}\|_2=O\{(n\rho_n^{1/2})^{-1}\},
\qquad
\max_{i,l}\|\bds\beta_{nil}\|_2=O(1/n).
\]
Finally $\|\be_i^\top\bR_\bX^{(1)}\|_2=\|\br_{\bx_i}^{(1)}\|_2$, so
\eqref{eq:fs-finalbound} is \eqref{eq:fs-stronger}, and since
$\log n\le(\log n)^{2+\eps}$ for every $\eps>0$ and all large $n$, the
requirement \eqref{eq:fs-target} of \Cref{def:linearization} holds as
well.
\end{proof}

\begin{remark}[Relation to \Cref{thm:OnestepFisher} and to the induction]\label{rem:fs-role}
\Cref{prop:first-step} and \Cref{thm:OnestepFisher} are two readings of the
same computation. \Cref{thm:OnestepFisher} records the one-step stochastic
expansion in score form, $\bQ^\top\widehat\bx_i^{(1)}-\bx_{0i}
=\bG_{in}^{-1}\bs_{in}+\br_{\bx_i}^{(1)}$; \Cref{prop:first-step} makes
the leading term explicit as a deterministic linear functional of
$(\bE,\bW)$ and checks the coefficient-size and rowwise-remainder
requirements of \Cref{def:linearization} one by one. This is what
supplies the base case $t=1$ of the induction in \Cref{sec:iterative}:
together with \Cref{res:linearization-rate}, it yields
$\|\widehat\bX^{(1)}\bQ-\bX_0\|_{2\to\infty}=\widetilde O_p\{(\log n)^{1/2}/\sqrt n\}$
from the linearization alone, with no further appeal to the ASE-specific
decomposition in \Cref{lem:X0}. Note also that the rate
\eqref{eq:fs-stronger} is one full power of $\log n$ sharper than the
$(\log n)^{2+\eps}$ tolerance of \Cref{def:linearization}; the slack is
what absorbs the $\log n$ losses incurred at each subsequent iteration.
\end{remark}

\section{Iterative Stochastic Expansion}\label{sec:iterative}

\paragraph{Use of Assumption~\ref{assump:3}.}
The condition $\rho_n^{-1}=O\{(\log n)^a\}$ for every $a>0$
implies that, for any fixed $b>0$ and every $a>0$,
\[
 \rho_n^{-b}=O\{(\log n)^a\}.
\]
We use this property to absorb inverse powers of $\rho_n$ in the
covariate remainder estimates into arbitrarily small logarithmic
powers. Choosing the intermediate exponents smaller than the desired
$\epsilon$ preserves the final numerator $(\log n)^{2+\epsilon}$.



\begin{proof}[Proof of Theorem \ref{thm:efficiency}]
We follow the least-squares expansion, then assemble the network update.
Fix the desired $\epsilon>0$. In intermediate estimates choose
$0<\eta<\epsilon/8$. Assumption~\ref{assump:3} implies, for every
fixed $a>0$ and $b>0$,
\[
 \rho_n^{-b}=O\{(\log n)^a\}.
\]
Thus losses consisting of a fixed inverse power of $\rho_n$ can be
absorbed into an arbitrarily small logarithmic exponent. Also,
$n\rho_n/(\log n)^K\to\infty$ for every fixed $K>0$.
These facts will be used explicitly below; no positive limit of
$\rho_n$ is assumed.
Result~\ref{res:linearization-rate} gives
\[
 \|\bigl(\widehat{\mathbf X}^{(t)}\mathbf Q-\mathbf X_0\bigr)\|_{2\to\infty}
 =\tOp\{\sqrt{\log n/n}\},\qquad
 \|\bigl(\widehat{\mathbf X}^{(t)}\mathbf Q-\mathbf X_0\bigr)\|_F=\tOp(\sqrt{\log n}).
\]
Since $\sigma_d(\mathbf X_0)\gtrsim\sqrt{n\rho_n}$ and
$\|\bigl(\widehat{\mathbf X}^{(t)}\mathbf Q-\mathbf X_0\bigr)\|_2\le\|\bigl(\widehat{\mathbf X}^{(t)}\mathbf Q-\mathbf X_0\bigr)\|_F
=o_p(\sqrt{n\rho_n})$ with polynomial tails, singular-value perturbation
implies
\[
 \left\|\{(\widehat{\mathbf X}^{(t)}\mathbf Q)^\top
 (\widehat{\mathbf X}^{(t)}\mathbf Q)\}^{-1}\right\|_2
 =\tOp\{(n\rho_n)^{-1}\}.
\]

\paragraph{Weighted linearization terms.}
The linearization represents row $j$ of $\bigl(\widehat{\mathbf X}^{(t)}\mathbf Q-\mathbf X_0\bigr)$ as
$\sum_k\boldsymbol\alpha_{njk}E_{jk}
 +\sum_a\boldsymbol\beta_{nja}W_{aj}
 +\mathbf r_{\mathbf x_j}^{(t)}$, where the deterministic coefficients
have norms $O((n\sqrt{\rho_n})^{-1})$ and $O(n^{-1})$, respectively.
We first obtain
\begin{equation}\label{eq:51-weighted-signal}
 \|\mathbf X_0^\top\bigl(\widehat{\mathbf X}^{(t)}\mathbf Q-\mathbf X_0\bigr)\|_2
 =\tOp\{(\log n)^{2+\eta}\},\qquad \eta>0.
\end{equation}
The graph part of this product is
\[
 \sum_{j=1}^n\sum_{k=1}^n
 E_{jk}\mathbf x_{0j}\boldsymbol\alpha_{njk}^\top
 =\sum_{j\le k}E_{jk}\mathbf C_{jk},
\]
where we combine the coefficients of the symmetric noise entries:
\[
 \mathbf C_{jk}=
 \begin{cases}
 \mathbf x_{0j}\boldsymbol\alpha_{njk}^\top
 +\mathbf x_{0k}\boldsymbol\alpha_{nkj}^\top,&j<k,\\
 \mathbf x_{0j}\boldsymbol\alpha_{njj}^\top,&j=k.
 \end{cases}
\]
Each outer product satisfies
\[
 \|\mathbf x_{0j}\boldsymbol\alpha_{njk}^\top\|_F
 =\|\mathbf x_{0j}\|_2\|\boldsymbol\alpha_{njk}\|_2
 \lesssim\sqrt{\rho_n}\frac1{n\sqrt{\rho_n}}=\frac1n.
\]
Thus $\max_{j\le k}\|\mathbf C_{jk}\|_F\lesssim n^{-1}$.
To apply the first inequality of \citep[Lemma~S2.1]{xie2024entrywise}, enumerate the
$N=n(n-1)/2$ independent off-diagonal entries $E_{jk}$ with $j<k$ and form
$\mathbf V\in\mathbb R^{N\times d^2}$ whose corresponding rows are
$\operatorname{vec}(\mathbf C_{jk})^\top$. Its coefficient norms obey
\[
 \|\mathbf V\|_{2\to\infty}
 =\max_{j<k}\|\mathbf C_{jk}\|_F\lesssim\frac1n,
 \qquad
 \|\mathbf V\|_F^2
 =\sum_{j<k}\|\mathbf C_{jk}\|_F^2
 \lesssim\frac{N}{n^2}\lesssim1.
\]
The random variables $E_{jk}$ are centered, satisfy $|E_{jk}|\le1$,
and have variance at most $\rho_n$. \citep[Lemma~S2.1]{xie2024entrywise} therefore applies with
$\sigma=1$ and $\rho=\rho_n$, and gives, for $s\ge1$,
\[
 \mathbb P\left(
 \left\|\sum_{j<k}E_{jk}\mathbf C_{jk}\right\|_F
 >3s^2\|\mathbf V\|_{2\to\infty}
 +\sqrt6\sqrt{\rho_n}s\|\mathbf V\|_F
 \right)\le28e^{-s^2}.
\]
Here vectorization preserves the Frobenius norm. For any prescribed
$c>0$, take $s=\sqrt{(c+1)\log n}$. The threshold is at most
\[
 C_c\left\{\frac{\log n}{n}+\sqrt{\rho_n\log n}\right\},
\]
and the failure probability is at most $28n^{-(c+1)}$.
The full weighted signal sum also includes
\[
 -\sum_jP_{jj}\mathbf C_{jj},\qquad
 \left\|\sum_jP_{jj}\mathbf C_{jj}\right\|_F
 \le\sum_j\rho_n(C/n)\lesssim\rho_n.
\]
Since $\rho_n\le1$, this is absorbed by
$\sqrt{\rho_n\log n}$ for all sufficiently large $n$.
Since the operator norm is bounded by the Frobenius norm, we obtain
\[
 \left\|\sum_{j,k}E_{jk}\mathbf x_{0j}\boldsymbol\alpha_{njk}^\top\right\|_2
 =\tOp\left\{\sqrt{\rho_n\log n}+\frac{\log n}{n}\right\}.
\]
For the Gaussian part, the coefficient matrices are deterministic and
satisfy
\[
 \|\mathbf x_{0j}\boldsymbol\beta_{njl}^\top\|_F^2
 =\|\mathbf x_{0j}\|_2^2\|\boldsymbol\beta_{njl}\|_2^2
 \lesssim\frac{\rho_n}{n^2}.
\]
Since the $W_{lj}$ are independent standard Gaussians, centering and
independence eliminate the cross terms, giving
\begin{align*}
 \mathbb E\left\|\sum_{j=1}^n\sum_{l=1}^p
 W_{lj}\mathbf x_{0j}\boldsymbol\beta_{njl}^\top\right\|_F^2
 &=\sum_{j=1}^n\sum_{l=1}^p
 \|\mathbf x_{0j}\boldsymbol\beta_{njl}^\top\|_F^2\\
 &\lesssim np\frac{\rho_n}{n^2}
 =\frac{p\rho_n}{n}\lesssim\rho_n.
\end{align*}
To obtain a polynomial-tail bound, we also use the Gaussian distribution.
For each $a,b\in[d]$, the corresponding matrix entry
\[
 \sum_{j=1}^n\sum_{l=1}^p
 W_{lj}(\mathbf x_{0j})_a(\boldsymbol\beta_{njl})_b
\]
is centered Gaussian with variance at most $C\rho_n$, because its
variance is bounded by the sum of the squared Frobenius norms above.
Its absolute value therefore exceeds $\sqrt{2C\rho_n u}$ with
probability at most $2e^{-u}$. A union bound over the $d^2$ entries and
$\|\mathbf M\|_2\le\|\mathbf M\|_F
\le d\max_{a,b}|M_{ab}|$ show that, for $u\ge1$,
\[
 \mathbb P\left(
 \left\|\sum_{j=1}^n\sum_{l=1}^p
 W_{lj}\mathbf x_{0j}\boldsymbol\beta_{njl}^\top\right\|_2
 >C_d\sqrt{\rho_n u}\right)\le2d^2e^{-u}.
\]
Taking $u=(c+1)\log n$ for any prescribed $c>0$, and using that $d$
is fixed, yields
\[
 \left\|\sum_{j=1}^n\sum_{l=1}^p
 W_{lj}\mathbf x_{0j}\boldsymbol\beta_{njl}^\top\right\|_2
 =\tOp\{\sqrt{\rho_n\log n}\}.
\]
Finally,
\[
 \|\mathbf X_0^\top\mathbf R_{\mathbf X}^{(t)}\|_2
 \le\sum_j\|\mathbf x_{0j}\|_2\|\mathbf r_{\mathbf x_j}^{(t)}\|_2
 \le n\sqrt{\rho_n}\|\mathbf R_{\mathbf X}^{(t)}\|_{2\to\infty}
 =\tOp\{(\log n)^{2+\eta}\}.
\]
These three bounds prove \eqref{eq:51-weighted-signal}.

We also need
\begin{equation}\label{eq:51-weighted-noise}
 \|\mathbf W\bigl(\widehat{\mathbf X}^{(t)}\mathbf Q-\mathbf X_0\bigr)\|_{2\to\infty}
 =\tOp\{(\log n)^{2+\eta}\}.
\end{equation}
Here is the concentration argument for the linear terms; it does not
assume independence between $\mathbf W$ and $\bigl(\widehat{\mathbf X}^{(t)}\mathbf Q-\mathbf X_0\bigr)$.
For the graph linear term, first split each row as
$\sum_{a\ne j}\boldsymbol\alpha_{nja}E_{ja}
-P_{jj}\boldsymbol\alpha_{njj}$. The diagonal correction has maximum
row norm $O(\sqrt{\rho_n}/n)$ and Frobenius norm
$O(\sqrt{\rho_n/n})$. Bernstein's inequality for the off-diagonal
part and a union bound, followed by addition of this correction, show that its maximum row norm is $\tOp\{\sqrt{\log n/n}\}$, so its
Frobenius norm is $\tOp(\sqrt{\log n})$.
Conditional on $\mathbf E$, each row of $\mathbf W$ applied to this
term is Gaussian with total second moment bounded by this squared
Frobenius norm. Gaussian concentration and a union bound over $p=O(n)$
therefore give $\tOp(\log n)$.
For the covariate linear term, fix a row $l\in[p]$ and a coordinate
$q\in[d]$, and write $\beta_{nja}^{[q]}=(\boldsymbol\beta_{nja})_q$.
The corresponding scalar sum is
\[
 \sum_{j=1}^n\sum_{a=1}^p W_{lj}W_{aj}\beta_{nja}^{[q]},
 \qquad |\beta_{nja}^{[q]}|\le C/n.
\]
Split this sum according to whether the two Gaussian variables are
from the same covariate row:
\[
 \sum_{j,a}W_{lj}W_{aj}\beta_{nja}^{[q]}
 =\sum_j\beta_{njl}^{[q]}W_{lj}^2
 +\sum_j\sum_{a\ne l}\beta_{nja}^{[q]}W_{lj}W_{aj}.
\]
Gaussian norm concentration and a union bound over $p=O(n)$ imply
\[
 \max_{l\in[p]}\sum_{j=1}^n W_{lj}^2=\tOp(n).
\]
For example, for a standard Gaussian vector in $\mathbb R^n$,
its norm exceeds $\sqrt n+\sqrt{2u}$ with probability at most $e^{-u}$.
Taking $u=(c+2)\log n$ and then a union bound proves the displayed
statement for each prescribed $c>0$.
Consequently, the diagonal contribution satisfies
\[
 \left|\sum_j\beta_{njl}^{[q]}W_{lj}^2\right|
 \le\frac Cn\sum_jW_{lj}^2=\tOp(1),
\]
uniformly in $l$ and $q$.
For the off-diagonal contribution, condition on
$(W_{l1},\ldots,W_{ln})$. The remaining $W_{aj}$, $a\ne l$,
are independent standard Gaussians, so the conditional sum is centered
Gaussian with variance
\[
 \sum_j\sum_{a\ne l}(\beta_{nja}^{[q]})^2W_{lj}^2
 \le\frac{C^2p}{n^2}\sum_jW_{lj}^2.
\]
On the event $\sum_jW_{lj}^2\le K_cn$, this variance is at most
$C_c p/n\lesssim C_c$. Conditional Gaussian concentration with
$u=(c+2)\log n$, followed by integration over the conditioned row
and addition of the row-norm event's failure probability, therefore
bounds the off-diagonal sum by $C_c\sqrt{\log n}$ with polynomially
small failure probability. A union bound over $p=O(n)$ rows and the
fixed $d$ coordinates preserves arbitrarily high polynomial probability.
This yields, in particular, the bound needed below:
\[
 \max_{l\in[p]}\left\|\sum_{j=1}^n\sum_{a=1}^p
 W_{lj}W_{aj}\boldsymbol\beta_{nja}\right\|_2
 =\tOp(\log n).
\]
The remaining term is $\mathbf W\mathbf R_{\mathbf X}^{(t)}$.
No independence is assumed. For a desired intermediate exponent
$\eta>0$, apply approximate linearization with exponent $\eta/2$:
\begin{align*}
 \|\mathbf W\mathbf R_{\mathbf X}^{(t)}\|_{2\to\infty}
 &\le\|\mathbf W\|_{2\to\infty}\sqrt n
       \|\mathbf R_{\mathbf X}^{(t)}\|_{2\to\infty}\\
 &=\tOp\left\{\frac{(\log n)^{2+\eta/2}}{\sqrt{\rho_n}}\right\}
 =\tOp\{(\log n)^{2+\eta}\}.
\end{align*}
The last step uses $\rho_n^{-1/2}=O\{(\log n)^{\eta/2}\}$,
as implied by Assumption~\ref{assump:3}; no monotonicity of
$\rho_n$ is required.

Combining the graph linear,
covariate linear, and remainder bounds proves
\eqref{eq:51-weighted-noise}.

\paragraph{The inverse cross-product difference.}
Expanding the cross-product gives
\begin{align*}
 &(\widehat{\mathbf X}^{(t)}\mathbf Q)^\top
 (\widehat{\mathbf X}^{(t)}\mathbf Q)-\mathbf X_0^\top\mathbf X_0\\
 &\qquad=\mathbf X_0^\top\bigl(\widehat{\mathbf X}^{(t)}\mathbf Q-\mathbf X_0\bigr)
 +(\bigl(\widehat{\mathbf X}^{(t)}\mathbf Q-\mathbf X_0\bigr))^\top\mathbf X_0
 +(\bigl(\widehat{\mathbf X}^{(t)}\mathbf Q-\mathbf X_0\bigr))^\top\bigl(\widehat{\mathbf X}^{(t)}\mathbf Q-\mathbf X_0\bigr).
\end{align*}
By \eqref{eq:51-weighted-signal} and
$\|\bigl(\widehat{\mathbf X}^{(t)}\mathbf Q-\mathbf X_0\bigr)\|_2^2\le\|\bigl(\widehat{\mathbf X}^{(t)}\mathbf Q-\mathbf X_0\bigr)\|_F^2=\tOp(\log n)$,
its norm is $\tOp\{(\log n)^{2+\eta}\}$.
For invertible matrices, the exact identity
$\mathbf A^{-1}-\mathbf B^{-1}
=\mathbf A^{-1}(\mathbf B-\mathbf A)\mathbf B^{-1}$ follows by
expanding the product on the right. Applied to the two cross-product
matrices, it gives
\begin{align*}
 \{(\widehat{\mathbf X}^{(t)}\mathbf Q)^\top
 (\widehat{\mathbf X}^{(t)}\mathbf Q)\}^{-1}
 -(\mathbf X_0^\top\mathbf X_0)^{-1}
 &=\{(\widehat{\mathbf X}^{(t)}\mathbf Q)^\top
 (\widehat{\mathbf X}^{(t)}\mathbf Q)\}^{-1}\left\{\mathbf X_0^\top\mathbf X_0
 -(\widehat{\mathbf X}^{(t)}\mathbf Q)^\top
 (\widehat{\mathbf X}^{(t)}\mathbf Q)\right\}
 (\mathbf X_0^\top\mathbf X_0)^{-1}.
\end{align*}
By submultiplicativity of the operator norm,
\begin{align*}
 &\left\|\{(\widehat{\mathbf X}^{(t)}\mathbf Q)^\top
 (\widehat{\mathbf X}^{(t)}\mathbf Q)\}^{-1}
 -(\mathbf X_0^\top\mathbf X_0)^{-1}\right\|_2\\
 &\quad\le
 \left\|\{(\widehat{\mathbf X}^{(t)}\mathbf Q)^\top
 (\widehat{\mathbf X}^{(t)}\mathbf Q)\}^{-1}\right\|_2
 \left\|\mathbf X_0^\top\mathbf X_0
 -(\widehat{\mathbf X}^{(t)}\mathbf Q)^\top
 (\widehat{\mathbf X}^{(t)}\mathbf Q)\right\|_2
 \left\|(\mathbf X_0^\top\mathbf X_0)^{-1}\right\|_2.
\end{align*}
The first factor is $\tOp\{(n\rho_n)^{-1}\}$ by the singular-value
bound proved above. The middle factor is
$\tOp\{(\log n)^{2+\eta}\}$ by the preceding expansion, and the last
factor satisfies
\[
 \left\|(\mathbf X_0^\top\mathbf X_0)^{-1}\right\|_2
 =\sigma_d(\mathbf X_0)^{-2}\lesssim(n\rho_n)^{-1}.
\]
Multiplying these three bounds yields
\begin{equation}\label{eq:51-inverse}
 \left\|\{(\widehat{\mathbf X}^{(t)}\mathbf Q)^\top
 (\widehat{\mathbf X}^{(t)}\mathbf Q)\}^{-1}
 -(\mathbf X_0^\top\mathbf X_0)^{-1}\right\|_2
 =\tOp\left\{\frac{(\log n)^{2+\eta}}{(n\rho_n)^2}\right\}.
\end{equation}
This bound retains the contribution of the linearization remainder;
it need not have the smaller numerator $\log n$.

\paragraph{The covariate refresh.}
Substitute $\mathbf B=\mathbf Z_0\mathbf X_0^\top+\mathbf W$ into the
least-squares refresh and add and subtract the oracle inverse.
The exact remainder is
\begin{align*}
 \mathbf R_{\mathbf Z}^{(t)}
 &=\mathbf Z_0\mathbf X_0^\top\bigl(\widehat{\mathbf X}^{(t)}\mathbf Q-\mathbf X_0\bigr)
 (\mathbf X_0^\top\mathbf X_0)^{-1}
 +\mathbf W\bigl(\widehat{\mathbf X}^{(t)}\mathbf Q-\mathbf X_0\bigr)(\mathbf X_0^\top\mathbf X_0)^{-1}\\
 &\quad+(\mathbf Z_0\mathbf X_0^\top+\mathbf W)
 \widehat{\mathbf X}^{(t)}\mathbf Q
 \left[\{(\widehat{\mathbf X}^{(t)}\mathbf Q)^\top
 (\widehat{\mathbf X}^{(t)}\mathbf Q)\}^{-1}
 -(\mathbf X_0^\top\mathbf X_0)^{-1}\right].
\end{align*}
The first two terms have maximum row norms
$\tOp\{(\log n)^{2+\eta}/(n\rho_n)\}$ by
\eqref{eq:51-weighted-signal}--\eqref{eq:51-weighted-noise}.
For the last term, $\|\mathbf X_0\|_2\le\sqrt{n\rho_n}$ and
$\|\widehat{\mathbf X}^{(t)}\mathbf Q\|_2=\tOp(\sqrt{n\rho_n})$
show that its signal prefactor has maximum row norm $\tOp(n\rho_n)$.
Also, Gaussian concentration gives
$\|\mathbf W\mathbf X_0\|_{2\to\infty}
=\tOp\{\sqrt{n\rho_n\log n}\}$, and thus
\[
 \|\mathbf W\widehat{\mathbf X}^{(t)}\mathbf Q\|_{2\to\infty}
 \le\|\mathbf W\mathbf X_0\|_{2\to\infty}
 +\|\mathbf W\bigl(\widehat{\mathbf X}^{(t)}\mathbf Q-\mathbf X_0\bigr)\|_{2\to\infty}
 =\tOp(n\rho_n).
\]
Here $n\rho_n/(\log n)^K\to\infty$ for every fixed $K$
absorbs both $\sqrt{n\rho_n\log n}$ and $(\log n)^{2+\eta}$.
Multiplying by \eqref{eq:51-inverse} proves the asserted covariate
remainder rate, with $\eta<\epsilon$.
The leading Gaussian term
$\mathbf W\mathbf X_0(\mathbf X_0^\top\mathbf X_0)^{-1}$
has maximum row norm
$\tOp\{\sqrt{\log n/(n\rho_n)}\}$: each row has covariance
$(\mathbf X_0^\top\mathbf X_0)^{-1}$ with norm $O((n\rho_n)^{-1})$,
and we use a union bound over $p=O(n)$ rows.
Moreover,
\[
 \frac{(\log n)^{2+\eta}/(n\rho_n)}
 {\sqrt{\log n/(n\rho_n)}}
 =\frac{(\log n)^{3/2+\eta}}{\sqrt{n\rho_n}}\longrightarrow0.
\]
Hence
\[
 \|\widehat{\mathbf Z}^{(t)}\mathbf Q-\mathbf Z_0\|_{2\to\infty}
 =\tOp\{\sqrt{\log n/(n\rho_n)}\}.
\]

\paragraph{The network update.}
The exact decomposition \eqref{lem:taylor} gives
\begin{align*}
 \mathbf r_{\mathbf x_i}^{(t+1)}
 &=\mathbf G_{in}^{-1}\{
 \mathbf G_{in}(\mathbf Q^\top\widehat{\mathbf x}_i^{(t)}-\mathbf x_{0i})
 +\mathbf r_{i1}^{(t)}\}
 +\mathbf R_{i2}^{(t)}\mathbf r_{i1}^{(t)}
 +\mathbf R_{i2}^{(t)}\frac1{n+p}
 \frac{\partial\ell_{in}}{\partial\mathbf x_i}
 (\mathbf x_{0i};\mathbf Y_0,\mathbf Z_0).
\end{align*}
Lemma~\ref{lem:score-remainder} bounds the first brace by
$\tOp\{(\log n)^{2+\epsilon}/(n\sqrt{\rho_n})\}$ and gives
$\|\mathbf r_{i1}^{(t)}\|_2=\tOp\{\sqrt{\log n/n}\}$.
Lemma~\ref{lem:Ri2} gives $\|\mathbf G_{in}^{-1}\|_2\lesssim1$ and
$\|\mathbf R_{i2}^{(t)}\|_2
=\tOp\{\sqrt{\log n/(n\rho_n)}\}$.
The oracle-score concentration used for the first update gives
\[
 \left\|\frac1{n+p}\frac{\partial\ell_{in}}{\partial\mathbf x_i}
 (\mathbf x_{0i};\mathbf Y_0,\mathbf Z_0)\right\|_2
 =\tOp\{\sqrt{\log n/n}\}.
\]
Both product terms are therefore
$\tOp\{\log n/(n\sqrt{\rho_n})\}$. The triangle inequality proves
the asserted network remainder rate.
All products and sums are bounded on finite intersections of input
events, preserving arbitrarily high polynomial probability.
\paragraph{Consistency for each fixed iterate.}
The base case is supplied by the one-step expansion and the uniform
linearization verification above. The remainder bounds hold uniformly
in the row index; choosing the failure exponent larger before taking
an $n$-row union bound preserves arbitrarily high polynomial probability.
For every fixed finite iteration index, induction and the oracle
score bound consequently give
\[
 \|\widehat\bX^{(t+1)}\bQ-\bX_0\|_{2\to\infty}
 =\widetilde O_p\left\{\sqrt{\frac{\log n}{n}}
       +\frac{(\log n)^{2+\epsilon}}{n\sqrt{\rho_n}}\right\}
 =\widetilde O_p\left\{\sqrt{\frac{\log n}{n}}\right\}
 \longrightarrow_p0.
\]
Indeed, Assumption~\ref{assump:3} implies
$\rho_n^{-1/2}=(\log n)^{o(1)}$, and hence
\[
 \sqrt n\frac{(\log n)^{2+\epsilon}}{n\sqrt{\rho_n}}
 =\frac{(\log n)^{2+\epsilon+o(1)}}{\sqrt n}\longrightarrow0.
\]
Likewise the covariate maximum row error
$\widetilde O_p\{\sqrt{\log n/(n\rho_n)}\}$ tends to zero.
\end{proof}

\bibliography{reference,reference_contextual}
\end{document}